\documentclass[12pt]{article}
\usepackage[utf8]{inputenc}
\usepackage[margin=2cm]{geometry}
\usepackage{epsfig,amsfonts,amssymb,framed,amsmath,xcolor}

\usepackage{epsfig,amsfonts,amssymb}

\usepackage{slashed,upgreek}
\usepackage{hyperref}
\input epsf.sty
\usepackage{hhline}
\usepackage{bm}
\usepackage{array}
\usepackage{cite}

\newcommand{\be}{\begin{eqnarray}\displaystyle}
\newcommand{\ee}{\end{eqnarray}}

\newcommand{\bi}{\begin{itemize}}
\newcommand{\ei}{\end{itemize}}

\newcommand{\bse}{\begin{subequations}}
\newcommand{\ese}{\end{subequations}}

\newcommand{\f}{\frac}

\newcommand{\p}{\partial}

\newcommand{\non}{\nonumber}

\def\ZZZ{{\hbox{ Z\kern-1.6mm Z}}}
\def\RRR{{\hbox{ R\kern-2.4mm R}}}
\def\CCC{{\hbox{ C\kern-2.0mm C}}}
\def\zzz{{\hbox{z\kern-1mm z}}}

\newcommand{\qeq}{{\hbox{=\kern-2.3mm ? \kern.5mm }}}
\renewcommand{\qeq}{=}

\newcommand{\eps}{\epsilon}

\newcommand{\ve}{\varepsilon}

\newcommand{\BB}{{\cal B}}
\newcommand{\DD}{{\cal D}}

\newcommand{\AAA}{{\cal A}}

\newcommand{\BM}{{\bf M}}
\newcommand{\MM}{{\cal M}}

\newcommand{\wt}{\widetilde}
\newcommand{\wh}{\widehat}

\newcommand{\SSS}{{\cal S}}

\newcommand{\refb}[1]{(\ref{#1})}
\newcommand{\sectiono}[1]{\section{#1}\setcounter{equation}{0}}

\def\one{{\hbox{ 1\kern-.8mm l}}}
\def\zero{{\hbox{ 0\kern-1.5mm 0}}}

\def\hat{\wh}

\def\tilde{\wt}

\newcommand{\bea}[1]{\begin{eqnarray}\label{#1} }
\newcommand{\eea}{\end{eqnarray}}

\numberwithin{equation}{section}

\begin{document}

\baselineskip 24pt

\begin{center}
{\Large \bf  Testing Subleading Multiple Soft Graviton Theorem for
CHY Prescription}

\end{center}

\vskip .6cm
\medskip

\vspace*{4.0ex}

\baselineskip=18pt

\centerline{\rm Subhroneel Chakrabarti$^{a}$, Sitender Pratap 
Kashyap$^{a}$,  Biswajit Sahoo$^{a}$, Ashoke Sen$^{a}$, 
Mritunjay Verma$^{a,b}$}

\vspace*{4.0ex}

\centerline{\large \it $^a$Harish-Chandra Research Institute, HBNI}
\centerline{\large \it  Chhatnag Road, Jhusi,
Allahabad 211019, India}



\medskip

\centerline{\large \it  $^b$International Centre for Theoretical Sciences}
\centerline{\large \it  
Hesarghatta,
Bengaluru - 560 089, India.}

\vspace*{1.0ex}
\centerline{\small E-mail:  subhroneelchak, sitenderpratap, biswajitsahoo, sen, mritunjayverma @hri.res.in}

\vspace*{5.0ex}

\centerline{\bf Abstract} \bigskip

In arXiv:1707.06803 
we derived the subleading multiple soft graviton theorem in a generic quantum theory of
gravity for arbitrary number of soft external gravitons and arbitrary number of finite
energy external states carrying arbitrary mass and spin. 
In this paper we verify this explicitly
using the CHY formula for tree level scattering amplitudes of
arbitrary number of gravitons in Einstein gravity. We pay special care to fix the signs of the
amplitudes and resolve an apparent discrepancy between our general results in
arXiv:1707.06803 and previous results on soft graviton theorem from CHY formula.

\vspace*{5.0ex}

\vfill \eject

\baselineskip 18pt

\tableofcontents

\sectiono{Introduction and summary}

Soft graviton theorem relates the scattering amplitude of finite energy
external states and low energy gravitons to a similar amplitude in which the low energy
gravitons are absent \cite{weinberg1,weinberg2,jackiw1,jackiw2,
1103.2981,1404.4091,1404.7749,1405.1015,
1405.1410,1405.2346,
1405.3413,1405.3533,1406.6574,1406.6987,1406.7184,1407.5936,
1407.5982,1408.4179,1410.6406,1412.3699,1504.01364,1507.08882,
1509.07840,1604.00650,1604.03893,1611.02172,1611.07534,1611.03137,
1702.03934,1703.00024,1706.00759,ademollo,shapiro,
1406.4172,1406.5155,1411.6661,1502.05258,1505.05854,1507.08829,
1511.04921,
1512.00803,1601.03457,1604.03355,1610.03481,1503.04816,1504.05558,
1504.05559,1507.00938,1604.02834,1607.02700,1702.02350,1705.06175,1707.06803,
1505.08130}. 
In 3+1 
space-time dimensions they are also related to  
asymptotic symmetries\cite{1312.2229,1401.7026,
1411.5745,1506.05789,1509.01406,1605.09094,1608.00685,1612.08294,1701.00496,
1612.05886,1703.05448}.

Our main interest in this paper will be multiple soft graviton theorem -- the case where multiple
gravitons carry soft momentum. Even though, at the leading order in soft momenta, this was
studied in the original paper of Weinberg\cite{weinberg1,weinberg2}, much of the analysis
at the subleading order has been restricted to the case of two soft 
gravitons\cite{1503.04816,1504.05558,1504.05559,1507.00938,1604.02834,
1607.02700,1702.02350,
1705.06175}. 
In a
previous paper\cite{1707.06803} we gave a general formula 
for amplitudes with multiple soft gravitons
in terms of amplitudes without soft gravitons in any quantum theory of gravity, for
arbitrary mass and spin of the external states, following the general method developed
in \cite{1702.03934,1703.00024,1706.00759}. 
This result may be stated as follows. Let us suppose that we have 
$n$ external finite energy particles carrying  polarizations and
momenta $(\ve_a, p_a)$ for $a=1,\cdots, n$, and $m$ soft gravitons carrying polarizations
and momenta $(\ve_{n+r}, \tau\, k_{n+r})$ for $r=1,\cdots, m$. 
$\ve_a$ for $1\le a\le n$ can be any tensor or spinor representation of the Lorentz group.
Then for small $\tau$ the
amplitude $\BM_{n+m}$ may be expressed in terms of the amplitude $\BM_n$ without the
soft gravitons as follows:\footnote{In \cite{1707.06803} the amplitudes $\BM_n$ included the
momentum conserving delta-functions in their definition. In this paper we shall use the CHY
formula for $\BM_n$ and will not include the momentum conserving delta functions in the
definition of  $\BM_n$. As has been discussed below \refb{edefJ}, the soft graviton
theorem takes the same form as in \cite{1707.06803} for amplitudes without delta functions.
}
\be \label{efullgenintro}
{\bf M}_{n+m} &=& \tau^{-m} \, \Bigg\{ \prod_{r=1}^m
S^{(0)}_{n+r} \Bigg\} \ \BM_n
+ \tau^{-m+1} \,
\sum_{s=1}^m \Bigg\{\prod_{r=1\atop r\ne s}^m \, S^{(0)}_{n+r} \Bigg\}\ 
\left[S^{(1)}_{n+s} \BM_n\right]
\nonumber \\ &&  \hskip -.9in
+ \tau^{-m+1}\, \sum_{r,u=1\atop r<u}^m 
\Bigg\{ \prod_{s=1\atop s\ne r,u}^m \, S^{(0)}_{n+s} \Bigg\} 
\ \Bigg\{\sum_{a=1}^n \ \{p_a\cdot (k_{n+r}+k_{n+u})\}^{-1}\ 
\ \MM(p_a; \ve_{n+r}, k_{n+r}, \ve_{n+u}, k_{n+u}) \Bigg\}  \
\BM_n \,  , \nonumber \\
&& + \, O(\tau^{-m+2})\, .
\ee
where
\be \label{defs0}
S^{(0)}_{n+r} = \sum_{a=1}^n (p_a\cdot k_{n+r})^{-1}
\ \ve_{(n+r),\mu\nu} \, p_a^\mu  p_{a}^\nu  \, ,
\ee
\be \label{edefs1}
[S^{(1)}_{n+s}\BM_n]=
\sum_{a=1}^n (p_a\cdot k_{n+s})^{-1} 
\, \ve_{n+s,\sigma\mu}  \,  k_{n+s, \rho} \, 
p_a^\mu\
\left[ p_a^\sigma 
{\p \over \p p_{a\rho}} - p_a^\rho 
{\p  \over \p p_{a\sigma}}
+   J^{\rho\sigma}\right]
\BM_n\, ,
\ee
\be \label{edefmiintro}
&& \MM(p_a; \ve, k, \ve', k') \nonumber \\
&=&
(p_a\cdot k)^{-1}  (p_a\cdot k')^{-1} 
 \ \Bigg\{-(k\cdot k') \ ( p_a\cdot \ve\cdot p_a )\ 
( p_a\cdot \ve'\cdot p_a)
 + \ 2 \ ( p_a\cdot k' )\ ( p_a\cdot \ve\cdot p_a) \ ( p_a\cdot \ve'\cdot k) 
 \nonumber \\ && \hskip -.3in
   + 2 \ ( p_a\cdot k )\ ( p_a\cdot \ve'\cdot p_a )\ ( p_a\cdot \ve\cdot k')
  - 2  \ ( p_a\cdot k )\ ( p_a\cdot k' ) \ (p_a\cdot \ve\cdot \ve'\cdot p_a)\Bigg\}
 \nonumber \\ && 
 \hskip -.3in +\ (k\cdot k')^{-1}
 \Bigg\{-(k'\cdot\ve\cdot\ve'\cdot p_a)(k'\cdot p_a) 
 -(k\cdot\ve'\cdot\ve\cdot p_a)(k\cdot p_a)  \nonumber\\
&& \hskip -.3in +\ (k'\cdot\ve\cdot\ve'\cdot p_a)(k\cdot p_a)
+(k\cdot\ve'\cdot\ve\cdot p_a)(k'\cdot p_a)  - \ve^{\gamma\delta}\ve_{2\gamma\delta}(k \cdot p_a)( k'\cdot p_a)  \nonumber\\
&&  \hskip -.3in -\ 2(p_a\cdot\ve\cdot k')(p_a\cdot\ve'\cdot k) 
+ (p_a\cdot\ve'\cdot p_a)(k'\cdot\ve\cdot k')  
+ (p_a\cdot\ve\cdot p_a)(k\cdot\ve'\cdot k)\Bigg\}\, ,
\ee
and $J^{\mu\nu}$ is the spin angular momentum which acts on the polarization vectors
$\ve_1,\cdots \ve_n$ in $\BM_n$. 
$J^{\mu\nu}$ is normalized such that acting on a covariant vector 
$\ve_\rho$, it gives
\be
(J^{\mu\nu}\, \ve)_\rho = \delta^\mu_\rho \, \ve^\nu - \delta^\nu_\rho \,\ve^\mu\, .
\ee

Our goal in this paper will be to test this formula using the Cachazo-He-Yuan
(CHY) formula\cite{1306.6575,1307.2199,1309.0885,1409.8256,1412.3479,yuan,supplement} 
for 
tree level 
graviton scattering amplitude in Einstein's theory in any dimensions.
Our final formula, given in eq.\refb{eafinform}, turns out to be in perfect agreement with
\refb{edefmiintro}. This provides a non-trivial test of the multiple soft 
graviton theorem
derived in \cite{1707.06803} and also of the CHY formula for graviton scattering amplitudes
in Einstein theory.

Refs.\cite{1607.02700,1702.02350} derived the double soft graviton 
theorem from CHY scattering 
equation.\footnote{Double and multiple soft theorem for other soft particles from CHY scattering
equations has been studied in
\cite{1503.04816,supplement,1708.05016}.} 
In \cite{1707.06803} we noticed that this differs from our general formula by a sign.
In this paper we show by careful analysis that the sign that we get from the double soft
limit of the CHY scattering formula agrees with that of \cite{1707.06803} provided we 
normalize the amplitude so that the single soft theorem comes with the correct sign. We
also find a few extra terms in the intermediate stages of analysis that were missed in the
analysis of \cite{1607.02700,1702.02350}, but they cancel in the final result.

The rest of the paper is organized as follows. In section \ref{schy} we review the CHY
formula for scattering amplitude of $n$ gravitons. In section \ref{ssingle} we study the
limit of this formula in which one graviton becomes soft, and derive the single soft graviton
theorem. In this section we also fix the normalization of the amplitudes to ensure that the
single soft graviton theorem comes with the conventional sign. 
In section \ref{sdouble} we study the limit of the CHY formula when two gravitons become
soft and show that the result is consistent with the general formula found in \cite{1707.06803}
including the sign. 
One distinguishing feature of our analysis is that we work with general choice of gauge for 
various polarization tensors since the gauge choice that simplified previous analyses is not
available for multiple soft gravitons.\footnote{The only exception is that we do impose 
$\ve_\mu^\mu=0$ gauge condition on the polarization tensor $\ve$ of the soft gravitons. Using
on-shell condition this also implies $k^\mu\ve_{\mu\nu}=0$ where $k$ is the momentum of the 
soft graviton. \label{fo1}}
Finally in section \ref{smultiple} we analyze CHY formula in the limit
in which multiple gravitons become soft, and show that the result agrees with the general
formula derived in \cite{1707.06803}. Appendix \ref{sa} describes the 
counting of different types
of solutions of CHY scattering equations for multiple soft gravitons.

\def\cdot{.}

\sectiono{Cachazo-He-Yuan prescription for amplitude calculation} \label{schy}

For computation of tree level scattering amplitude of 
massless particles 
in arbitrary dimension, Cachazo-He-Yuan gave a prescription known as CHY prescription. 
Our goal in this section will be to describe this formula. However since this makes
use of delta functions with holomorphic arguments, we shall first set up appropriate
conventions to deal with these delta functions.

\subsection{Conventions}

In writing down the CHY formula for scattering amplitudes, one makes use of delta functions
for holomorphic variables defined via the equation
\be \label{ea1}
\int d\sigma\ \delta (f(\sigma)) \ F(\sigma) = \sum_{(\alpha)} (f'(\sigma_{(\alpha)}))^{-1} 
F(\sigma_{(\alpha)})
\ee
where $f$ and $F$ are arbitrary functions and $\sigma_{(\alpha)}$ are the zeroes of the function
$f(\sigma)$. In contrast to the usual delta function, the weight factor is 
$(f'(\sigma_{(\alpha)}))^{-1}$ instead of $(|f'(\sigma_{(\alpha)})|)^{-1}$.  Therefore we can also
represent this as a contour integral
\be \label{ea2}
\sum_{\alpha} \ointop_{\sigma_{(\alpha)}} d\sigma\  (f(\sigma))^{-1} \ F(\sigma)\, ,
\ee
where $\ointop_{\sigma_{(\alpha)}}$ denotes an anti-clockwise contour around $\sigma_{(\alpha)}$,
including the $(2\pi i)^{-1}$ factor.

The absence of absolute value in the right hand side of \eqref{ea1} can some time cause
confusion when we have multiple integration. Consider for example the integral
\be
\int dx\, dy\, \delta(x-y) \,\delta(x+y) \, F(x,y)\, .
\ee
Let us do the $y$ integration first. If we use the second delta function to do this integral
then the result is
\be
\int dx \, \delta(2x) \, F(x, -x) = {1\over 2} F(0,0)\, .
\ee
On the other hand if we use the first delta function to carry out the $y$ integral first then
the result is
\be
- \int dx \delta(2x) F(x, x) = -{1\over 2} F(0,0)\, .
\ee
Therefore the two results do not agree. 

This ambiguity can be resolved if we regard the 
delta functions as grassmann odd objects so that exchanging their positions costs a
sign. We also  regard
the integration measure as 
wedge products so that changing the order of doing the integration
also costs a sign. For a given order of integration and delta functions we shall follow
the convention that the last integration will be done first using the last delta function, and
successive integrations will follow the same order. If we want to use a different delta
function for the last integration, we need to first bring that particular delta function to the
end picking up appropriate minus sign and then carry out the integration using \eqref{ea1}
or \eqref{ea2}.

It is easy to see that with this convention, under a change of variables the product of
delta functions pick up the inverse of the Jacobian without absolute value
\be
\prod_{i=1}^n \delta\left(\sum_j A_{ij}\sigma_j\right) = (\det A)^{-1} \prod_{i=1}^n \delta(\sigma_i)\, ,
\ee
where on both sides the delta functions in the product are arranged from left to right in the
order of increasing $i$.

\subsection{CHY formula}

We are now ready to describe the CHY proposal\cite{1307.2199,1309.0885,1409.8256,
1412.3479,yuan,supplement}. 
According to this
proposal, an $n$-point tree amplitude of massless particles can be derived from a sum over discrete
set of points in the 
moduli space of an $n$-punctured Riemann sphere $\mathfrak{M}_{0,n}$. The 
position of the punctures corresponding to these points in $\mathfrak{M}_{0,n}$
are determined from the solutions of scattering 
equations:
\be
\sum_{\substack{b=1\\b\neq a}}^{n}\dfrac{p_{a}\cdot p_{b}}{\sigma_{a}-\sigma_{b}} = 0 \hspace{15mm} \forall a\in \lbrace 1,2,...,n\rbrace \, , \label{seq}
\ee
where $\lbrace \sigma_{a}\rbrace$ are the holomorphic 
coordinates of the punctures and $\lbrace p_{a}\rbrace$ 
are momenta of massless particles. Using the $SL(2,C)$ 
invariance of $\mathfrak{M}_{0,n}$ we can fix the positions of three punctures and  
use any $(n-3)$ of the $n$ 
scattering equations as constraint relations.
The CHY formula is given as,
\be \label{edefchy}
\BM_{n}=\int \Big[ \prod_{c=1\atop c\neq p,q,r}^n
d\sigma_{c}\Big] (\sigma_{pq}\sigma_{qr}\sigma_{rp})(\sigma_{ij}\sigma_{jk}\sigma_{ki})\Big[ \prod_{a=1\atop a\neq i,j,k}^n\delta \Big( \sum_{b=1\atop
b\neq a}^{n} \dfrac{p_{a}\cdot p_{b}}{\sigma_{a}-\sigma_{b}} \Big) \Big] 
\, I_{n}\, ,
\label{chy}
\ee
where $\sigma_{ab}\equiv\sigma_{a}-\sigma_{b}$.
Here we used SL(2,C) invariance to fix three punctures $p,q,r$, and made use of the
linear dependence of the scattering equations to remove 
three scattering equations for $i,j,k$ particles.
The expression \eqref{chy} is independent of the
choice of $p,q,r$ and $i,j,k$ if 
\be 
I_n \to I_n \, \prod_{a=1}^n \,(c\, \sigma_a + d)^4 \quad \hbox{under} \quad
\sigma_a \to (a\, \sigma_a+b) \, (c\, \sigma_a+d)^{-1}\, .
\ee
Using this requirement, together with
multi-linearity in all the polarization vectors of n-gravitons and gauge invariance of $\BM_{n}$, CHY gave the following
form of the integrand 
$I_{n}$ 
for Einstein gravity\cite{1307.2199}: 
\be
I_{n}= 4 \, (-1)^{n}\,
(\sigma_{s}-\sigma_{t})^{-2}\, \det(\Psi_{st}^{st})\label{integrand}
\ee
where $\Psi$ is a $2n\times 2n$ anti-symmetric matrix defined below and 
$\Psi_{st}^{st}$ is obtained by removing $s$-th and $t$-th row 
from first n rows and removing $s$-th and $t$-th columns from first n columns of 
$\Psi$. The $(-1)^n$ factor was not present in the original formula, but we have
included it in order to get soft graviton theorems with conventional signs for the leading
soft graviton theorem. This factor has also
appeared recently in \cite{1708.04514}. The overall $n$ independent normalization and sign
of the amplitude will be irrelevant for our analysis.
$\Psi$ has the form:
\be
\Psi = 
\left(\begin{array}{cc} A & -C^{T}\\ C & B \end{array}\right)\label{matrix}
\ee 
where $A,B,C$ are $n\times n$ matrices defined as,
\be \label{eadefabc}
A_{ab}=\Bigg\lbrace\begin{array}{cc}\dfrac{p_{a}\cdot p_{b}}{\sigma_{a}-\sigma_{ b}} 
& a\neq b\\ 0 & a=b \end{array}\non\\
 B_{ab}=\Bigg\lbrace\begin{array}{cc}\dfrac{\epsilon_{a}.\epsilon_{b}}{\sigma_{a}-
 \sigma_{ b}} & a\neq b\\ 0 & a=b \end{array}\non\\ \non\\
C_{ab}=\left\lbrace\begin{array}{cc}\dfrac{\epsilon_{a}.p_{b}}{\sigma_{a}-
\sigma_{ b}} & a\neq b\\ -\sum\limits_{c=1\atop c\neq a}^{n}\dfrac{\epsilon_{a}.p_{c}}
{\sigma_{a}-\sigma_{c}} & a=b\, . \label{defABC}
 \end{array} \right .
\ee
Here for each $a$, $\eps_a$ is a space-time vector. In the expression for $I_n$, each
$\eps_a$ appears quadratically. In the final formula, we replace $\eps_{a,\mu} \eps_{a,\nu}$
by $\ve_{a,\mu\nu}$ and identify $\ve_a$ as the polarization tensor for the $a$-th 
graviton.

Since the total number of independent scattering equations is equal to the total number of
independent variables, the integration over $\sigma_a$'s reduce to a sum over discrete set
of points, it can be seen using simple counting that the total number of solutions to the
scattering equation is $(n-3)!$ \cite{1306.6575}.

In the rest of the paper we shall use the  CHY formula \refb{edefchy} 
to study soft limits -- limits
in which one or more external momenta become small. Our general strategy will be as follows.
We first represent the integration over $\sigma_a$'s as contour integrals using \refb{ea2},
taking the contours to lie close to the solutions of the scattering equations. 
In this region we can make appropriate
approximation on the integrand using the soft limit. Once we make this approximation, we now
deform the contours away from the solutions of the scattering equation, possibly through regions
where the original approximation fails, but Cauchy's theorem guarantees that the value of the
integral is unchanged. As we shall see, this gives an effective method for approximating the
CHY formula in the soft limit.

\sectiono{Single soft-graviton theorem} \label{ssingle}

In this section we shall review the derivation of subleading
single soft graviton theorem from CHY prescription given in \cite{1404.7749,1405.3533,
1407.5936,1407.5982}, taking
special care about the signs. 
We shall see that the $(-1)^n$ factor in \eqref{integrand}
is necessary for getting the conventional signs. 
We also
work with general polarizations of external
gravitons 
without making any gauge choice (other than the one given in footnote \ref{fo1})
as in \cite{1407.5982}
-- this will be necessary for generalizing the 
analysis to multiple soft
graviton theorem where special gauge choices of the type used {\it e.g.} 
in \cite{1405.3533}
is not possible. Single soft graviton theorem for general gauge choice has been
derived in \cite{1407.5982}.

\subsection{Statement of the theorem}

We shall begin by stating the
subleading single soft graviton theorem in the standard form so that we know what we
want to prove. We consider the  scattering amplitude of
$(n+1)$ 
massless gravitons with $(n+1)$'th particle momentum
soft and expand the amplitude about the soft momentum. 
Let $p_1,\cdots p_n$ denote the momenta of the
finite energy particles, and $p_{n+1}=\tau\, k_{n+1}$ denote the momentum of the
soft particle. We shall take the soft limit by taking $\tau\to 0$ limit at fixed 
$k_{n+1}$. We also denote by $\ve_i$ the polarization tensor of the $i$-th graviton.
Then the subleading soft theorem stated in \refb{efullgenintro} takes the form  
\be \label{esoftonegrav}
\BM_{n+1}(p_{1},p_{2},...,p_{n},\tau k_{n+1}) = \Big[\frac{1}{\tau}S^{(0)}_{n+1}
+S^{(1)}_{n+1}+O(\tau) \Big]\BM_{n}(p_{1},p_{2},...,p_{n})
\ee
where, 
\be \label{eas0}
S^{(0)}_{n+1}=\sum_{a=1}^{n}\dfrac{\varepsilon_{(n+1),\mu\nu}\, p_{a}^{\mu}\,
p_{a}^{\nu}}{k_{n+1}\cdot p_{a}}\, ,\\
S^{(1)}_{n+1}= \sum_{a=1}^{n}\dfrac{\varepsilon_{(n+1),\mu\nu}\, p_{a}^{\mu}\, k_{(n+1),\rho}
\, \hat{J}_{a}^{\rho\nu}}{k_{n+1}\cdot p_{a}} \, .\label{eas1}
\ee
Here, $\hat{J}_{a}^{\mu\nu}$ is the total angular momentum acting on the $a$-th state,
defined as,
\be
\hat{J}_{a}^{\mu\nu}=p_{a}^{\nu}\dfrac{\p}{\p p_{a,\mu}}-p_{a}^{\mu}\dfrac{\p}{\p p_{a,\nu}}+J_{a}^{\mu\nu}
\ee
where $J_{a}^{\mu\nu}$ is the spin angular momentum of the $a^{th}$ hard graviton,
defined as
\be \label{edefJ}
(J^{\mu\nu} \, \ve)_{\rho\sigma} \equiv (J^{\mu\nu})_{\rho\sigma}^{~~\alpha\beta}\ve_{
\alpha\beta}
= \delta^\mu_\rho \ve^\nu_{~\sigma} - \delta^\nu_\rho \ve^\mu_{~\sigma}
+\delta^\mu_\sigma \ve^{~\nu}_{\rho} - \delta^\nu_\sigma \ve^{~\mu}_{\rho}
\, .
\ee

There is one subtle point that must be mentioned here.
The full amplitude contains a momentum conserving delta function, and the derivation
of the multiple soft graviton theorem in \cite{1707.06803} included this momentum conserving delta
function in the definition of the amplitude. In that case 
the external momenta can be taken as independent and
the derivatives with respect to momenta, present in the definition of $J^{\mu\nu}$, are
well defined. On the other hand, the CHY
formula given in \refb{edefchy} 
does not have this delta function. It is however easy to see that as long
as \refb{edefchy}  satisfies the soft graviton theorem, the full amplitude including the delta
function also satisfies the soft graviton theorem\cite{1404.4091}.   
For this we multiply both sides of \refb{esoftonegrav} by $\delta^{(D)}(p_1+\cdots +
p_n+\tau \, k_{n+1})$, and manipulate the right hand side as follows:
\be \label{err1}
&& \tau^{-1} \, \sum_{a=1}^{n}\dfrac{\varepsilon_{(n+1),\mu\nu}\, p_{a}^{\mu}\,
p_{a}^{\nu}}{k_{n+1}\cdot p_{a}} \, \BM_n \, \left(
\delta^{(D)}(p_1+\cdots +p_n)  + \tau\, 
k_{n+1,\rho} \, {\p\over \p p_{a,\rho}}\, \delta^{(D)}(p_1+\cdots +p_n)\right) \nonumber \\ &&
+ \sum_{a=1}^{n}\dfrac{\varepsilon_{(n+1),\mu\nu}\, p_{a}^{\mu}\, k_{(n+1),\rho}
}{k_{n+1}\cdot p_{a}}  
\left\{ p_{a}^{\nu}\dfrac{\p}{\p p_{a,\rho}}-p_{a}^{\rho}\dfrac{\p}{\p p_{a,\nu}}+J_{a}^{\rho\nu}
\right\} \BM_n \, \delta^{(D)}(p_1+\cdots +p_n)
+ O(\tau) \nonumber \\ &=&
\tau^{-1} \, \sum_{a=1}^{n}\dfrac{\varepsilon_{(n+1),\mu\nu}\, p_{a}^{\mu}\,
p_{a}^{\nu}}{k_{n+1}\cdot p_{a}} \, \BM_n \, 
\delta^{(D)}(p_1+\cdots +p_n)   \nonumber \\ &&
+ \sum_{a=1}^{n}\dfrac{\varepsilon_{(n+1),\mu\nu}\, p_{a}^{\mu}\, k_{(n+1),\rho}
}{k_{n+1}\cdot p_{a}}  
\left\{ p_{a}^{\nu}\dfrac{\p}{\p p_{a,\rho}}-p_{a}^{\rho}\dfrac{\p}{\p p_{a,\nu}}+J_{a}^{\rho\nu}
\right\} \left\{ \BM_n \, \delta^{(D)}(p_1+\cdots +p_n)\right\}
\nonumber \\ &&
+ O(\tau) 
\, ,
\ee
where we have used the fact that
\be
\sum_{a=1}^n \, \dfrac{\varepsilon_{(n+1),\mu\nu}\, p_{a}^{\mu}\, k_{(n+1),\rho}
}{k_{n+1}\cdot p_{a}}  p_{a}^{\rho}\dfrac{\p}{\p p_{a,\nu}} \delta^{(D)}(p_1+\cdots +p_n)
=0\, ,
\ee
as a consequence of the condition $\ve_\mu^\mu=0$ 
that we impose on the 
polarization tensors
of the soft gravitons. Since the right hand side of \refb{err1}
now takes the form of the right hand side of
the soft graviton theorem with the momentum conserving delta function included
in the definition of the amplitude, we see that once we prove eq.\refb{esoftonegrav}
for the CHY
amplitudes, it also holds for the full amplitude including the momentum conserving delta 
functions. A similar result holds also for the multiple soft graviton theorem given in
\refb{efullgenintro}.

While proving \refb{esoftonegrav}, or more generally \refb{efullgenintro} for the CHY amplitude
\refb{edefchy}, we shall treat all 
the external momenta as independent while computing the derivatives
of the amplitude with respect to the external momenta. However while verifying the equality of
the two sides of the equation we shall make use of the conservation of total momenta, since
the final identity we are interested in proving is not \refb{esoftonegrav} or 
\refb{efullgenintro}, but the ones obtained from these by multiplying both sides of the equation
by momentum conserving delta functions.

\subsection{Single soft limit of CHY formula}

We now begin the analysis of the single soft limit of CHY formula.
Let us introduce some compact notations,\footnote{For single soft gravitons we only need
$f^n_{n+1}$, but later we shall make use of the definition of $f^n_{n+r}$ for general $r$.}
\be \label{eadefvar}
f_{a}^{n}\equiv \sum_{b=1 \atop b\neq a}^{n}
\dfrac{p_{a}.p_{b}}{\sigma_{a}-\sigma_{b}} \ \ \hbox{for $1\le a\le n$}, \quad f_{n+r}^{n}\equiv 
\sum_{b=1 }^{n}
\dfrac{k_{n+r}.p_{b}}{\sigma_{n+r}-\sigma_{b}} \, ,\\
\int D\sigma\equiv \int \Big[\prod_{c=1 \atop c\neq p,q,r}^{n}d\sigma_{c} \Big]
(\sigma_{pq}\sigma_{qr}\sigma_{rp})(\sigma_{ij}\sigma_{jk}\sigma_{ki})\, ,
\ee
and rewrite the CHY formula for the $(n+1)$-point scattering amplitude as,
\be
&&\hspace*{-.4in}\BM_{n+1}(p_{1},p_{2},...,p_{n},\tau \, k_{n+1})\non\\
&=& \int D\sigma\int d\sigma_{n+1} \Big[ \prod_{\substack{a=1{}\\a\neq i,j,k}}^{n}\delta \Big( \sum_{\substack{b=1{}\\b\neq
a}}^{n}\frac{p_{a}\cdot p_{b}}{\sigma_{ab}}+\tau \, \dfrac{p_{a}\cdot k_{n+1}}{\sigma_{a(n+1)}} \Big)\Big]\delta\Big( \tau \sum_{b=1}^{n}\frac{k_{n+1}\cdot p_{b}}{\sigma_{(n+1)b}} \Big)\,
 I_{n+1}\, .
 \label{ssoft}
\ee
While carrying out various manipulations below, we shall not be careful about orders
of the delta functions. However it should be understood that while doing the final
integration over the $\sigma_a$'s, the delta functions must be brought to the same order
in which they were arranged initially. With this understanding we expand
the product of delta functions inside the integral \eqref{ssoft}
in a Taylor series in 
$\tau$ as:
\be
&&\hspace*{-.4in}\prod_{a=1\atop a\neq i,j,k}^{n}\delta \Big( \sum_{b=1\atop b\neq
a}^{n}\frac{p_{a}.p_{b}}{\sigma_{ab}}+\tau \,
\dfrac{p_{a}.k_{n+1}}{\sigma_{a(n+1)}} \Big) \non\\[.4cm]
&=& \prod_{a=1\atop a\neq i,j,k}^{n}\delta \Big( \sum_{b=1 \atop b\neq
a}^{n}\frac{p_{a}.p_{b}}{\sigma_{ab}}\Big)+\tau \sum_{l=1}^{n} \dfrac{p_{l}\cdot k_{n+1}}{\sigma_{l(n+1)}}\delta' \Big( \sum_{c=1\atop c\neq
l}^{n}\frac{p_{l}\cdot p_{c}}{\sigma_{lc}}\Big)\prod_{a=1\atop a\neq i,j,k,l}^{n}\delta \Big( \sum_{b=1\atop b\neq
a}^{n}\frac{p_{a}\cdot p_{b}}{\sigma_{ab}}\Big)+O(\tau^{2}) \non\\[.4cm]
&\equiv& \delta^{(0)}+\tau \, \delta^{(1)}+O(\tau^{2})\label{delta}\, .
\ee
The integrand can also be expanded in powers of $\tau$ as,
\be
I_{n+1} &=&I_{n+1}\Bigl|_{\tau =0}\ +\ \tau\, \frac{\p I_{n+1}}{\p \tau}\biggl|_{\tau =0}+O(\tau^{2}) 
\equiv I_{n+1}^{(0)}+\tau\, I_{n+1}^{(1)}+O(\tau^{2})\label{In1}\, .
\ee
Now substituting \eqref{delta} and \eqref{In1} in \eqref{ssoft} and keeping terms up to subleading order in $\tau$, we obtain
\be
\BM_{n+1}(\{p_{a}\},\tau \, k_{n+1}) &=& \frac{1}{\tau} \int 
D\sigma\int d\sigma_{n+1}\Big[\delta^{(0)}+\tau \, \delta^{(1)} \Big]
\, \delta(f_{n+1}^{n})\, 
(I_{n+1}^{(0)}+\tau\, I_{n+1}^{(1)})\non\\
&=& \frac{1}{\tau}\int D\sigma\, \delta^{(0)} \int d\sigma_{n+1}\,
\delta(f_{n+1}^{n})\, I_{n+1}^{(0)}+\int D\sigma\int d\sigma_{n+1}\, 
\delta^{(1)}\, \delta(f_{n+1}^{n})\, I_{n+1}^{(0)}\non \\
&&+ \int D\sigma\, \delta^{(0)}\, \int d\sigma_{n+1}\, \delta(f_{n+1}^{n})\, I_{n+1}^{(1)}\, .
\label{expansion}
\ee
Now we shall analyze the three terms appearing in 
equation \eqref{expansion} one by one.

\subsubsection{First term} \label{esfirst}

Using the expressions \eqref{integrand}, \eqref{matrix} and \eqref{defABC}, we get,
on putting $\tau=0$,
\be
I_{n+1}^{(0)} =- C_{n+1,n+1}^{2}I_{n} = -\Big( \sum_{a=1}^{n}\dfrac{\epsilon_{n+1}\cdot p_{a}}{\sigma_{n+1}-\sigma_{a}} \Big)^{2}I_{n}\, .
\ee
Here, the negative sign comes from the $(-1)^{n}$ factor in the 
definition of $I_{n}$. 
Using this, the first term on the right hand side of \eqref{expansion} can be written as
\be \label{ealead1}
\AAA_1&\equiv& \frac{1}{\tau}\int D\sigma\ \delta^{(0)}\int d\sigma_{n+1}\ \delta(f_{n+1}^{n})\ I_{n+1}^{(0)}\non \\
&=& -\frac{1}{\tau}\int D\sigma \prod_{a=1\atop a\neq i,j,k}^{n}\delta \Big( \sum_{b=1,b\neq
a}^{n}\frac{p_{a}.p_{b}}{\sigma_{ab}}\Big)\int d\sigma_{n+1}\delta\Big( \sum_{b=1}^{n}\frac{k_{n+1}\cdot p_{b}}{\sigma_{(n+1)b}} \Big)\Big( \sum_{a=1}^{n}\dfrac{\epsilon_{n+1}\cdot p_{a}}{\sigma_{n+1}-\sigma_{a}} \Big)^{2}I_{n}\non\\
&=&-\frac{1}{\tau}\int D\sigma \prod_{a=1\atop 
a\neq i,j,k}^{n}\delta \Big( \sum_{b=1\atop 
b\neq a}^{n}\frac{p_{a}.p_{b}}{\sigma_{ab}}\Big) I_{n}\oint d\sigma_{n+1}\Big( \sum\limits_{b=1}^{n}\frac{k_{n+1}\cdot p_{b}}{\sigma_{n+1}-\sigma_{b}} \Big)^{-1}
\Big( \sum_{a=1}^{n}\dfrac{\epsilon_{n+1}\cdot p_{a}}{\sigma_{n+1}-\sigma_{a}} \Big)^{2}\, . \non\\
\ee
In going to the last line, we have used the contour integral representation of the 
delta function as described in \eqref{ea2}, with the 
$\sigma_{n+1}$ integration contour wrapping  
the solutions of the scattering equation of 
the $(n+1)$-th particle. We now deform the contour and do the contour integration 
about the poles of rest of the integrand. 
The deformed contour will wrap the other poles in clockwise direction, and therefore
the residue theorem will have an extra minus sign.
We shall also need to take care of poles at infinity if they exist.  If we denote
collectively by $\{R_i\}$ all these poles, 
the first term becomes
\be
\AAA_1= \frac{1}{\tau}\int D\sigma \prod_{\substack{a=1{}\\a\neq i,j,k}}^{n}\delta \Big( \sum_{\substack{b=1{}\\b\neq a}}^{n}\frac{p_{a}.p_{b}}{\sigma_{ab}}\Big) I_{n}
\oint_{\{R_i\}} d\sigma_{n+1}\dfrac{1}{\Big( \sum\limits_{b=1}^{n}\frac{k_{n+1}.
p_{b}}{\sigma_{n+1}-\sigma_{b}} \Big)}\Big( \sum_{a=1}^{n}\dfrac{\epsilon_{n+1}.p_{a}}
{\sigma_{n+1}-\sigma_{a}} \Big)^{2}\, .\label{ea1exp}
\ee

Using momentum conservation $\sum_{a=1}^n p_a=-\tau \, k_{n+1}$ and the
on-shell conditions $k_{n+1}^2=0$, $\eps_{n+1}.k_{n+1}=0$, we can see that for
large $\sigma_{n+1}$, the
last term in \eqref{ea1exp} falls off as $(\sigma_{n+1})^{-4}$ and the last but one term
grows as $(\sigma_{n+1})^2$. Therefore there is no pole at $\infty$.
The rest of the poles inside the deformed contour are the simple 
poles at $\sigma_{n+1}=\sigma_{a}$ coming from the combination of the last two
terms in \eqref{ea1exp}. 
Computing the residue at each such pole, we obtain the result 
\be \label{eaaa1}
\AAA_1&=&\frac{1}{\tau}\sum_{a=1}^{n}\dfrac{(\epsilon_{n+1}\cdot p_{a})^{2}}{k_{n+1}\cdot p_{a}}\ \BM_{n}\, .
\ee
Replacing $\eps_{n+1,\mu} \eps_{n+1,\nu}$
by $\ve_{n+1,\mu\nu}$ in the above expression, we see that it agrees with the leading soft theorem 
\eqref{esoftonegrav} including the sign.

\subsubsection{Second Term} \label{essecond}
The second term on the right hand side of \eqref{expansion} is given by
\be \label{ea2first}
\AAA_2&\equiv& \int D\sigma\int d\sigma_{n+1}\ \delta^{(1)}\ 
\delta(f_{n+1}^{n})\ I_{n+1}^{(0)}\non \\
&=& -\int D\sigma \int d\sigma_{n+1}\sum_{l=1}^{n} 
\dfrac{p_{l}\cdot k_{n+1}}{\sigma_{l(n+1)}}\delta' \Big( \sum_{\substack{c=1{}\\c\neq
l}}^{n}\frac{p_{l}.p_{c}}{\sigma_{lc}}\Big)
\Big[\prod_{\substack{a=1{}\\a\neq i,j,k,l}}^{n}\delta 
\Big( \sum_{\substack{b=1{}\\b\neq
a}}^{n}\frac{p_{a}.p_{b}}{\sigma_{ab}}\Big)\Big]\non\\
&&\delta\Big( \sum_{b=1}^{n}\frac{k_{n+1}.p_{b}}
{\sigma_{(n+1)b}} \Big)\Big( \sum_{d=1}^{n}\dfrac{\epsilon_{n+1}
\cdot p_{d}}{\sigma_{n+1}-\sigma_{d}} \Big)^{2}I_{n}\, .
\ee
Let us focus on the $\sigma_{n+1}$ integration. The relevant integral is
\be \label{eainteg}
 \int d\sigma_{n+1} \dfrac{p_{l}.k_{n+1}}{\sigma_{l} - \sigma_{n+1}}
\delta\Big( \sum_{b=1}^{n}\frac{k_{n+1}.p_{b}}{\sigma_{n+1} -\sigma_b} \Big)
\Big( \sum_{d=1}^{n}\dfrac{\epsilon_{n+1}\cdot p_{d}}{\sigma_{n+1}-\sigma_{d}} \Big)^{2}\, .
\ee
We can analyze this by deforming the $\sigma_{n+1}$ integration contour
in the same way as before. The only extra complication is that along with the simple poles at $\sigma_{n+1}=\sigma_a$ ($a\not=\ell$),
we now also have a double pole at $\sigma_\ell$ and we have to be a little more careful in
evaluating the residue. This can be done, yielding the result
\be
\sum_{b=1\atop 
b\neq l}^{n}\dfrac{1}{\sigma_{l}-\sigma_{b}}\Big[ 2\big(\epsilon_{n+1}.p_{l}\big)\big(\epsilon_{n+1}.p_{b}\big)-
\big(\epsilon_{n+1}.p_{l}\big)^{2}\dfrac{k_{n+1}.p_{b}}{k_{n+1}.p_{l}} - \big(\epsilon_{n+1}.p_{b}\big)^{2}\dfrac{k_{n+1}.p_{l}}{k_{n+1}.p_{b}}\Big]\, .
\ee
Substituting this into \refb{ea2first} we get
\be \label{ea2second}
\AAA_2&=& -\int D\sigma \  I_{n}\  \sum_{l=1}^{n} \Big[\prod_{a=1\atop
a\neq i,j,k,l}^{n}\delta \Big( \sum_{b=1\atop b\neq
a}^{n}\frac{p_{a}.p_{b}}{\sigma_{ab}}\Big)\Big]
\delta'\Big( \sum_{c=1\atop c\neq
l}^{n}\frac{p_{l}\cdot p_{c}}{\sigma_{lc}}\Big)
\sum_{b=1\atop b\neq l}^{n}\dfrac{1}{\sigma_{l}-\sigma_{b}}\non \\
&&\Big[2\big(\epsilon_{n+1}\cdot p_{l}\big)\big(\epsilon_{n+1}\cdot p_{b}\big)-
\big(\epsilon_{n+1}\cdot p_{l}\big)^{2}\dfrac{k_{n+1}\cdot p_{b}}{k_{n+1}\cdot p_{l}} - \big(\epsilon_{n+1}\cdot p_{b}\big)^{2}\dfrac{k_{n+1}\cdot p_{l}}{k_{n+1}.p_{b}}\Big]\, .\label{second}
\ee

The derivative of the delta function can be obtained by 
using the representation \eqref{ea2} of the delta function. Instead of contour integral
of $1/f$ we have contour integral of $-1/f^2$.
We shall see, however, that we do not need to make use of this explicit representation.

\subsubsection{Third term} \label{sthth}

We now turn to the evaluation of the third term on the right hand side of
\eqref{expansion}
\be\label{eadefa3}
\AAA_{3}=\int D\sigma\,  \delta^{(0)}\int d\sigma_{n+1}\, \delta(f_{n+1}^{n})\, 
I_{n+1}^{(1)}\, .
\ee
We begin by introducing some notations. We denote by $\tilde\Psi$ the matrix $\Psi^{st}_{st}$
defined in \refb{matrix} and the 
paragraph above it for $n$ finite energy external states, with some
fixed choice of $s$ and $t$ in the range $1\le s,t\le n$. We shall denote by $\hat\Psi$
the same matrix for $n+1$ external states with the $(n+1)$-th state describing a soft
graviton carrying momentum $\tau\, k_{n+1}$. 
$\tilde\Psi$ is a $(2n-2)\times (2n-2)$ matrix and $\hat\Psi$ is a $2n\times 2n$ matrix. 
We shall denote by $\tilde P$ the Pfaffian
of $\tilde\Psi$ and by $\hat P$ the Pfaffian of $\hat\Psi$:
\be \label{edefPf}
\tilde P
={1\over 2^{n-1} \, (n-1)!}\ 
\eps^{\alpha_1\cdots \alpha_{2n-2}} \tilde\Psi_{\alpha_1\alpha_2}\cdots \tilde
\Psi_{\alpha_{2n-3}\alpha_{2n-2}}\, , \qquad
\hat P
={1\over 2^n\, n!} \eps^{\alpha_1\cdots \alpha_{2n}} \hat\Psi_{\alpha_1\alpha_2}\cdots 
\hat\Psi_{\alpha_{2n-1} \alpha_{2n}}\, . \non\\
\ee 
Furthermore we shall denote by $\hat P_{\alpha\beta}=-\hat P_{\beta\alpha}$ for
$\alpha<\beta$  the Pfaffian of the matrix obtained
by eliminating from $\hat P$ the $\alpha$-th row and column and $\beta$-th row and
column. Similarly $\hat P_{\alpha\beta\gamma\delta}$ is defined to be 
totally anti-symmetric
in $\alpha,\beta,\gamma,\delta$, which,
for $\alpha<\beta<\gamma<\delta$, is given by
the Pfaffian of the matrix obtained by eliminating the $\alpha,\beta,\gamma,\delta$-th rows
and columns of the matrix $\hat \Psi$.
Similar definitions and properties hold for $\tilde P_{\alpha\beta}$ and
$\tilde P_{\alpha\beta\gamma\delta}$.
Then we have
\be
I_{n+1} = 4 (-1)^{n+1}\,  (\sigma_s -\sigma_t)^{-2} \, \det \hat\Psi = 4 (-1)^{n+1}\,  
(\sigma_s -\sigma_t)^{-2} \, \hat P^2\, ,
\ee
and therefore
\be \label{eadefin1}
I_{n+1}^{(1)} \equiv \dfrac{\p I_{n+1}}{\p \tau}\bigg|_{\tau=0}= 8 (-1)^{n+1}\,  (\sigma_s -\sigma_t)^{-2} \,\, \hat P\, {\p \hat P\over
\p\tau}\bigg|_{\tau=0} \, .
\ee
Now from the definition \refb{edefPf} of the Pfaffian it follows that
\be \label{ederP}
{\p \hat P\over \p\tau} = 
\sum_{\alpha,\beta=1\atop \alpha<\beta, \alpha,\beta\ne s,t}^{2n+2} 
(-1)^{\alpha-\beta+1}\dfrac{\p \hat\Psi_{\alpha\beta}}{\p \tau}
\hat P_{\alpha\beta}\bigg|_{\tau=0}\, .
\ee
Here we have adopted the notation that in $\hat\Psi_{\alpha\beta}$
and $\hat P_{\alpha\beta}$ we shall let the indices run over $1\le \alpha,\beta
\le (n+1)$, $\alpha,\beta\ne s,t$,
and $(n+2) \le \alpha,\beta\le 2n+2$. Therefore the last rows and columns will be called
$(2n+2)$-th rows and columns, even though they are actually $2n$-th rows and columns.
We have also implicitly assumed that the integers 
$s$ and $t$ are consecutive -- otherwise there will be extra minus signs when 
one of $\alpha$ or $\beta$ falls between $s$ and $t$. Our final result will be valid even when
$s$ and $t$ are not consecutive.

Now using the explicit form of the $(n+1)\times(n+1)$ matrices $A, B, C $ given in
\refb{eadefabc} for $n+1$ gravitons, with $p_{n+1}=\tau\, k_{n+1}$, we get,
\be \label{eabcder}
&& \dfrac{\p A_{ab}}{\p \tau}=0 \ \ \hbox{for $1\le a,b\le n$}, \quad
\dfrac{\p A_{a(n+1)}}{\p \tau}= \dfrac{p_{a}.k_{n+1}}{\sigma_{a}-\sigma_{n+1}}
\ \ \hbox{for $1\le a\le n$}, \non\\
&&  \dfrac{\p B_{ab}}{\p \tau}=0 \ \ \hbox{for $1\le a,b\le n+1$}\, , \non\\
&& \dfrac{\p C_{ab}}{\p \tau}= 0 \ \ \hbox{for $1\le a,b\le n$, $a\ne b$}, \quad
\dfrac{\p C_{a(n+1)}}{\p \tau}=\dfrac{\epsilon_{a}.k_{n+1}}{\sigma_{a}-\sigma_{n+1}}
\ \ \hbox{for $1\le a\le n$}\, , \non\\ &&
\dfrac{\p C_{(n+1)b}}{\p \tau}= 0  \ \ \hbox{for $1\le b\le n$}\, , \quad
\dfrac{\p C_{aa}}{\p \tau}=-\dfrac{\epsilon_{a}.k_{n+1}}{\sigma_{a}-\sigma_{n+1}}  
\ \ \hbox{for $1\le a\le n$},
\quad {\p C_{(n+1)(n+1)}\over \p\tau} = 0\, . \non\\
\ee
On the other hand using \refb{ederP} and
\refb{matrix}, we get
\be \label{eapder}
\dfrac{\p \hat P}{\p \tau}&=& 
\sum_{a,b=1\atop a<b, a,b\ne s,t}^{n+1} (-1)^{a-b+1}\dfrac{\p A_{ab}}{\p \tau}\hat{ P}_{ab}
+ \sum_{a,b=1\atop a<b}^{n+1}(-1)^{a-b+1}
\dfrac{\p B_{ab}}{\p \tau}\hat{ P}_{( n+1+a)( n+1+b)} 
\non\\ && 
+\sum_{a,b=1\atop 
a\ne s,t}^{n+1} (-1)^{a- n-b}\dfrac{\p (-C^{T})_{ab}}{\p \tau}\hat{ P}_{a( n+1+b)}
\, .
\ee
Using \refb{eabcder}, \refb{eapder} we can rewrite  \refb{eadefin1} as,
\be \label{eai1ag}
I_{n+1}^{(1)} &=&  8\, (-1)^{n+1} \,  (\sigma_s -\sigma_t)^{-2} \, \hat P \bigg\{
\sum_{a=1\atop a\ne s,t}^n (-1)^{a-n} {\p A_{a(n+1)}\over \p \tau} \hat P_{a(n+1)}\non\\ && + 
\sum_{a=1}^n (-1)^a {\p C_{a (n+1)}\over \p\tau} \hat P_{(n+1)(n+1+a)}
+\sum_{a=1\atop a\ne s,t}^n (-1)^{n+1} {\p C_{aa}\over \p\tau} \hat P_{a(n+1+a)}
\bigg\}\bigg|_{\tau=0}\, . 
\ee

We shall now evaluate the various components $\hat P_{\alpha\beta}$ that appear in
the expression \refb{eai1ag}. 
For this it will be useful to express $\hat\Psi$ in the
matrix form:
\be \label{edefhatpsi}
\hat \Psi = \begin{pmatrix}
A_{ab} &  {\tau p_a.k_{n+1}\over \sigma_a - \sigma_{n+1}} & - C^T_{ad}
& {p_a. \eps_{n+1}\over \sigma_a-\sigma_{n+1}} \cr {\tau k_{n+1}.p_b\over
\sigma_{n+1}-\sigma_b} & 0 & {\tau k_{n+1}.\eps_d\over \sigma_{n+1}-\sigma_d}
& - C_{(n+1)(n+1)} \cr
C_{cb} & {\tau \eps_c.k_{n+1}\over \sigma_c - \sigma_{n+1}} & B_{cd} &
{\eps_c .\eps_{n+1}\over \sigma_c - \sigma_{n+1}} \cr
{\eps_{n+1}.p_b\over \sigma_{n+1}-\sigma_b} & C_{(n+1)(n+1)} & {\eps_{n+1}.\eps_d\over
\sigma_{n+1}-\sigma_d} & 0
\end{pmatrix} \, ,
\ee
where the indices $a,b,c,d$ for the matrices $A,B,C$ run from 1 
to $n$ (with $a,b\ne s,t$), 
and the $(n+1)$-th
components of these matrices have been written down explicitly. Also useful will be the
explicit form of $C_{(n+1)(n+1)}$:
\be
C_{(n+1)(n+1)} = - \sum_{c=1}^n {\eps_{n+1}.p_c\over \sigma_{n+1}-\sigma_c}\, .
\ee

We begin with $\hat P_{a ( n+1)}$. This is the Pfaffian
of the matrix
\refb{edefhatpsi} without the $a$-th and $(n+1)$-th rows and columns. Expanding this 
about the last column we get
\be \label{eapanp1}
\hat P_{a ( n+1)} &=& \sum_{e=1\atop e\ne s,t,a}^n (-1)^{e} 
{\eps_{n+1}.p_e\over \sigma_{e}-\sigma_{n+1}} \hat P_{ae(n+1) (2n+2)}
+  \sum_{d=1}^n (-1)^{n+d}
{\eps_{n+1}.\eps_d\over\sigma_d-  \sigma_{n+1}} 
\hat P_{a(n+1)(n+1+d)(2n+2)} \, . \non\\
\ee
We now notice that for
$\tau=0$, the matrix obtained by eliminating the $(n+1)$-th and 
$(2n+2)$-th rows and columns of $\hat\Psi$,  
is in fact the matrix $\tilde\Psi$. Therefore we can rewrite
\refb{eapanp1} as
\be \label{eapanp1n}
\hat P_{a ( n+1)}|_{\tau=0} &=& \sum_{e=1\atop e\ne s,t,a}^n (-1)^{e+1} 
{\eps_{n+1}.p_e\over \sigma_{n+1}-\sigma_e} \tilde P_{a e}
+  \sum_{d=1}^n (-1)^{n+d+1}
{\eps_{n+1}.\eps_d\over \sigma_{n+1}-\sigma_d} 
\tilde P_{a(n+d)}\, .
\ee
Similarly we have
\be
\hat P_{(n+1)(n+1+a)}|_{\tau=0} &=& \sum_{e=1\atop e\ne s,t}^n (-1)^{e} 
{\eps_{n+1}.p_e\over \sigma_{n+1}-\sigma_e} \tilde P_{e(n+a)}
+  \sum_{d=1\atop d\ne a}^n (-1)^{n-d+1}
{\eps_{n+1}.\eps_d\over \sigma_{n+1}-\sigma_d} 
\tilde P_{(n+a)(n+d)} \, . \non\\
\ee 
To analyze $\hat P_{a (n+1+a)}$ we expand it in the $(n+1)$-th row. For $\tau=0$
only the $(2n+2)$-th column contributes and gives
\be
\hat P_{a (n+1+a)}|_{\tau=0} &=& 
- (-1)^{n-1} C_{(n+1)(n+1)} \hat P_{a (n+1) (n+1+a) (2n+2)}|_{\tau=0}
 \non\\ &=& (-1)^{n-1}
\left(\sum_{c=1}^n {\eps_{n+1}.p_c\over \sigma_{n+1}-\sigma_c}\right) \tilde P_{a (n+a)}\, .
\ee
In order to evaluate the right hand side of \refb{eai1ag}
we also need $\hat P|_{\tau=0}$.
This is evaluated by expanding the Pfaffian of the matrix $\hat\Psi$ given in 
\refb{edefhatpsi} in the
$(n+1)$-th row. The only contribution comes from the last element in the row, giving the 
result
\be
\hat P|_{\tau=0} = -(-1)^{n} C_{(n+1)(n+1)}|_{\tau=0} \, \tilde P
=(-1)^{n} \, \sum_{c=1}^n {\eps_{n+1}.p_c\over \sigma_{n+1}-\sigma_c} \, \tilde P\, . 
\ee 

Substituting these results in \refb{eai1ag} and using \refb{eabcder} we get the result for
$I^{(1)}_{n+1}$ which can be substituted into \refb{eadefa3}.
Using the definition of $f^n_{n+1}$ given in \refb{eadefvar} and the definition 
\refb{ea2} of the delta function, we now get
\be\label{eaaa3}
\AAA_{3}&=& - 8  \,  \int D\sigma \, \delta^{(0)}\ointop d\sigma_{n+1}
\left(\sum_{b=1}^n {k_{n+1}.p_b\over \sigma_{n+1}-\sigma_b}\right)^{-1}
(\sigma_s -\sigma_t)^{-2} \, \tilde P
\, \left(\sum_{g=1}^n {\eps_{n+1}.p_g\over \sigma_{n+1}-\sigma_g}\right)  
\non\\ && \hskip -.5in
\bigg[\sum_{a=1\atop a\ne s,t}^n (-1)^{a-n}
{p_a. k_{n+1}\over \sigma_a -
\sigma_{n+1}} \bigg\{\sum_{e=1\atop e\ne s,t,a}^n (-1)^{e+1} 
{\eps_{n+1}.p_e\over  \sigma_{n+1}-\sigma_e} \tilde P_{a e}
+  \sum_{d=1}^n (-1)^{n+1+d}
{\eps_{n+1}.\eps_d\over \sigma_{n+1}-\sigma_d} 
\tilde P_{a(n+d)}
\bigg\} \non\\
&& \hskip -.5in + \sum_{a=1}^n (-1)^{a}  {\eps_a.k_{n+1}\over \sigma_a -\sigma_{n+1}}
\bigg\{
\sum_{e=1\atop e\ne s,t}^n (-1)^{e} 
{\eps_{n+1}.p_e\over \sigma_{n+1}-\sigma_e} \tilde P_{e(n+a)}
+  \sum_{d=1\atop d\ne a}^n (-1)^{n-d+1}
{\eps_{n+1}.\eps_d\over \sigma_{n+1}-\sigma_d} 
\tilde P_{(n+a)(n+d)}
\bigg\} \non\\ &&  + \sum_{a=1\atop a\ne s,t}^n 
\left({\eps_a.k_{n+1}\over \sigma_{n+1}-\sigma_a}\right)
\left(\sum_{c=1}^n {\eps_{n+1}.p_c\over \sigma_{n+1}-\sigma_c}\right)
\tilde P_{a(n+a)}
\bigg]\, . 
\ee
The $\sigma_{n+1}$ contour winds anti-clockwise around the zeroes of 
$\left(\sum_{b=1}^n {k_{n+1}.p_b\over \sigma_{n+1}-\sigma_b}\right)$. We can now deform
the contour towards infinity, picking up residues from $\sigma_{n+1}=\sigma_a$, -- it is easy
to see that there are no poles at $\infty$. For the coefficients of $\tilde P_{ab}$, 
$\tilde P_{(n+a)(n+b)}$ and $\tilde P_{a(n+b)}$ for $a\ne b$, the integrand has 
single poles at $\sigma_{n+1}=\sigma_a$ and $\sigma_{n+1}=\sigma_b$, and the
result can be computed easily by picking up the residues. Finally for the coefficient of
$\tilde P_{a(n+a)}$ the integrand has a double pole at
$\sigma_{n+1}=\sigma_a$, therefore to evaluate the residue we have to expand it in a Laurent
series around the pole and pick the coefficient of the $(\sigma_{n+1}-\sigma_a)^{-1}$ term.
After some algebra, one finds:
\be \label{efinA3}
&& \AAA_3 = 8  \, \int D\sigma \delta^{(0)}\, (\sigma_s -\sigma_t)^{-2} \, \tilde P \non\\
&& \Bigg[\sum_{a,b=1\atop a\ne b; a,b\ne s,t}^n \, (-1)^{a+b-n}\, (\sigma_a-\sigma_b)^{-1} \,
\left\{ {\eps_{n+1}.p_a\over k_{n+1}.p_a}
- {\eps_{n+1}.p_b\over k_{n+1}.p_b}\right\} \, k_{n+1}.p_a \, \eps_{n+1}.p_b \, \tilde P_{ab}
\non\\
&& + \sum_{a,b=1\atop a\ne b}^n \, (-1)^{a+b-n}\, (\sigma_a-\sigma_b)^{-1} \,
\left\{ {\eps_{n+1}.p_a\over k_{n+1}.p_a}
- {\eps_{n+1}.p_b\over k_{n+1}.p_b}\right\} \, k_{n+1}.\eps_a \, 
\eps_{n+1}.\eps_b \, \tilde P_{(n+a)
(n+b)}\non \\ &&
\hskip -.3in 
+ \sum_{a,b=1\atop a\ne b; a\ne s,t}^n \, (-1)^{a+b}\, (\sigma_a-\sigma_b)^{-1} \, 
 \left\{ {\eps_{n+1}.p_a\over k_{n+1}.p_a}
- {\eps_{n+1}.p_b\over k_{n+1}.p_b}\right\} \, \{k_{n+1}.p_a \, \eps_{n+1}.\eps_b 
- k_{n+1}.\eps_b \, \eps_{n+1}.p_a \}
\, \tilde P_{a(n+b)}\non \\ && +
\sum_{a=1\atop a\ne s,t}^n \sum_{c=1\atop c\ne a}^n (\sigma_a-\sigma_c)^{-1}
 \left\{
{\eps_{n+1}.p_a\over k_{n+1}.p_a}
- {\eps_{n+1}.p_c\over k_{n+1}.p_c}\right\}  \left\{ k_{n+1}.\eps_a \ \eps_{n+1}.p_c
- k_{n+1}.p_c \ \eps_{n+1}.\eps_a 
\right\}\ \tilde P_{a(n+a)}
\Bigg]
\, .\non\\
\ee

\subsection{Comparison with soft graviton theorem} \label{scomp}
The first term $\AAA_1$ given in \refb{eaaa1}
exactly matches the leading soft graviton result given by the first term on the right hand
side of \refb{esoftonegrav} once we make the identification $\ve_{a,\mu\nu} =\eps_{a,\mu} 
\, \eps_{a,\nu}$. 
To check that the second and the third terms
$\AAA_2$ and $\AAA_3$ given respectively in \eqref{ea2second} and \eqref{efinA3}
agree with the known subleading soft graviton theorem, we
first express \eqref{eas1} as
\be 
S^{(1)}_{n+1}=S^{(1)}_{\rm orbital} + S^{(1)}_{\rm spin}
\ee
where  
\be \label{edefspor}
S^{(1)}_{\rm orbital}&=&\sum_{a=1}^{n}\dfrac{\epsilon_{n+1}\cdot p_{a}}
{k_{n+1}.p_{a}}\Big[ \big(\epsilon_{n+1}\cdot p_{a}\big)k_{n+1,\rho}\frac{\p}
{\p p_{a\rho}}-\big(k_{n+1}\cdot p_{a}\big)\epsilon_{(n+1),\nu}\frac{\p}{\p p_{a\nu}} 
\Big] ,
\non\\
S^{(1)}_{\rm spin}&=&\sum_{a=1}^{n} {\eps_{n+1}.p_a\over k_{n+1}\cdot p_{a}}
\eps_{(n+1),\nu} k_{(n+1),\mu} J_{a}^{\mu\nu}\, .
\ee
Therefore we have, from \refb{edefchy}
\be \label{e3.41}
S^{(1)}_{n+1}\, \BM_{n} &=& 
\{S^{(1)}_{\rm orbital} + S^{(1)}_{\rm spin}\}  
\int D\sigma \prod_{a=1\atop a\neq i,j,k}^{n}\delta 
\Big( \sum_{b=1\atop b\neq a}^{n}\frac{p_{a}.p_{b}}{\sigma_{ab}}\Big) I_{n}\non \\
&=& \int D\sigma \sum_{l=1}^{n}\left\{S^{(1)}_{\rm orbital}\, 
\delta\Big( \sum_{c=1\atop c\neq l}^{n}
\frac{p_{l}.p_{c}}{\sigma_{l}-\sigma_{c}} \Big)\right\} 
\prod_{a=1\atop a\neq i,j,k,l}^{n}\delta \Big( \sum_{b=1\atop b\neq a}^{n}
\frac{p_{a}.p_{b}}{\sigma_{ab}}\Big) \, I_{n}\non \\
&& +\int D\sigma \prod_{a=1\atop a\neq i,j,k}^{n}\delta 
\Big( \sum_{b=1\atop b\neq a}^{n}\frac{p_{a}.p_{b}}{\sigma_{ab}}\Big)\; 
\{S^{(1)}_{\rm orbital} + S^{(1)}_{\rm spin}\}
I_{n} \, .\label{ang}
\ee

First let us compute the operation of $S^{(1)}_{\rm orbital}$ on the delta function,
\be
&& S^{(1)}_{\rm orbital}\delta\Big( \sum_{c=1\atop c\neq l}^{n}
\frac{p_{l}\cdot p_{c}}{\sigma_{l}-\sigma_{c}} \Big) \non\\
&=& \delta'\Big( \sum_{c=1\atop c\neq l}^{n}
\frac{p_{l}\cdot p_{c}}{\sigma_{l}-\sigma_{c}} \Big)\sum_{a=1}^{n}\dfrac{(\epsilon_{n+1}.p_{a})}{k_{n+1}.p_{a}}\Big[ \big(\epsilon_{n+1}.p_{a}\big)k_{n+1,\rho}\frac{\p}{\p p_{a\rho}}-\big(k_{n+1}.p_{a}\big)\epsilon_{n+1,\rho}\frac{\p}{\p p_{a\rho}} \Big]\sum_{c=1
\atop c\neq l}^{n}
\frac{p_{l}\cdot p_{c}}{\sigma_{l}-\sigma_{c}}\non\\
&=& \delta'\Big( \sum_{c=1\atop c\neq l}^{n}
\frac{p_{l}\cdot p_{c}}{\sigma_{l}-\sigma_{c}} \Big)
\sum_{a=1\atop a\neq l}^{n}\dfrac{1}{\sigma_{l}-\sigma_{a}}\Big[-2\big(\epsilon_{n+1}.p_{l}\big)
\big(\epsilon_{n+1}.p_{a}\big)+
\big(\epsilon_{n+1}.p_{l}\big)^{2}\dfrac{k_{n+1}.p_{a}}{k_{n+1}.p_{l}} \non \\
&&+ \big(\epsilon_{n+1}.p_{a}\big)^{2}\dfrac{k_{n+1}.p_{l}}{k_{n+1}.p_{a}}\Big]\, .
\ee
With this, the first term on the right hand side of equation \eqref{ang} becomes,
\be \label{e3.42}
&& -\int D\sigma \sum_{l=1}^{n} \prod_{a=1\atop a\neq i,j,k,l}^{n}
\delta \Big( \sum_{b=1\atop b\neq a}^{n}\frac{p_{a}.p_{b}}{\sigma_{ab}}\Big)
\delta'\Big( \sum_{c=1\atop c\neq
l}^{n}\frac{p_{l}\cdot p_{c}}{\sigma_{lc}} \Big)
\sum_{a=1\atop a\neq l}^{n}\dfrac{1}{\sigma_{l}-\sigma_{a}}\non \\
&&\Big[2\big(\epsilon_{n+1}\cdot p_{l}\big)
\big(\epsilon_{n+1}\cdot p_{a}\big)-
\big(\epsilon_{n+1}\cdot p_{l}\big)^{2}\dfrac{k_{n+1}\cdot p_{a}}{k_{n+1}\cdot p_{l}} - \big(\epsilon_{n+1}\cdot p_{a}\big)^{2}\dfrac{k_{n+1}\cdot p_{l}}{k_{n+1}\cdot p_{a}}\Big]\
 I_{n}\, .
\ee
This exactly agrees with $\AAA_2$ given in equation \eqref{second}.

It remains to show that 
the second term on the right hand side 
of equation \eqref{ang} agrees with $\AAA_3$. Using the definition of $I_n$ given in
\refb{integrand} and that $\det (\Psi^{st}_{st})=\det \tilde\Psi = \tilde P^2$, we get
\be
I_{n} = 4 \, (-1)^{n}\,
(\sigma_{s}-\sigma_{t})^{-2}\, \tilde P^2 \, .\label{eadefinnew}
\ee
In order to calculate the action of $S^{(1)}$ on this, it will be useful to 
record how  $S^{(1)}_{\rm spin}$ defined in \refb{edefspor}
acts on the `square root' $\eps$ of the graviton polarization
tensor $\ve$. This is done by identifying the action of $J^{\mu\nu}$ on $\eps$ as
\be \label{edefJeps}
\left(J^{\mu\nu} \, \eps\right)_{\rho} \equiv (J^{\mu\nu})_{\rho}^{~~\alpha}\eps_{\alpha}
= \delta^\mu_\rho \eps^\nu - \delta^\nu_\rho \eps^\mu
\, .
\ee
This allows us to express $S^{(1)}_{n+1}$ as
\be \label{es1half}
S^{(1)}_{n+1}=\sum_{c=1}^n {\eps_{n+1}.p_c\over k_{n+1}.p_c} \ \eps_{n+1}^\mu \ 
k_{n+1}^\nu \ \left( p_{c\mu} {\p\over \p p_{c\nu}} - p_{c\nu} {\p\over \p p_{c\mu}}
+ \eps_{c\mu} {\p\over \p \eps_{c\nu}} - \eps_{c\nu} {\p\over \p \eps_{c\mu}}
\right)\, ,
\ee
and
therefore
\be \label{eas1in}
S^{(1)}_{n+1}\, I_n
= 8 \, (-1)^{n}\,
(\sigma_{s}-\sigma_{t})^{-2}\, \tilde P \ S^{(1)}_{n+1} \tilde P\, .
\ee
Using \refb{es1half} and \refb{eadefabc} we get
\be \label{eas1abc}
&& S^{(1)}_{n+1} A_{ab}= {1\over \sigma_a-\sigma_b}
\left[{\eps_{n+1}.p_a\over k_{n+1}.p_a} -{\eps_{n+1}.p_b\over k_{n+1}.p_b}
\right]  \ \eps_{n+1}^\mu \ 
k_{n+1}^\nu \left[p_{a\mu} p_{b\nu} - p_{a\nu} p_{b\mu}\right] , 
\non\\ && 
S^{(1)}_{n+1} B_{ab}= {1\over \sigma_a-\sigma_b}
\left[{\eps_{n+1}.p_a\over k_{n+1}.p_a} -{\eps_{n+1}.p_b\over k_{n+1}.p_b}
\right]  \ \eps_{n+1}^\mu \ 
k_{n+1}^\nu \left[\eps_{a\mu} \eps_{b\nu} - \eps_{a\nu} \eps_{b\mu}\right] , \non\\ &&  
S^{(1)}_{n+1} C_{ab} =  {1\over \sigma_a-\sigma_b} \ \eps_{n+1}^\mu \ 
k_{n+1}^\nu
\left[{\eps_{n+1}.p_a\over k_{n+1}.p_a} -{\eps_{n+1}.p_b\over k_{n+1}.p_b} \right]
\left[\eps_{a\mu}\, p_{b\nu}- \eps_{a\nu} p_{b\mu} \right]
 \ \ \hbox{for $a\ne b$}\, , \non\\
&& S^{(1)}_{n+1} C_{aa} = - \sum_{b=1\atop b\ne a}^n {1\over \sigma_a-\sigma_b}
\ \eps_{n+1}^\mu \ 
k_{n+1}^\nu
\left[{\eps_{n+1}.p_a\over k_{n+1}.p_a} -{\eps_{n+1}.p_b\over k_{n+1}.p_b} \right]
\left[\eps_{a\mu}\, p_{b\nu}- \eps_{a\nu} p_{b\mu} \right]
\, . \non\\
\ee
We now use a formula similar to \refb{eapder} with $(n+1)$ replaced by $n$ and $\p_\tau$
replaced by $S^{(1)}_{n+1}$:
\be
S^{(1)}_{n+1} \tilde P &=& 
\sum_{a,b=1\atop a<b, a,b\ne s,t}^{n} (-1)^{a-b+1} (S^{(1)}_{n+1} A_{ab}) \, \tilde{ P}_{ab}
+ \sum_{a,b=1\atop a<b}^{n}(-1)^{a-b+1}
(S^{(1)}_{n+1} B_{ab}) \, \tilde{ P}_{( n+a)( n+b)} 
\non\\ && 
+\sum_{a,b=1\atop 
a\ne s,t}^{n} (-1)^{a- n-b+1} (-S^{(1)}_{n+1} C_{ba}) \, \tilde{ P}_{a( n+b)} \, ,
\ee
and use \refb{eas1in}, \refb{eas1abc}  to express
the second term on the right hand side of eq.\eqref{ang} as
\be  \label{eA3equiv}
&& \int D\sigma \prod_{a=1\atop a\neq i,j,k}^{n}\delta 
\Big( \sum_{b=1\atop b\neq a}^{n}\frac{p_{a}.p_{b}}{\sigma_{ab}}\Big) 
\ 8 \, (-1)^{n}\,
(\sigma_{s}-\sigma_{t})^{-2}\, \tilde P \non\\
&& \hskip -.3in \Bigg[\sum_{a,b=1\atop a\ne b, a,b\ne s,t}^{n} (-1)^{a-b} \, \tilde{ P}_{ab}\,
(\sigma_a-\sigma_b)^{-1}
\ \left\{ {\eps_{n+1}.p_a\over k_{n+1}.p_a}-
{\eps_{n+1}.p_b\over k_{n+1}.p_b} \right\} \ \eps_{n+1}.p_b \ 
k_{n+1}.p_{a} 
\non\\ &&  \hskip -.3in 
+ \sum_{a,b=1\atop a\ne b}^{n}(-1)^{a-b}
\, \tilde{ P}_{( n+a)( n+b)} \, (\sigma_a-\sigma_b)^{-1}
\ \left\{ {\eps_{n+1}.p_a\over k_{n+1}.p_a}-
{\eps_{n+1}.p_b\over k_{n+1}.p_b} \right\}
 \ \eps_{n+1}.\eps_b\ 
k_{n+1}.\eps_{a}
\non\\ &&  \hskip -.3in 
+\sum_{a,b=1\atop 
a\ne s,t; a\ne b}^{n} (-1)^{a- n-b} \, \tilde{ P}_{b( n+a)}  {1\over \sigma_a-\sigma_b} 
\ \left\{ {\eps_{n+1}.p_a\over k_{n+1}.p_a}-
{\eps_{n+1}.p_b\over k_{n+1}.p_b} \right\}
\left\{\eps_{n+1}.\eps_a \ k_{n+1}.p_b - \eps_{n+1}.p_b \ k_{n+1}.\eps_a\right\}
 \non\\
&&  \hskip -.3in -\sum_{a=1\atop 
a\ne s,t}^{n} (-1)^{n}  \, \tilde{ P}_{a( n+a)} 
\sum_{b=1\atop b\ne a}^n {1\over \sigma_a-\sigma_b}
\ \left\{ {\eps_{n+1}.p_a\over k_{n+1}.p_a}-
{\eps_{n+1}.p_b\over k_{n+1}.p_b} \right\}
\left\{\eps_{n+1}.\eps_a \ k_{n+1}.p_b - \eps_{n+1}.p_b \ k_{n+1}.\eps_a\right\}\Bigg]\, .\non \\
\ee
This is identical to \refb{efinA3}.

This finishes the 
proof that the CHY prescription for graviton scattering amplitude is consistent with the subleading
single soft graviton theorem.

\sectiono{Double soft graviton theorem} \label{sdouble}

In this section we shall consider the limit of the graviton scattering amplitude when two of the 
gravitons carry soft momenta. We shall reproduce the 
result of \cite{1607.02700,1702.02350} 
with opposite sign and
in general gauge. During this analysis we also find some additional terms that were left out
in the analysis of \cite{1607.02700,1702.02350}, which nevertheless cancel at the end.

\subsection{Double soft limit}
We shall first follow \cite{1503.04816,supplement} to 
analyze the solutions to the scattering equations of 
$(n+2)$ particles with momenta 
$ \lbrace p_{1},p_{2},...,p_{n},\tau \, k_{n+1}, \tau\, k_{n+2}\rbrace$
in the double soft limit  $\tau\rightarrow 0$ keeping $k_{n+1},k_{n+2}$
fixed. 
The scattering equations of first $n$  particles, $(n+1)$-th particle and
$(n+2)$-th particle are given respectively by:
\be
\sum_{b=1\atop 
b\neq a}^{n}\dfrac{p_{a}.p_{b}}{\sigma_{a}-\sigma_{b}} + \dfrac{\tau \,  
p_{a}.k_{n+1}}{\sigma_{a}-\sigma_{n+1}} + \dfrac{\tau \,  p_{a}.k_{n+2}}
{\sigma_{a}-\sigma_{n+2}} &=& 0 
\hspace{8mm} \forall a\in \lbrace 1,2,...,n\rbrace\, , \label{e4.1}\\
\sum_{b=1}^{n}\dfrac{ k_{n+1}.p_{b}}{\sigma_{n+1}-\sigma_{b}}
+\dfrac{\tau\,
k_{n+1}.k_{n+2}}{\sigma_{n+1}-\sigma_{n+2}}&=&0\, , \label{e4.2}\\
\sum_{b=1}^{n} \dfrac{ k_{n+2}.p_{b}}{\sigma_{n+2}-\sigma_{b}} 
+\dfrac{\tau\, k_{n+2}.k_{n+1}}{\sigma_{n+2}-\sigma_{n+1}}&=& 0\label{e4.3}\, ,
\ee
where we have removed overall factors of $\tau$ from the last two equations.
We also have momentum conservation relation,
\be \label{e4.4}
p_{1}^{\mu}+p_{2}^{\mu}+...+p_{n}^{\mu}+\tau \, (k_{n+1}^{\mu}+k_{n+2}^{\mu})=0\, .
\ee
The solution to the scattering equations can be divided into 
two classes\cite{1503.04816,supplement}: 
degenerate
solutions where $\sigma_{n+1}-\sigma_{n+2}\sim \tau$ and 
non-degenerate solutions for which $\sigma_{n+1}-\sigma_{n+2}\sim 1$ in the $\tau\to  0$
limit. For both classes of solutions $\sigma_a$'s for $1\le a\le n$ remain finite distance
away from each other
and from $\sigma_{n+1}$, $\sigma_{n+2}$ as $\tau\to 0$. In order to show that there are
no other types of solutions we shall now show that the total number of degenerate and
non-degenerate solutions add up to $(n-1)!$ -- the actual number of solutions for scattering
equation of $(n+2)$ particles\cite{1307.2199}.

At $\tau=0$, eq.\refb{e4.1} describes the scattering equations for $n$ 
particles which have $(n-3)!$  solutions for 
$\lbrace \sigma_{1},\sigma_{2},...,\sigma_{n} \rbrace$. Generically these solutions are
non-degenerate, with non-coincident $\sigma_a$'s. 
Therefore to count the number of non-degenerate and
degenerate solutions in the $\tau\to 0$ limit,
we have to multiply $(n-3)!$ by the number of solutions for 
$\sigma_{n+1}$ and $\sigma_{n+2}$ for fixed $\sigma_1,\cdots,\sigma_n$.
 
Let us first count the number of non-degenerate solutions. For fixed $\sigma_1,\cdots,\sigma_n$,
we can ignore the last term in 
\refb{e4.2} in $\tau\to 0$ limit and express this 
as a polynomial equation for $\sigma_{n+1}$:
\be
\sum_{a=1}^n \, k_{k+1}. p_a\
\prod_{b=1\atop b\ne a}^n (\sigma_{n+1}-\sigma_b) = 0\, .
\ee
Naively this is of degree $(n-1)$, but momentum conservation \refb{e4.4}
makes the coefficient of the
$(\sigma_{n+1})^{n-1}$ term $-\tau\,  k_{n+1}\cdot k_{n+2}$ which
vanish in the $\tau\to 0$ limit. Therefore in this limit this is a
polynomial equation of degree $(n-2)$ and gives $(n-2)$ solutions for $\sigma_{n+1}$.
Similarly eq.\refb{e4.3} gives $(n-2)$ solutions for $\sigma_{n+2}$. Therefore the total number
of non-degenerate solutions is given by
\be \label{enonondeg}
(n-3)! \times (n-2)^2 = (n-2)! \times (n-2)\, .
\ee

For counting the number of degenerate solutions, we 
first define $\rho,\xi$ through \cite{1503.04816,supplement},
\be \label{edefrhoxi}
\sigma_{n+1}=\rho - \f{\xi}{2}\, , \hspace{10mm} \sigma_{n+2}=\rho + \f{\xi}{2}\, ,
\ee
and add and subtract  \refb{e4.2} and \refb{e4.3}  to write them as,
\be \label{eafir}
\sum_{b=1}^{n}\dfrac{k_{n+1}.p_{b}}{\rho - \f{\xi}{2}-\sigma_{b}}+\sum_{b=1}^{n} 
\dfrac{k_{n+2}.p_{b}}{\rho +\f{\xi}{2}-\sigma_{b}}=0\, ,\\
\sum_{b=1}^{n}\dfrac{k_{n+1}.p_{b}}{\rho - \f{\xi}{2}-\sigma_{b}}-\sum_{b=1}^{n} 
\dfrac{k_{n+2}.p_{b}}{\rho +\f{\xi}{2}-\sigma_{b}} -\dfrac{2\tau\, 
k_{n+2}.k_{n+1}}{\xi} =0\, . \label{easec}
\ee
Expanding $\xi$ as 
$\xi = \tau\, \xi_{1}+\tau^{2}\, 
\xi_{2}+O(\tau^{3})$, the second equation may be written as,
\be \label{exi1}
\f{1}{\xi_{1}}=\f{1}{k_{n+1}.k_{n+2}}\sum_{b=1}^{n}\dfrac{k_{n+1}.p_{b}}{\rho-\sigma_{b}} = - \f{1}{k_{n+1}.k_{n+2}}\sum_{b=1}^{n}\dfrac{k_{n+2}.p_{b}}{\rho-\sigma_{b}}\, .
\ee
On the other hand \refb{eafir} in the $\tau\to 0$ limit gives
\be
\sum_{b=1}^{n}\dfrac{ (k_{n+1}+k_{n+2}).p_{b}}{\rho -\sigma_{b}}=0\non\\
\Rightarrow \sum_{b=1}^{n}(k_{n+1}+k_{n+2}).p_{b}\prod_{a=1,a\neq b}^{n}(\rho-\sigma_{b})=0
\ee
This is a polynomial equation in $\rho$ of degree $(n-2)$, since the $\rho^{n-1}$ 
coefficient vanishes by momentum conservation \refb{e4.4}
as $\tau\rightarrow0$. Hence it gives $(n-2)$ 
solutions for $\rho$ for a given set of $\lbrace\sigma_{1},\sigma_{2},...,\sigma_{n}\rbrace$. 
For each such solution, $\xi_1$ is fixed uniquely from \refb{exi1}. 
Therefore we will get total of 
\be \label{enodeg}
(n-3)!(n-2)=(n-2)!
\ee
degenerate solutions. Adding this to \refb{enonondeg} we get
$(n-1)!$ solutions which is the expected total number of solutions of the scattering equations
for $(n+2)$ particles. This shows that we have not missed any solution.

\subsection{Contribution from non-degenerate solutions}

The contribution from the non-degenerate solutions to subleading order in $\tau$ may
be written as,
\be
&& \int D\sigma\int d\sigma_{n+1} d\sigma_{n+2} \Big[ 
\prod_{a=1\atop a\neq i,j,k}^{n}\delta \Big( \sum_{b=1\atop b\neq
a}^{n}\frac{p_{a}.p_{b}}{\sigma_{ab}}+\tau \dfrac{p_{a}.k_{n+1}}{\sigma_{a(n+1)}} + \tau \dfrac{p_{a}.k_{n+2}}{\sigma_{a(n+2)}}\Big)\Big]\non\\
&&\delta\Big( \tau \sum_{b=1}^{n}\frac{k_{n+1}.p_{b}}{\sigma_{(n+1)b}}+\tau^{2}\dfrac{k_{n+1}.k_{n+2}}{\sigma_{(n+1)(n+2)}} \Big)\delta\Big( \tau \sum_{b=1}^{n}\frac{k_{n+2}.p_{b}}{\sigma_{(n+2)b}}+\tau^{2}\dfrac{k_{n+2}.k_{n+1}}{\sigma_{(n+2)(n+1)}} \Big) I_{n+2} \, ,
\label{n+2pointamp}
\ee
where it is understood that we pick contributions from those zeroes of the $\delta$-functions
for which $\sigma_{n+1}-\sigma_{n+2}\sim 1$ as $\tau\to 0$.
The product of first $(n-3)$ delta functions can be expanded in powers of $\tau$ as:
\be \label{edefdel1}
&&\prod_{a=1\atop a\neq i,j,k}^{n}\delta \Big( \sum_{b=1\atop b\neq
a}^{n}\frac{p_{a}.p_{b}}{\sigma_{ab}}+\tau \, \dfrac{p_{a}.k_{n+1}}{\sigma_{a(n+1)}} 
+ \tau \,\dfrac{p_{a}.k_{n+2}}{\sigma_{a(n+2)}}\Big)\non\\
&=& \prod_{a=1 \atop a\neq i,j,k}^{n}\delta \Big( \sum_{b=1\atop b\neq
a}^{n}\frac{p_{a}.p_{b}}{\sigma_{ab}}\Big)+\tau \, \sum_{l=1}^{n}
\Big[\dfrac{p_{l}.k_{n+1}}{\sigma_{l(n+1)}} + \dfrac{p_{l}.k_{n+2}}{\sigma_{l(n+2)}} \Big]
\delta' \Big( \sum_{c=1\atop c\neq l}^{n}\frac{p_{l}.p_{c}}{\sigma_{lc}}\Big)
\prod_{a=1\atop 
a\neq i,j,k,l}^{n}\delta \Big( \sum_{b=1\atop b\neq
a}^{n}\frac{p_{a}.p_{b}}{\sigma_{ab}}\Big)\non\\
&&\equiv \delta^{(0)}+\tau \, \delta^{(1)}\, .
\ee
On the other hand the product of last two delta functions have the form:
\be
\delta\Big( \tau \sum_{b=1}^{n}\frac{k_{n+1}.p_{b}}{\sigma_{(n+1)b}}+\tau^{2}\dfrac{k_{n+1}.k_{n+2}}{\sigma_{(n+1)(n+2)}} \Big)=\f{1}{\tau}\Big[\delta(f_{n+1}^{n})+\tau \, \frac{k_{n+1}.k_{n+2}}{\sigma_{(n+1)(n+2)}}\delta'(f_{n+1}^{n})+O(\tau^{2})\Big]\, ,\\
\delta\Big( \tau \sum_{b=1}^{n}\frac{k_{n+2}.p_{b}}{\sigma_{(n+2)b}}+\tau^{2}\,
\dfrac{k_{n+2}.k_{n+1}}{\sigma_{(n+2)(n+1)}} \Big)=\f{1}{\tau}\, 
\Big[\delta(f_{n+2}^{n})+\tau \, \frac{k_{n+2}.k_{n+1}}{\sigma_{(n+2)(n+1)}}\delta'(f_{n+2}^{n})+O(\tau^{2})\Big]\, ,
\ee
where $f^n_{n+r}$ has been defined in \refb{eadefvar}.
Finally the integrand $I_{n+2}$ may be expanded as,
\be
I_{n+2}&=&I_{n+2}|_{\tau=0}+\tau \f{\p I_{n+2}}{\p \tau}|_{\tau=0}+O(\tau^{2})
\equiv I_{n+2}^{(0)}+\tau \, I_{n+2}^{(1)}+O(\tau^{2})\, .
\ee
Substituting all these expansions in the $(n+2)$-point amplitude \eqref{n+2pointamp}, we get
\be \label{e4.18}
&& \f{1}{\tau^{2}}\int D\sigma \int d\sigma_{n+1}\, d\sigma_{n+2} \big[\delta^{(0)}+\tau\,
 \delta^{(1)}\big]\Big[\delta(f_{n+1}^{n})+\tau\, \frac{k_{n+1}.k_{n+2}}{\sigma_{(n+1)(n+2)}}\, 
 \delta'(f_{n+1}^{n})\Big]\non\\
&&\Big[\delta(f_{n+2}^{n})+\tau\, \frac{k_{n+2}.k_{n+1}}{\sigma_{(n+2)(n+1)}}\,
\delta'(f_{n+2}^{n})\Big]
\, \big[I_{n+2}^{(0)}+\tau \, I_{n+2}^{(1)}\big]\non\\
&=& \f{1}{\tau^{2}}\int D\sigma \, \delta^{(0)}\, \int d\sigma_{n+1}\, d\sigma_{n+2}\, 
\delta(f_{n+1}^{n})\, \delta(f_{n+2}^{n})\, I_{n+2}^{(0)}\non\\
&& +\f{1}{\tau}\int D\sigma \, \int d\sigma_{n+1}\, d\sigma_{n+2}\, \delta^{(1)}\, 
\delta(f_{n+1}^{n})\, \delta(f_{n+2}^{n})\, I_{n+2}^{(0)}\non\\
&&+ \f{1}{\tau}\, \int D\sigma \, \delta^{(0)}\, \int d\sigma_{n+1}\, d\sigma_{n+2}\, 
\delta(f_{n+1}^{n})\, \delta(f_{n+2}^{n})\, I_{n+2}^{(1)}\non\\
&&+\f{1}{\tau}\, \int D\sigma \, \delta^{(0)}\, \int d\sigma_{n+1}\, d\sigma_{n+2}\, 
\dfrac{k_{n+1}.k_{n+2}}{\sigma_{n+1}-\sigma_{n+2}}\, \delta'(f_{n+1}^{n})\,
\delta(f_{n+2}^{n})\, I_{n+2}^{(0)}\non\\
&&+\f{1}{\tau}\, \int D\sigma \, \delta^{(0)}\, \int d\sigma_{n+1}\, d\sigma_{n+2}\, 
\delta(f_{n+1}^{n})\dfrac{k_{n+2}.k_{n+1}}{\sigma_{n+2}-\sigma_{n+1}}\, \delta'(f_{n+2}^{n})\,
I_{n+2}^{(0)}\, . \label{nondegenerate}
\ee
For evaluating this, it will be convenient to 
introduce two soft parameters 
$\tau_1$ and $\tau_2$ instead of a single parameter $\tau$ and take the external
momenta to be $p_1,\cdots p_n, \tau_1\, k_{n+1}, \tau_2\, k_{n+2}$. 
$I_{n+2}$ is then given by
\be
I_{n+2} =  4\, (-1)^{n+2}\,  (\sigma_s -\sigma_t)^{-2} \, \hat P^2
\ee
where $\hat P$ is the Pfaffian of the matrix
\be \label{edefhatpsinew}
\hat \Psi = \begin{pmatrix}
A_{ab} &  {\tau_1 p_a.k_{n+1}\over \sigma_a - \sigma_{n+1}} 
&  {\tau_2 p_a.k_{n+2}\over \sigma_a - \sigma_{n+2}}
& - C^T_{ad}
& {p_a. \eps_{n+1}\over \sigma_a-\sigma_{n+1}}
& {p_a. \eps_{n+2}\over \sigma_a-\sigma_{n+2}} \cr 
{\tau_1 k_{n+1}.p_b\over
\sigma_{n+1}-\sigma_b} & 0 & 0 & {\tau_1 k_{n+1}.\eps_d\over \sigma_{n+1}-\sigma_d}
& - C_{(n+1)(n+1)}  & - {\tau_1 \eps_{n+2}. k_{n+1}\over \sigma_{n+2}
-\sigma_{n+1}}\cr
{\tau_2 k_{n+2}.p_b\over
\sigma_{n+2}-\sigma_b} & 0 & 0 & {\tau_2 k_{n+2}.\eps_d\over \sigma_{n+2}-\sigma_d}
& - {\tau_2 \eps_{n+1}. k_{n+2}\over \sigma_{n+1}
-\sigma_{n+2}}& - C_{(n+2)(n+2)}  \cr
C_{cb} & {\tau_1 \eps_c.k_{n+1}\over \sigma_c - \sigma_{n+1}} 
& {\tau_2 \eps_c.k_{n+2}\over \sigma_c - \sigma_{n+2}} & B_{cd} &
{\eps_c .\eps_{n+1}\over \sigma_{c} - \sigma_{n+1}} &
{\eps_c .\eps_{n+2}\over \sigma_c - \sigma_{n+2}} \cr
{\eps_{n+1}.p_b\over \sigma_{n+1}-\sigma_b} & C_{(n+1)(n+1)} & {\tau_2 
\eps_{n+1}.k_{n+2}\over \sigma_{n+1} - \sigma_{n+2}} & {\eps_{n+1}.\eps_d\over
\sigma_{n+1}-\sigma_{d}} & 0 & {\eps_{n+1} .\eps_{n+2}\over \sigma_{n+1} - \sigma_{n+2}} \cr
{\eps_{n+2}.p_b\over \sigma_{n+2}-\sigma_b} & {\tau_1 
\eps_{n+2}.k_{n+1}\over \sigma_{n+2} - \sigma_{n+1}} & C_{(n+2)(n+2)} & {\eps_{n+2}.\eps_d\over
\sigma_{n+2}-\sigma_{d}} & {\eps_{n+2} .\eps_{n+1}\over \sigma_{n+2} - \sigma_{n+1}} & 0 
\end{pmatrix} \, .
\ee
In \eqref{edefhatpsinew} 
we have set terms with two or more powers of $\tau$ to be zero. The values of 
$C_{(n+1)(n+1)}$ and $C_{(n+2)(n+2)}$ to linear order in $\tau$ are:
\be\label{eacvar}
&& \hskip -.3in C_{(n+1)(n+1)} = - \sum_{a=1}^n {\eps_{n+1}.p_a\over \sigma_{n+1}-\sigma_a}
- {\tau_2 \, \eps_{n+1}.k_{n+2}\over \sigma_{n+1}-\sigma_{n+2}}\, , \quad
C_{(n+2)(n+2)} = - \sum_{a=1}^n {\eps_{n+2}.p_a\over \sigma_{n+2}-\sigma_a}
- {\tau_1 \, \eps_{n+2}.k_{n+1}\over \sigma_{n+2}-\sigma_{n+1}}\, . \non\\
\ee
At the end of the computation we shall take the limit $\tau_1,\tau_2\to \tau\to 0$.

We shall now analyze the different terms on the right hand side of \refb{nondegenerate}.
However,
before we proceed to evaluate the integrals, we would like to remind the reader of the
general strategy for analyzing these integrals 
that was described in the last paragraph of section \ref{schy}. The 
expansion
\refb{nondegenerate} of the amplitude in powers of $\tau$ is valid when the integration
contours wrap around the non-degenerate
solutions for which $\sigma_{n+1}-\sigma_{n+2}\sim 1$.
Once we have made this approximation we can deform the contours of $\sigma_{n+1}$ and
$\sigma_{n+2}$ even
to regions where  $|\sigma_{n+1}-\sigma_{n+2}|<<1$. Treating \refb{nondegenerate}
as the integrand still gives the correct result by Cauchy's theorem even though the original
integrand is no longer approximated by \eqref{nondegenerate}. A similar remark will hold for the
contribution from degenerate solutions --  we shall make our approximation in the
region where the integration contour is close to the degenerate
solution of the scattering equation, and
once the approximation is made, we shall be free to 
deform the contour away from the region where the
approximation is valid.

\medskip

\noindent{\bf First two terms}

\medskip

In \eqref{nondegenerate}, the first two terms on the right hand side may be 
analyzed as in the case of 
single soft graviton by carrying out the integration over $\sigma_{n+2}$ and $\sigma_{n+1}$ 
independently. For this we note from \refb{edefhatpsinew} that
\be \label{ehp00}
\hat P|_{\tau_1=\tau_2=0} = - C_{(n+1)(n+1)} \, C_{(n+2)(n+2)}\, \tilde P
= - \left(\sum_{a=1}^n {\eps_{n+1}. p_a\over \sigma_{n+1}-\sigma_a}\right)
\, \left(\sum_{a=1}^n {\eps_{n+2}. p_a\over \sigma_{n+2}-\sigma_a}\right) \, \tilde P\, ,
\ee
where $\tilde P$ is the Pfaffian of the matrix $\tilde\Psi$ for $n$-particle scattering amplitude
without soft graviton. This gives
\be
I_{n+2}^{(0)} = \left(\sum_{a=1}^n {\eps_{n+1}. p_a\over \sigma_{n+1}-\sigma_a}\right)^2
\, \left(\sum_{a=1}^n {\eps_{n+2}. p_a\over \sigma_{n+2}-\sigma_a}\right)^2 \, I_n\, .
\ee
We can now proceed to evaluate the first two terms on the right hand side of
\refb{nondegenerate} as in sections \ref{esfirst} and \ref{essecond}.
The first term is given by
\be 
\BB_0 &\equiv& \f{1}{\tau^{2}}\int D\sigma \, \delta^{(0)}  \, I_n \, 
\ointop_{\{A_i\}} d\sigma_{n+1} \, \left(\sum_{a=1}^{n}
\dfrac{k_{n+1}.p_{a}}{\sigma_{n+1}-\sigma_{a}}\right)^{-1} 
\left(\sum_{a=1}^n {\eps_{n+1}. p_a\over \sigma_{n+1}-\sigma_a}\right)^2\non  \\  &&
\hskip .3in \ointop_{\{B_i\}} d\sigma_{n+2} \, 
\left(\sum_{a=1}^{n}
\dfrac{k_{n+2}.p_{a}}{\sigma_{n+2}-\sigma_{a}}\right)^{-1}
\, \left(\sum_{a=1}^n {\eps_{n+2}. p_a\over \sigma_{n+2}-\sigma_a}\right)^2\, ,
\ee
where   $\{A_i\}$ and $\{B_i\}$ are defined as
the set of points satisfying
\be \label{edefaibi}
\sum_{a=1}^{n}
\dfrac{k_{n+1}.p_{a}}{\rho-\sigma_{a}}=0 \quad \hbox{at} \quad \rho=A_i\, ,
\qquad \sum_{a=1}^{n} 
\dfrac{k_{n+2}.p_{a}}{\rho-\sigma_{a}} =0  \quad \hbox{at} \quad \rho=B_i\, .
\ee
We can now carry out integration over $\sigma_{n+1}$ and $\sigma_{n+2}$ independently
by deforming the contours to $\infty$. The contribution comes from residues at
the poles at $\sigma_{a}$ -- which can be evaluated following the procedure described
in section \ref{esfirst} -- and the contribution from $\infty$ can be evaluated after using
momentum conservation. The result is
\be 
\BB_0 &\equiv& \f{1}{\tau^{2}}\int D\sigma \, \delta^{(0)}  \, I_n \, \left[
 \sum_{a=1}^n {(\eps_{n+1}.p_a)^2 \over k_{n+1}.p_a} + \tau \,
(\eps_{n+1}.k_{n+2})^2 \, (k_{n+1}.k_{n+2})^{-1}
\right] \non\\ && \hskip  .3in
\left[
 \sum_{a=1}^n {(\eps_{n+2}.p_a)^2 \over k_{n+2}.p_a} + \tau \,
(\eps_{n+2}.k_{n+1})^2 \, (k_{n+1}.k_{n+2})^{-1}
\right]\, .
\ee
Using $\BM_n=\int D\sigma\, \delta^{(0)} \, I_n$, this may be rewritten as
\be 
\BB_0 &=& \tau^{-2}\, S^{(0)}_{n+1}\, S^{(0)}_{n+2} \, {\bf M}_n + \tau^{-1}\,  
\bigg[ S^{(0)}_{n+1} \, (\eps_{n+2}.k_{n+1})^2 \, (k_{n+1}.k_{n+2})^{-1} \non\\ &&
\hskip .3in + S^{(0)}_{n+2}\, (\eps_{n+1}.k_{n+2})^2 \, (k_{n+1}.k_{n+2})^{-1}\bigg] \BM_{n} 
+ O(\tau^0)\, ,
\non\\
\ee
where we define
\be \label{edefs0s1}
S^{(0)}_r \equiv \sum_{a=1}^n {(\eps_r.p_a)^2 \over p_a.k_r}\, , \quad 
S^{(1)}_r \equiv \sum_{a=1}^{n}\dfrac{\eps_{r,\mu}\, \eps_{r,\nu}\, p_{a}^{\mu}\, k_{r,\rho}
\, \hat{J}_{a}^{\rho\nu}}{k_{r}\cdot p_{a}} \, .
\ee

For evaluating the second term on the right hand side of \refb{nondegenerate} we note that 
$\delta^{(1)}$ defined in 
\refb{edefdel1} contains sum of two sets of terms -- one set involving $1/\sigma_{l (n+1)}$ and
the other set involving $1/\sigma_{l(n+2)}$. First consider the set involving $1/\sigma_{l (n+1)}$.
In this case we can carry out the integration over $\sigma_{n+2}$ first following the procedure
of section \ref{esfirst} and arrive at the product of $S^{(0)}_{n+2}$ times an integral of the form
given in \refb{ea2first}. The contribution from the pole at $\sigma_{n+2}=\infty$ can be
ignored
since that will produce an extra factor of $\tau$ and give a subsubleading contribution.
The integration over $\sigma_{n+1}$ is proportional to \refb{ea2first} and
can be analyzed as in section \ref{essecond}, leading to the result
\refb{ea2second} multiplied by $S^{(0)}_{n+2}$. Using the equality between \refb{ea2second},
\refb{e3.42}  and the first term on the right hand side of \refb{e3.41}, this
may be expressed as
\be \label{es0diff}
\tau^{-1} \, S^{(0)}_{n+2} \int D\sigma\, (S^{(1)}_{n+1} \, \delta^{(0)}) \, I_n
= \tau^{-1}\, S^{(0)}_{n+2} S^{(1)}_{n+1} {\bf M}_n 
- \tau^{-1}\,  S^{(0)}_{n+2} \int D\sigma \delta^{(0)} \, S^{(1)}_{n+1} I_n\, ,
\ee
where in the left hand side of this equation it is understood that only the orbital part of
$S^{(1)}_{n+1}$ acts on $\delta^{(0)}$.
The contribution from the second term in $\delta^{(1)}$ introduced in \refb{edefdel1}
is given by an expression similar to that in \refb{es0diff} but with $(n+1)$ and $(n+2)$
interchanged. Therefore the total contribution from the second term on the right
hand side of \refb{nondegenerate} 
may be expressed as
\be \label{eadefbb1}
&&\hskip -.3in \BB_1={1\over \tau}\, 
\left\{  S^{(0)}_{n+1} S^{(1)}_{n+2} + S^{(0)}_{n+2} S^{(1)}_{n+1}
\right\} \ {\bf M}_n -{1\over \tau}\, S^{(0)}_{n+2} \int D\sigma \delta^{(0)} \, S^{(1)}_{n+1} I_n
- {1\over \tau}\, S^{(0)}_{n+1} \int D\sigma \delta^{(0)} \, S^{(1)}_{n+2} I_n\, . \non\\
\ee
We shall now turn to the analysis of the last three terms in \refb{nondegenerate}.

\medskip

\noindent{\bf Third Term:}

\medskip

We now consider the third term on the right hand side of \refb{nondegenerate}:
\be \label{eathth}
\f{1}{\tau}\int D\sigma \, \delta^{(0)}\, 
\int d\sigma_{n+1}\, d\sigma_{n+2}\, 
\delta(f_{n+1}^{n})\, \delta(f_{n+2}^{n})\, I_{n+2}^{(1)}\, .
\ee
To evaluate this we have to 
first evaluate $I_{n+2}^{(1)}$.  
We have
the analog of \refb{eadefin1}:
\be \label{eadefin1new}
I_{n+2}^{(1)} \equiv  \left({\p I_{n+2}\over \p\tau_1}
 +{\p I_{n+2}\over \p\tau_2}\right)\bigg|_{\tau_1=\tau_2=0}
 = 8 (-1)^{n+2}\,  (\sigma_s -\sigma_t)^{-2} \,\, \hat P\,  \left[{\p \hat P\over
\p\tau_1} + {\p \hat P\over
\p\tau_2}\right]\bigg|_{\tau_1=\tau_2=0} \, ,
\ee
where $\hat P$ is the Pfaffian of the matrix $\hat\Psi$ given in \refb{edefhatpsinew}.

Let us now examine the $\tau_1$ derivative of $\hat P$. 
Since we have to set $\tau_2=0$ at the end, we can make this replacement even before
taking the $\tau_1$ derivative. In this case the Pfaffian $\hat P$ of the
matrix \refb{edefhatpsinew} reduces to
\be  \label{eat1exp}
(-1)^n \, C_{(n+2)(n+2)} \, Pf \begin{pmatrix}
A_{ab} &  {\tau_1 p_a.k_{n+1}\over \sigma_a - \sigma_{n+1}} & - C^T_{ad}
& {p_a. \eps_{n+1}\over \sigma_a-\sigma_{n+1}} \cr {\tau_1 k_{n+1}.p_b\over
\sigma_{n+1}-\sigma_b} & 0 & {\tau_1 k_{n+1}.\eps_d\over \sigma_{n+1}-\sigma_d}
& - C_{(n+1)(n+1)} \cr
C_{cb} & {\tau_1 \eps_c.k_{n+1}\over \sigma_c - \sigma_{n+1}} & B_{cd} &
{\eps_c .\eps_{n+1}\over \sigma_c - \sigma_{n+1}} \cr
{\eps_{n+1}.p_b\over \sigma_{n+1}-\sigma_b} & C_{(n+1)(n+1)} & {\eps_{n+1}.\eps_d\over
\sigma_{n+1}-\sigma_d} & 0
\end{pmatrix} \, .
\ee 
One can now recognize this matrix as the same matrix that appears in \refb{edefhatpsi}. 
Therefore when the 
$\tau_1$ derivative acts on the Pfaffian of this matrix, then 
at $\tau_1=0$ it will have the same form as that in \refb{efinA3} after $\sigma_{n+1}$ 
integration.
The extra factor of $C_{(n+2)(n+2)}$ at $\tau_1=0$, multiplied by a similar factor coming
from the $\hat P$ factor in \refb{eadefin1new}, will generate a factor of $S^{(0)}_{n+2}$ after 
integration over $\sigma_{n+2}$. There is an additional contribution from $\sigma_{n+2}=\infty$, but this is
subsubleading and may be ignored.
Using the equality of \refb{efinA3}, \refb{eA3equiv} and  second
term on the right hand side of \refb{ang}, the net contribution of 
this term to the right hand side of \refb{e4.18} can be shown to have the form:
\be
S^{(0)}_{n+2} \, \int D\sigma \, \delta^{(0)} \, S^{(1)}_{n+1} \, I_n \, .
\ee
A similar term with $(n+1)$ and $(n+2)$ exchanged comes from the $\p \hat P/\p \tau_2$
term in \refb{eadefin1new}. The sum of these two give
\be 
\BB_2 \, \equiv \, \f{1}{\tau}S^{(0)}_{n+2} \, \int D\sigma \, \delta^{(0)} S^{(1)}_{n+1} I_n 
+ \f{1}{\tau}S^{(0)}_{n+1} \, \int D\sigma \, \delta^{(0)} S^{(1)}_{n+2} I_n \, .
\ee
This cancels the last two terms in \refb{eadefbb1}.

This however is not the complete contribution from the third term in \refb{nondegenerate},
since we still have to include the contribution where $\tau_1$ derivative acts on the
$C_{(n+2)(n+2)}$ factor in \refb{eat1exp}, and a similar contribution with $(n+1)$ 
and $(n+2)$ exchanged. Now from \refb{eacvar} we have
\be
{\p C_{(n+2)(n+2)}\over \p \tau_1} = - {\eps_{n+2}.k_{n+1}\over \sigma_{n+2}-\sigma_{n+1}}\, .
\ee
Therefore at $\tau_1=\tau_2=0$, the contribution to $\p \hat P/\p \tau_1$, with
the derivative acting on the first term in \refb{eat1exp}, reduces to
\be
 {\eps_{n+2}.k_{n+1}\over \sigma_{n+2}-\sigma_{n+1}} \, C_{(n+1)(n+1)}|_{\tau_1=\tau_2=0}\, 
 \tilde P
 = -  {\eps_{n+2}.k_{n+1}\over \sigma_{n+2}-\sigma_{n+1}} \, \sum_{a=1}^n {\eps_{n+1}.p_a\over
 \sigma_{n+1}-\sigma_a} \, \tilde P\, ,
\ee
where $\tilde P$ is the Pfaffian of the matrix $\begin{pmatrix} A_{ab} & - (C^T)_{ad}\cr C_{cb} &
B_{cd} \end{pmatrix}$. Substituting this into \refb{eadefin1new} and setting 
$\tau_1=\tau_2=0$,
using \refb{ehp00}, and adding also the extra contribution from the $\p \hat P/\p \tau_2$ term
in \refb{eadefin1new},
we get the net extra contribution to $I^{(1)}_{n+2}$ to be
\be \label{eexx1}
&&  8\, (-1)^{n} \, (\sigma_s -\sigma_t)^{-2} \,  \left(\sum_{a=1}^n 
{\eps_{n+1}.p_a\over
 \sigma_{n+1}-\sigma_a}\right)  \, \left(\sum_{a=1}^n 
{\eps_{n+2}.p_a\over
 \sigma_{n+2}-\sigma_a}\right) \, \tilde P^2\, 
\non\\ &&
\left[ {\eps_{n+2}.k_{n+1}\over \sigma_{n+2}-\sigma_{n+1}} \, \sum_{a=1}^n 
{\eps_{n+1}.p_a\over
 \sigma_{n+1}-\sigma_a} +{\eps_{n+1}.k_{n+2}\over \sigma_{n+1}-\sigma_{n+2}} \, 
 \sum_{a=1}^n {\eps_{n+2}.p_a\over
 \sigma_{n+2}-\sigma_a}
 \right]\, .
\ee
Noting that $4\, (-1)^{n} \, (\sigma_s -\sigma_t)^{-2} \, \tilde P^2 = I_n$, 
the contribution of \refb{eexx1} to \refb{eathth} is given by
\be \label{eab3sec}
&& {2\over \tau} \, \int D\sigma\, \delta^{(0)}\, I_n\,  
\int d\sigma_{n+1} \, d\sigma_{n+2} \, \delta(f_{n+1}^{n}) \, \delta(f_{n+2}^{n}) \, 
\left(\sum_{a=1}^n 
{\eps_{n+1}.p_a\over
 \sigma_{n+1}-\sigma_a}\right)  \, \left(\sum_{a=1}^n 
{\eps_{n+2}.p_a\over
 \sigma_{n+2}-\sigma_a}\right) 
\non\\
&& \left[ {\eps_{n+2}.k_{n+1}\over \sigma_{n+2}-\sigma_{n+1}} \, \sum_{a=1}^n 
{\eps_{n+1}.p_a\over
 \sigma_{n+1}-\sigma_a} +{\eps_{n+1}.k_{n+2}\over \sigma_{n+1}-\sigma_{n+2}} \, 
 \sum_{a=1}^n {\eps_{n+2}.p_a\over
 \sigma_{n+2}-\sigma_a}
 \right]\, .
\ee

We now regard the integrals over $\sigma_{n+1}$ and $\sigma_{n+2}$  as
contour integrals as in \eqref{ea2}. Let us examine the first term in the square bracket. It takes
the form
\be \label{easquare}
&& {2\over \tau} \, \int D\sigma\, \delta^{(0)}\, I_n \, 
\ointop_{\{A_i\}} d\sigma_{n+1} \ointop_{\{B_i\}} d\sigma_{n+2} \, \left(\sum_{a=1}^n 
{k_{n+1}.p_a\over
 \sigma_{n+1}-\sigma_a}\right)^{-1}  \, \left(\sum_{a=1}^n 
{k_{n+2}.p_a\over
 \sigma_{n+2}-\sigma_a}\right)^{-1} 
\non\\
&& \left(\sum_{a=1}^n 
{\eps_{n+1}.p_a\over
 \sigma_{n+1}-\sigma_a}\right)  \, \left(\sum_{a=1}^n 
{\eps_{n+2}.p_a\over
 \sigma_{n+2}-\sigma_a}\right) 
\left({\eps_{n+2}.k_{n+1}\over \sigma_{n+2}-\sigma_{n+1}}\right) \, \left(\sum_{a=1}^n 
{\eps_{n+1}.p_a\over
 \sigma_{n+1}-\sigma_a} \right)\, ,
\ee
where, $\{A_i\}$ and $\{B_i\}$ have been defined in \refb{edefaibi} and accordingly the $\sigma_{n+1}$ and $\sigma_{n+2}$ contours run anti-clockwise around the poles of
the first and second factors of the integrand respectively. We shall now deform the $\sigma_{n+2}$
contour away from the pole towards infinity. The only pole is at $\sigma_{n+2}=\sigma_{n+1}$
and a possible contribution from $\sigma_{n+2}=\infty$. Let us first examine the contribution
from the pole at $\sigma_{n+1}$. This is given by
\be  \label{edefbb3}
\BB_3 &\equiv& -{2\over \tau} \, \int D\sigma\, \delta^{(0)}\, I_n \,
\ointop_{\{A_i\}} d\sigma_{n+1}  \left(\sum_{a=1}^n 
{k_{n+1}.p_a\over
 \sigma_{n+1}-\sigma_a}\right)^{-1}  \, \left(\sum_{a=1}^n 
{k_{n+2}.p_a\over
 \sigma_{n+1}-\sigma_a}\right)^{-1} 
\non\\
&& \left(\sum_{a=1}^n 
{\eps_{n+1}.p_a\over
 \sigma_{n+1}-\sigma_a}\right)^2  \, \left(\sum_{a=1}^n 
{\eps_{n+2}.p_a\over
 \sigma_{n+1}-\sigma_a}\right) 
\left({\eps_{n+2}.k_{n+1}}\right) \, .
\ee
Next we analyze the contribution from $\sigma_{n+2}=\infty$. Using momentum
conservation equation $\sum_{a=1}^n p_a=-\tau (k_{n+1} + k_{n+2})$ this is given
by
\be \label{eextrainfinity}
{2\over \tau} \, \int D\sigma\, \delta^{(0)}\, I_n \, 
\ointop_{\{A_i\}} d\sigma_{n+1} \, \left(\sum_{a=1}^n 
{k_{n+1}.p_a\over
 \sigma_{n+1}-\sigma_a}\right)^{-1}  
\left(\sum_{a=1}^n 
{\eps_{n+1}.p_a\over
 \sigma_{n+1}-\sigma_a}\right)^2   \non\\ 
\times \ (\eps_{n+2}.k_{n+1})^2 \, (k_{n+1}.k_{n+2})^{-1} 
\, . 
\ee
Now we deform the $\sigma_{n+1}$ contour to $\infty$, picking up residues at 
$\sigma_a$. The result is
\be \label{efinextra}
\BB_4\equiv -{2\over \tau}  \, \BM_n\,   S^{(0)}_{n+1} \, (\eps_{n+2}.k_{n+1})^2 \, (k_{n+1}.k_{n+2})^{-1}\, .
\ee
The residue at $\sigma_{n+1}=\infty$ has an additional factor of $\tau$. Therefore its contribution is
subsubleading.

The contribution from the second term in the square bracket in \refb{eab3sec} 
can be obtained from  \refb{edefbb3}, \refb{efinextra} by exchange of 
$(n+1)$ and $(n+2)$ and is given by $\BB_5+\BB_6$, where
\be  \label{edefbb4}
\BB_5 &\equiv& -{2\over \tau} \, \int D\sigma\, \delta^{(0)}\, I_n \,  
\ointop_{\{B_i\}} d\sigma_{n+2}  \left(\sum_{a=1}^n 
{k_{n+2}.p_a\over
 \sigma_{n+2}-\sigma_a}\right)^{-1}  \, \left(\sum_{a=1}^n 
{k_{n+1}.p_a\over
 \sigma_{n+2}-\sigma_a}\right)^{-1} 
\non\\
&& \left(\sum_{a=1}^n 
{\eps_{n+2}.p_a\over
 \sigma_{n+2}-\sigma_a}\right)^2  \, \left(\sum_{a=1}^n 
{\eps_{n+1}.p_a\over
 \sigma_{n+2}-\sigma_a}\right) 
\left({\eps_{n+1}.k_{n+2}}\right) \, ,
\ee
and 
\be \label{efinextranew}
\BB_6\equiv -{2\over \tau}  \, \BM_n \, S^{(0)}_{n+2}  \, (\eps_{n+1}.k_{n+2})^2 \, (k_{n+1}.k_{n+2})^{-1}\, .
\ee

\noindent{\bf Fourth and fifth terms}

\medskip

The fourth term on the right hand side of \eqref{nondegenerate} is given by:
\be
&&\f{1}{\tau}\int D\sigma \delta^{(0)}\int d\sigma_{n+1}d\sigma_{n+2}
\dfrac{k_{n+1}.k_{n+2}}{\sigma_{n+1}-\sigma_{n+2}}\delta'(f_{n+1}^{n})\delta(f_{n+2}^{n})I_{n+2}^{(0)}\non\\
&=&-\f{1}{\tau}\int D\sigma \delta^{(0)}\ointop_{\{A_i\}} d\sigma_{n+1} \ointop_{\{B_i\}}
d\sigma_{n+2}
\dfrac{k_{n+1}.k_{n+2}}{\sigma_{n+1}-\sigma_{n+2}}\Big(\sum_{c=1}^{n}\dfrac{k_{n+1}.p_{c}}{\sigma_{n+1}-\sigma_{c}}\Big)^{-2}\Big(\sum_{a=1}^{n}\dfrac{k_{n+2}.p_{a}}{\sigma_{n+2}-\sigma_{a}}\Big)^{-1}\non\\
&& \Big(\sum_{b=1}^{n}\dfrac{\epsilon_{n+1}.p_{b}}{\sigma_{n+1}-\sigma_{b}}\Big)^{2}\Big(\sum_{b=1}^{n}
\dfrac{\epsilon_{n+2}.p_{b}}{\sigma_{n+2}-\sigma_{b}}\Big)^{2}I_{n}\, .
\ee
We now deform the $\sigma_{n+1}$ contour away from the poles $\{A_i\}$ towards infinity and
pick up residues at $\sigma_{n+1}=\sigma_{n+2}$ as well as the contribution from infinity. 
There are no poles at $\sigma_{n+1}=\sigma_a$. The
contribution from the pole at $\sigma_{n+2}$ is given by
\be  \label{eadefbb7}
\BB_7 &\equiv& \f{1}{\tau}\, \int D\sigma\, \delta^{(0)}\, I_n\, \oint_{\{B_i\}} 
d\sigma_{n+2} 
\Big(\sum_{c=1}^{n}\dfrac{k_{n+1}.p_{c}}{\sigma_{n+2}-\sigma_{c}}\Big)^{-2}
{\Big(\sum_{a=1}^{n}\dfrac{k_{n+2}.p_{a}}{\sigma_{n+2}-\sigma_{a}}\Big)^{-1}}\,
(k_{n+1}.k_{n+2})\non\\
&&\hskip 1in \Big(\sum_{b=1}^{n}\dfrac{\epsilon_{n+2}.p_{b}}{\sigma_{n+2}-\sigma_{b}}\Big)^{2}
\Big(\sum_{b=1}^{n}\dfrac{\epsilon_{n+1}.p_{b}}{\sigma_{n+2}-\sigma_{b}}\Big)^{2}\, .
\ee
On the other hand the contribution from infinity is given by, after using momentum conservation,
\be \label{eadefbb8}
 \hskip -.3in
\BB_8 &\equiv& -\f{1}{\tau}\int D\sigma \delta^{(0)} \, \ointop_{\{B_i\}}
d\sigma_{n+2} \
  \,  \f{(\eps_{n+1}. k_{n+2})^2 }{k_{n+1}.k_{n+2}}
\Big(\sum_{a=1}^{n}\dfrac{k_{n+2}.p_{a}}{\sigma_{n+2}-\sigma_{a}}\Big)^{-1} 
\Big(\sum_{b=1}^{n}
\dfrac{\epsilon_{n+2}.p_{b}}{\sigma_{n+2}-\sigma_{b}}\Big)^{2}I_{n} \non \\
 &=& \f{1}{\tau}\, \BM_n \, S^{(0)}_{n+2} \, (k_{n+1}.k_{n+2})^{-1}  
\,  (\eps_{n+1}. k_{n+2})^2 \, ,
\ee
where in the second step we have performed integration over $\sigma_{n+2}$, picking up
residues at $\sigma_b$. In this case, as $\sigma_{n+2}\to\infty$, the contribution to the
integrand goes as $\tau$, and therefore this does not contribute at the subleading order.

The contribution from the fifth term on the right hand side of \eqref{nondegenerate} can be
evaluated in a similar manner, giving contributions similar to \refb{eadefbb7} and 
\refb{eadefbb8}, with $n+1$ and $n+2$ exchanged:
\be  \label{eadefbb9}
\BB_9 &\equiv& \f{1}{\tau}\, \int D\sigma\, \delta^{(0)}\, I_n\, \oint_{\{A_i\}} 
d\sigma_{n+1} 
\Big(\sum_{c=1}^{n}\dfrac{k_{n+2}.p_{c}}{\sigma_{n+1}-\sigma_{c}}\Big)^{-2}
{\Big(\sum_{a=1}^{n}\dfrac{k_{n+1}.p_{a}}{\sigma_{n+1}-\sigma_{a}}\Big)^{-1}}\,
(k_{n+1}.k_{n+2})\non\\
&&\hskip 1in \Big(\sum_{b=1}^{n}\dfrac{\epsilon_{n+1}.p_{b}}{\sigma_{n+1}-\sigma_{b}}\Big)^{2}
\Big(\sum_{b=1}^{n}\dfrac{\epsilon_{n+2}.p_{b}}{\sigma_{n+1}-\sigma_{b}}\Big)^{2}\, ,
\ee
and 
\be \label{eadefbb10}
&& \hskip -.3in
\BB_{10} \equiv  \f{1}{\tau}\, \BM_n \, S^{(0)}_{n+1} \, (k_{n+1}.k_{n+2})^{-1}  
\,  (\eps_{n+2}. k_{n+1})^2  \, .
\ee

Adding the contributions $\BB_0$ to $\BB_{10}$ we get the total contribution from non-degenerate
solutions:
\be \label{eatotalnondeg}
\mathcal{B} &=& \{ \tau^{-2}S^{(0)}_{n+1} \, S^{(0)}_{n+2} \BM_n +\tau^{-1} S^{(0)}_{n+1} \, S^{(1)}_{n+2} \BM_n
+\tau^{-1} S^{(0)}_{n+2} \, S^{(1)}_{n+1} \BM_n\} \non\\ && \hskip -.8in
-{2\over \tau} \, \int D\sigma\, \delta^{(0)}\, I_n \,
\ointop_{\{A_i\}} d\rho\,  \left(\sum_{a=1}^n 
{k_{n+1}.p_a\over
 \rho-\sigma_a}\right)^{-1}  \, \left(\sum_{a=1}^n 
{k_{n+2}.p_a\over
 \rho-\sigma_a}\right)^{-1} 
\left(\sum_{a=1}^n 
{\eps_{n+1}.p_a\over
 \rho-\sigma_a}\right)^2  \, \left(\sum_{a=1}^n 
{\eps_{n+2}.p_a\over
 \rho-\sigma_a}\right) 
\left({\eps_{n+2}.k_{n+1}}\right) \non\\ && \hskip -.8in
-{2\over \tau} \, \int D\sigma\, \delta^{(0)}\, I_n \,  
\ointop_{\{B_i\}} d\rho \, \left(\sum_{a=1}^n 
{k_{n+2}.p_a\over
 \rho-\sigma_a}\right)^{-1}  \, \left(\sum_{a=1}^n 
{k_{n+1}.p_a\over
 \rho-\sigma_a}\right)^{-1} 
\left(\sum_{a=1}^n 
{\eps_{n+2}.p_a\over
 \rho-\sigma_a}\right)^2  \, \left(\sum_{a=1}^n 
{\eps_{n+1}.p_a\over
 \rho-\sigma_a}\right) 
\left({\eps_{n+1}.k_{n+2}}\right) \non\\ && \hskip -.8in
+\f{1}{\tau}\, \int D\sigma\, \delta^{(0)}\, I_n\, \oint_{\{B_i\}} 
d\rho \, 
\left(\sum_{c=1}^{n}\dfrac{k_{n+1}.p_{c}}{\rho-\sigma_{c}}\right)^{-2}
{\left(\sum_{a=1}^{n}\dfrac{k_{n+2}.p_{a}}{\rho-\sigma_{a}}\right)^{-1}}\,
(k_{n+1}.k_{n+2})  
\left(\sum_{b=1}^{n}\dfrac{\epsilon_{n+2}.p_{b}}{\rho-\sigma_{b}}\right)^{2}
\left(\sum_{b=1}^{n}\dfrac{\epsilon_{n+1}.p_{b}}{\rho-\sigma_{b}}\right)^{2}
\non\\ &&\hskip -.8in
+  \f{1}{\tau}\, \int D\sigma\, \delta^{(0)}\, I_n\, \oint_{\{A_i\}} 
d\rho \, 
\left(\sum_{c=1}^{n}\dfrac{k_{n+2}.p_{c}}{\rho-\sigma_{c}}\right)^{-2}
{\left(\sum_{a=1}^{n}\dfrac{k_{n+1}.p_{a}}{\rho-\sigma_{a}}\right)^{-1}}\,
(k_{n+1}.k_{n+2}) 
\left(\sum_{b=1}^{n}\dfrac{\epsilon_{n+1}.p_{b}}{\rho-\sigma_{b}}\right)^{2}
\left(\sum_{b=1}^{n}\dfrac{\epsilon_{n+2}.p_{b}}{\rho-\sigma_{b}}\right)^{2}\, .
\non\\
\ee
The term in the first line was derived in \cite{1702.02350}. 
The terms in the second and the third
line vanish in the gauge chosen in \cite{1702.02350}. 
The terms in the fourth and fifth lines were missed
in the analysis of \cite{1702.02350}. 
We shall see that these  terms cancel similar terms arising in the
analysis of the degenerate solutions. Using the gauge transformation laws of
$S^{(0)}_{n+1} \, S^{(1)}_{n+2} \BM_n$ and $S^{(0)}_{n+2} \, S^{(1)}_{n+1} \BM_n$ analyzed in
\cite{1707.06803}, one can verify that \refb{eatotalnondeg}  is invariant under the gauge transformations of the
soft graviton polarizations.

\subsection{Contribution from degenerate solutions} \label{s4.3}

When the integration contours of $\sigma_{n+1}$ and $\sigma_{n+2}$ wrap around the 
degenerate solutions for which 
$|\sigma_{n+1}-\sigma_{n+2}|\sim \tau$, we will carry out the
integration in $\rho$ and $\xi$ variables introduced in \refb{edefrhoxi}.
In terms of these variables the integration measure takes the form,
\be
\int d\sigma_{n+1}\, d\sigma_{n+2}\, \delta(f_{n+1}^{n})\, 
\delta(f_{n+2}^{n}) \, \cdots= -2\, \int d\rho \, d\xi\, \delta(f_{n+1}^{n}+f_{n+2}^{n}) \, 
\delta(f_{n+1}^{n}-f_{n+2}^{n})  \, \cdots
\ee
with the understanding that $\rho$ integration is done using the first delta function 
and $\xi$ integration is done using the second delta function. 
Therefore the contribution to the $(n+2)$-point amplitude from the degenerate solutions 
becomes,
\be \label{easubdeg}
&& \hskip -1in -\f{2}{\tau^{2}}\int D\sigma \, \delta^{(0)} \, \int d\rho \,
d\xi \, \delta\Big(\sum_{b=1}^{n}\dfrac{k_{n+1}.p_{b}}{\rho-\f{\xi}{2}-\sigma_{b}}+\sum_{b=1}^{n}\dfrac{k_{n+2}.p_{b}}{\rho+\f{\xi}{2}-\sigma_{b}}\Big)\non\\
&& \delta\Big(\sum_{b=1}^{n}\dfrac{k_{n+1}.p_{b}}{\rho-\f{\xi}{2}-\sigma_{b}} - \sum_{b=1}^{n}\dfrac{k_{n+2}.p_{b}}{\rho+\f{\xi}{2}-\sigma_{b}}-\dfrac{2\, \tau \, k_{n+1}.k_{n+2}}{\xi}\Big)
\ I_{n+2}\, ,
\ee
where it is understood that we sum over contributions from those zeroes of the second $\delta$-function for which
$\xi\sim\tau$. By carrying out the integration over $\xi$ using the second delta
function, this integral can be approximated  as 
\be
-\f{2}{\tau}\int D\sigma \, \delta^{(0)} \int d\rho\ \delta\Big(\sum_{b=1}^{n}\dfrac{k_{n+1}.p_{b}}{\rho-\sigma_{b}}
+\sum_{b=1}^{n}\dfrac{k_{n+2}.p_{b}}{\rho-\sigma_{b}}\Big)\, {\xi_1^2 \over  2 \, k_{n+1}.k_{n+2}}\, 
I_{n+2}|_{\xi=\tau \, \xi_1} +O(\tau^0)\, ,
\ee
where
\be
\xi_1 = 2 \, k_{n+1}.k_{n+2} \,  \left(
\sum_{b=1}^{n}\dfrac{k_{n+1}.p_{b}}{\rho-\sigma_{b}}-\sum_{b=1}^{n}\dfrac{k_{n+2}.p_{b}}
{\rho-\sigma_{b}}
\right)^{-1}  \, .
\ee
Now
in the degeneration limit we can evaluate $I_{n+2}$ by regarding this as the square of the 
Pfaffian of the matrix $\hat\Psi$ which now takes the form 
\be \label{edefhatpsismallxi}
\hat \Psi \simeq \begin{pmatrix}
A_{ab} &  {\tau \, p_a.k_{n+1}\over \sigma_a - \rho} 
&  {\tau \, p_a.k_{n+2}\over \sigma_a - \rho}
& - C^T_{ad}
& {p_a. \eps_{n+1}\over \sigma_a-\rho}
& {p_a. \eps_{n+2}\over \sigma_a-\rho} \cr 
{\tau \, k_{n+1}.p_b\over
\rho-\sigma_b} & 0 & -{\tau \, k_{n+1}.k_{n+2}\over \xi_1} & {\tau \, k_{n+1}.\eps_d\over \rho-\sigma_d}
& - C_{(n+1)(n+1)}  & -{\eps_{n+2}. k_{n+1}\over \xi_1}\cr
{\tau \, k_{n+2}.p_b\over
\rho-\sigma_b} & {\tau \, k_{n+1}.k_{n+2}\over \xi_1}  
& 0 & {\tau \, k_{n+2}.\eps_d\over \rho-\sigma_d}
&  {\eps_{n+1}. k_{n+2}\over \xi_1}& - C_{(n+2)(n+2)}  \cr
C_{cb} & {\tau \, \eps_c.k_{n+1}\over \sigma_c - \rho} 
& {\tau \, \eps_c.k_{n+2}\over \sigma_c - \rho} & B_{cd} &
{\eps_c .\eps_{n+1}\over \sigma_{c} - \rho} &
{\eps_c .\eps_{n+2}\over \sigma_c - \rho} \cr
{\eps_{n+1}.p_b\over \rho-\sigma_b} & C_{(n+1)(n+1)} &- { 
\eps_{n+1}.k_{n+2}\over \xi_1} & {\eps_{n+1}.\eps_d\over
\rho-\sigma_{d}} & 0 & -{\eps_{n+1} .\eps_{n+2}\over \tau \, \xi_1} \cr
{\eps_{n+2}.p_b\over \rho-\sigma_b} & {
\eps_{n+2}.k_{n+1}\over \xi_1} & C_{(n+2)(n+2)} & {\eps_{n+2}.\eps_d\over
\rho-\sigma_{d}} & {\eps_{n+2} .\eps_{n+1}\over \tau \, \xi_1} & 0 
\end{pmatrix} \, ,
\ee
where we now have
\be\label{eacvarnew}
&& \hskip -.3in C_{(n+1)(n+1)} \simeq - \sum_{a=1}^n {\eps_{n+1}.p_a\over \rho-\sigma_a}
+ {\eps_{n+1}.k_{n+2}\over \xi_1}\, , \quad
C_{(n+2)(n+2)} \simeq - \sum_{a=1}^n {\eps_{n+2}.p_a\over \rho-\sigma_a}
- {\eps_{n+2}.k_{n+1}\over \xi_1}\, . \non\\
\ee
Given the appearance of $1/\tau$ in the $(2n+3)$-$(2n+4)$-th element of the matrix, one
may wonder whether the Pfaffian may give a leading contribution of order $1/\tau$,
thereby upsetting the counting of powers of $\tau$. It is easy to see however that the
coefficient of this element in the expansion of the Pfaffian, given by the Pfaffian obtained by
eliminating the $(2n+3)$ and $(2n+4)$-th rows and columns of this matrix, actually goes as
$\tau$ and therefore does not upset the counting of powers of $\tau$. 
With this understanding
we can now compute the Pfaffian of $\hat\Psi$, arriving at the result\cite{supplement}:
\begin{eqnarray} \label{e4.58}
I_{n+2} &=& \Bigg[(\xi_1)^{-2} \epsilon_{n+1}\cdot \epsilon_{n+2} \ k_{n+1}\cdot k_{n+2} 
- (\xi_1)^{-2} \epsilon_{n+1}\cdot k_{n+2} \ \epsilon_{n+2}\cdot k_{n+1} 
\nonumber \\
&& \hskip 1in - C_{(n+1)(n+1)} \, C_{(n+2)(n+2)}
\bigg]^2  \, I_n + {\cal O}(\tau) \nonumber \\ 
&=& \bigg[(\xi_1)^{-2} \epsilon_{n+1}\cdot \epsilon_{n+2} \ k_{n+1}\cdot k_{n+2} 
- (\xi_1)^{-2} \epsilon_{n+1}\cdot k_{n+2} \ \epsilon_{n+2}\cdot k_{n+1} 
\nonumber \\ 
&& 
-\left\{\sum_{c=1}^n  {\epsilon_{n+1}\cdot p_c\over 
\rho -\sigma_c} - (\xi_1)^{-1} \, \epsilon_{n+1} \cdot k_{n+2} \right\}
\ \left\{\sum_{c=1}^n  {\epsilon_{n+2}\cdot p_c\over 
\rho -\sigma_c} 
+ (\xi_1)^{-1}  \epsilon_{n+2}\cdot k_{n+1}\right\}
 \Bigg]^2  \, I_n \nonumber \\
&& + {\cal O}(\tau) \, .
\end{eqnarray}
Substituting this into \refb{easubdeg} we get
\be \label{e4.59}
&&  \hskip -.3in -\f{4k_{n+1}.k_{n+2}}{\tau}\, \int D\sigma\, \delta^{(0)}\, I_n\, \int 
d\rho \
\delta\Big(\sum_{b=1}^{n}\dfrac{k_{n+1}.p_{b}}{\rho-\sigma_{b}}+\sum_{b=1}^{n}\dfrac{k_{n+2}.p_{b}}{\rho-\sigma_{b}}\Big) \Big(\sum_{a=1}^{n}
\dfrac{(k_{n+1}-k_{n+2}).p_{a}}{\rho-\sigma_{a}}\Big)^{-2}\non\\
&& \hskip -.3in
\Bigg[\Big(\sum_{c=1}^{n}
\dfrac{\epsilon_{n+1}.p_{c}}{\rho-\sigma_{c}}\Big) \Big(\sum_{d=1}^{n}\dfrac{\epsilon_{n+2}.p_{d}}{\rho-\sigma_{d}}\Big)
- \epsilon_{n+1}\cdot \epsilon_{n+2} \ 
{1\over 4 \, k_{n+1}.k_{n+2}} 
\Big(\sum_{a=1}^{n}
\dfrac{(k_{n+1}-k_{n+2}).p_{a}}{\rho-\sigma_{a}}\Big)^{2}
 \non \\
&& \hskip -.3in + \Big(\epsilon_{n+2}.k_{n+1} \sum_{c=1}^{n}
\dfrac{\epsilon_{n+1}.p_{c}}{\rho-\sigma_{c}}
- \epsilon_{n+1}.k_{n+2} \ \sum_{c=1}^{n}
\dfrac{\epsilon_{n+2}.p_{c}}{\rho-\sigma_{c}}\Big) {1\over 2 k_{n+1}.k_{n+2}} \,  \left(
\sum_{b=1}^{n}\dfrac{k_{n+1}.p_{b}}{\rho-\sigma_{b}}-\sum_{b=1}^{n}\dfrac{k_{n+2}.p_{b}}
{\rho-\sigma_{b}}
\right)
\Bigg]^2\, . \non\\
\ee
By making use of the $\delta$-function, we can rewrite this as
\be \label{eammaster}
&&  \f{k_{n+1}.k_{n+2}}{\tau} \, \int D\sigma\, \delta^{(0)}\, I_n\, \int 
d\rho \
\delta\Big(\sum_{b=1}^{n}\dfrac{k_{n+1}.p_{b}}{\rho-\sigma_{b}}+\sum_{b=1}^{n}\dfrac{k_{n+2}.p_{b}}{\rho-\sigma_{b}}\Big) \Big(\sum_{a=1}^{n}
\dfrac{k_{n+1}.p_{a}}{\rho-\sigma_{a}}\Big)^{-1}
\Big(\sum_{a=1}^{n}
\dfrac{k_{n+2}.p_{a}}{\rho-\sigma_{a}}\Big)^{-1}
\non\\
&&
\Bigg[\Bigg\{\Big(\sum_{c=1}^{n}
\dfrac{\epsilon_{n+1}.p_{c}}{\rho-\sigma_{c}}\Big) \Big(\sum_{d=1}^{n}\dfrac{\epsilon_{n+2}.p_{d}}{\rho-\sigma_{d}}\Big)
+\epsilon_{n+1}\cdot \epsilon_{n+2} \ 
{1\over \, k_{n+1}.k_{n+2}} 
\Big(\sum_{a=1}^{n}
\dfrac{k_{n+1}.p_{a}}{\rho-\sigma_{a}}\Big)
\Big(\sum_{a=1}^{n}
\dfrac{k_{n+2}.p_{a}}{\rho-\sigma_{a}}\Big)\Bigg\}^2
 \non \\
&& - 2  \ \Bigg\{\Big(\sum_{c=1}^{n}
\dfrac{\epsilon_{n+1}.p_{c}}{\rho-\sigma_{c}}\Big) \Big(\sum_{d=1}^{n}\dfrac{\epsilon_{n+2}.p_{d}}{\rho-\sigma_{d}}\Big)
\Bigg\}  (k_{n+1}.k_{n+2})^{-1}  \non\\
&&\Big(\epsilon_{n+2}.k_{n+1} \sum_{c=1}^{n}
\dfrac{\epsilon_{n+1}.p_{c}}{\rho-\sigma_{c}} \sum_{b=1}^{n}\dfrac{k_{n+2}.p_{b}}
{\rho-\sigma_{b}}
+ \epsilon_{n+1}.k_{n+2} \ \sum_{c=1}^{n}
\dfrac{\epsilon_{n+2}.p_{c}}{\rho-\sigma_{c}} \
\sum_{b}^{n}\dfrac{k_{n+1}.p_{b}}{\rho-\sigma_{b}}\Big)
\non\\
&& + \epsilon_{n+1}\cdot \epsilon_{n+2} \   (k_{n+1}.k_{n+2})^{-2} \
\Big(\sum_{a=1}^{n}
\dfrac{k_{n+1}.p_{a}}{\rho-\sigma_{a}}\Big)
\Big(\sum_{a=1}^{n}
\dfrac{k_{n+2}.p_{a}}{\rho-\sigma_{a}}\Big)\non\\
&&\Big(\epsilon_{n+2}.k_{n+1} \sum_{c=1}^{n}
\dfrac{\epsilon_{n+1}.p_{c}}{\rho-\sigma_{c}}
- \epsilon_{n+1}.k_{n+2} \ \sum_{c=1}^{n}
\dfrac{\epsilon_{n+2}.p_{c}}{\rho-\sigma_{c}}\Big) \,  \left(
\sum_{b=1}^{n}\dfrac{k_{n+1}.p_{b}}{\rho-\sigma_{b}}-\sum_{b=1}^{n}\dfrac{k_{n+2}.p_{b}}
{\rho-\sigma_{b}}
\right) \non\\
&& - (k_{n+1}.k_{n+2})^{-2} \Big(\epsilon_{n+2}.k_{n+1} \sum_{c=1}^{n}
\dfrac{\epsilon_{n+1}.p_{c}}{\rho-\sigma_{c}}
- \epsilon_{n+1}.k_{n+2} \ \sum_{c=1}^{n}
\dfrac{\epsilon_{n+2}.p_{c}}{\rho-\sigma_{c}}\Big)^2 \sum_{b=1}^{n}\dfrac{k_{n+1}.p_{b}}
{\rho-\sigma_{b}}\sum_{b=1}^{n}\dfrac{k_{n+2}.p_{b}}
{\rho-\sigma_{b}} 
\Bigg]\, . \non\\
\ee
Note that there are many other ways of writing this expression -- by replacing one or more factors of
$\sum_{b=1}^n {k_{n+1}.p_b\over \rho-\sigma_b}$ by $-\sum_{b=1}^n {k_{n+2}.p_b\over \rho-\sigma_b}$ and
vice versa. We have chosen the one that will be most convenient for our analysis, but the final result does not depend on this choice.

We now represent the delta function as a contour integration as in \eqref{ea2} 
\be 
\int 
d\rho \
\delta\Big(\sum_{b=1}^{n}\dfrac{k_{n+1}.p_{b}}{\rho-\sigma_{b}}+\sum_{b=1}^{n}\dfrac{k_{n+2}.p_{b}}{\rho-\sigma_{b}}\Big)  \, \cdots
= \ointop d\rho \, \Big(\sum_{b=1}^{n}\dfrac{k_{n+1}.p_{b}}{\rho-\sigma_{b}}+\sum_{b=1}^{n}\dfrac{k_{n+2}.p_{b}}{\rho-\sigma_{b}}\Big)^{-1} \, \cdots
\ee
and
deform the $\rho$ integration contour away from the zeroes of the argument of the
delta function. This will generate three kinds of terms, the residues at $\rho=\sigma_a$
for $1\le a\le n$, residues at the zeroes of  
$\Big(\sum\limits_{a=1}^{n}
\dfrac{k_{n+1}.p_{a}}{\rho-\sigma_{a}}\Big)$ and
$\Big(\sum\limits_{a=1}^{n}
\dfrac{k_{n+2}.p_{a}}{\rho-\sigma_{a}}\Big)$ -- called $\{A_i\}$ and $\{B_i\}$ in 
\refb{edefaibi} --
and residue at $\infty$. We shall now 
analyze these three kinds of terms one by one.

\def\ve{\varepsilon}
\def\eps{\epsilon}

\medskip

\noindent{\bf Residue at $\sigma_a$}:  This gives
\be
\DD_1 &\equiv& - \tau^{-1} \, \BM_n\, \sum_{a=1}^n \{
p_a.(k_{n+1}+k_{n+2})\}^{-1}
(p_a.k_{n+1})^{-1}  (p_a.k_{n+2})^{-1} \non\\
&& \Bigg[  (k_{n+1}.k_{n+2}) \ \Big\{ \epsilon_{n+1}.p_a \ \epsilon_{n+2}.p_a + (k_{n+1}.k_{n+2})^{-1}
\epsilon_{n+1}.\epsilon_{n+2}\ k_{n+1}.p_a \ k_{n+2}.p_a \Big\}^2
\non\\ &&
- 2\ \epsilon_{n+1}.p_a \ \epsilon_{n+2}.p_a \Big\{
\epsilon_{n+2}.k_{n+1} \ \epsilon_{n+1}.p_a\ k_{n+2}.p_a
+  \epsilon_{n+1}.k_{n+2} \ \epsilon_{n+2}.p_a\ k_{n+1}.p_a
\Big\}
\non\\ &&  
- (k_{n+1}.k_{n+2})^{-1}
\epsilon_{n+1}.\epsilon_{n+2}\ k_{n+1}.p_a \ k_{n+2}.p_a  \non\\ &&
\Big\{
\epsilon_{n+2}.k_{n+1} \ \epsilon_{n+1}.p_a\ k_{n+2}.p_a
- \epsilon_{n+2}.k_{n+1} \ \epsilon_{n+1}.p_a\ k_{n+1}.p_a \non\\ &&
+  \epsilon_{n+1}.k_{n+2} \ \epsilon_{n+2}.p_a\ k_{n+1}.p_a
- \epsilon_{n+1}.k_{n+2} \ \epsilon_{n+2}.p_a\ k_{n+2}.p_a
\Big\}\non\\ &&
- (k_{n+1}.k_{n+2})^{-1} \ k_{n+1}.p_a \ k_{n+2}.p_a \ 
\{ \epsilon_{n+2}.k_{n+1} \ \epsilon_{n+1}.p_a - \epsilon_{n+1}.k_{n+2} \ \epsilon_{n+2}.p_a
\}^2 
\Bigg]\,
\ee
where the overall minus sign comes from the reversal of the integration contour.
Expanding the square and using the convention
\be
\varepsilon_{n+1} = \epsilon_{n+1} \otimes \epsilon_{n+1}, \quad
\varepsilon_{n+2} = \epsilon_{n+2} \otimes \epsilon_{n+2}
\ee
we get
\be \label{eadefdd1}
\DD_1 &=& \tau^{-1} \, \BM_n\, \sum_{a=1}^n \{
p_a.(k_{n+1}+k_{n+2})\}^{-1} \, \MM(p_a;\ve_{n+1}, k_{n+1}, \ve_{n+2}, k_{n+2})\, ,
\ee
where
\be \label{edefMM}
&& \MM(p_a;\ve_{n+1}, k_{n+1}, \ve_{n+2}, k_{n+2}) \non\\
&& \hskip -.2in = -
(p_a.k_{n+1})^{-1}  (p_a.k_{n+2})^{-1} 
\bigg[ k_{n+1}.k_{n+2}\ p_a.\ve_{n+1}.p_a \ 
p_a.\ve_{n+2}.p_a   + 2 \ p_a.\ve_{n+1}.\ve_{n+2}.p_a\ k_{n+1}.p_a \ k_{n+2}.p_a 
\non\\ && -2 \ p_a.\ve_{n+2}.k_{n+1} \ p_a.\ve_{n+1}.p_a\ k_{n+2}.p_a
-2 \ p_a.\ve_{n+1}.k_{n+2} \ p_a.\ve_{n+2}.p_a\ k_{n+1}.p_a 
\non\\ &&
+ (k_{n+1}.k_{n+2})^{-1} \ (k_{n+1}.p_a) \ (k_{n+2}.p_a)
\bigg\{\ve_{n+1}.\ve_{n+2}\ k_{n+1}.p_a \ k_{n+2}.p_a \non \\ &&
- \ p_a.\ve_{n+1}.\ve_{n+2}.k_{n+1} \ k_{n+2}.p_a
+ \ p_a.\ve_{n+1}.\ve_{n+2}.k_{n+1} \ k_{n+1}.p_a \non\\ &&
- \ p_a.\ve_{n+2}.\ve_{n+1}.k_{n+2} \ k_{n+1}.p_a
+ \ p_a.\ve_{n+2}.\ve_{n+1}.k_{n+2} \ k_{n+2}.p_a
\non \\ && - k_{n+1}.\ve_{n+2}.k_{n+1} \ p_a.\ve_{n+1}.p_a - 
k_{n+2}.\ve_{n+1}.k_{n+2} \ p_a.\ve_{n+2}.p_a
  + 2\   k_{n+2}.\ve_{n+1}.p_a \ k_{n+1}.\ve_{n+2}.p_a \ \bigg\}
\bigg]\, .\non\\
&&
\ee

\medskip

\noindent{\bf Residue at $\{A_i\}$ and $\{B_i\}$}: 
The contribution to \eqref{eammaster} from the residues at $A_i$ and $B_i$ defined
in \refb{edefaibi} are
given by
\be \label{e4.66}
\DD_2
&\equiv&  - \f{k_{n+1}.k_{n+2}}{\tau}\, \int D\sigma\, \delta^{(0)}\, I_n\, \ointop_{\{A_i\}}
d\rho \
\Big(\sum_{b=1}^{n}\dfrac{k_{n+2}.p_{b}}{\rho-\sigma_{b}}\Big)^{-1} \Big(\sum_{a=1}^{n}
\dfrac{k_{n+1}.p_{a}}{\rho-\sigma_{a}}\Big)^{-1}
\Big(\sum_{a=1}^{n}
\dfrac{k_{n+2}.p_{a}}{\rho-\sigma_{a}}\Big)^{-1}
\non\\
&&
\Bigg[\Bigg\{\Big(\sum_{c=1}^{n}
\dfrac{\epsilon_{n+1}.p_{c}}{\rho-\sigma_{c}}\Big) \Big(\sum_{d=1}^{n}\dfrac{\epsilon_{n+2}.p_{d}}{\rho-\sigma_{d}}\Big)
\Bigg\}^2
 \non \\
&& - 2  \ \Bigg\{\Big(\sum_{c=1}^{n}
\dfrac{\epsilon_{n+1}.p_{c}}{\rho-\sigma_{c}}\Big) \Big(\sum_{d=1}^{n}\dfrac{\epsilon_{n+2}.p_{d}}{\rho-\sigma_{d}}\Big)
\Bigg\}  (k_{n+1}.k_{n+2})^{-1} \epsilon_{n+2}.k_{n+1} \sum_{c=1}^{n}
\dfrac{\epsilon_{n+1}.p_{c}}{\rho-\sigma_{c}} \sum_{b=1}^{n}\dfrac{k_{n+2}.p_{b}}
{\rho-\sigma_{b}}
\Bigg] \non\\ &&
- \f{k_{n+1}.k_{n+2}}{\tau}\, \int D\sigma\, \delta^{(0)}\, I_n\, \ointop_{\{B_i\}}
d\rho \
\Big(\sum_{b=1}^{n}\dfrac{k_{n+1}.p_{b}}{\rho-\sigma_{b}}\Big)^{-1} \Big(\sum_{a=1}^{n}
\dfrac{k_{n+1}.p_{a}}{\rho-\sigma_{a}}\Big)^{-1}
\Big(\sum_{a=1}^{n}
\dfrac{k_{n+2}.p_{a}}{\rho-\sigma_{a}}\Big)^{-1}
\non\\
&&
\Bigg[\Bigg\{\Big(\sum_{c=1}^{n}
\dfrac{\epsilon_{n+1}.p_{c}}{\rho-\sigma_{c}}\Big) \Big(\sum_{d=1}^{n}\dfrac{\epsilon_{n+2}.p_{d}}{\rho-\sigma_{d}}\Big)
\Bigg\}^2
 \non \\
&& - 2  \ \Bigg\{\Big(\sum_{c=1}^{n}
\dfrac{\epsilon_{n+1}.p_{c}}{\rho-\sigma_{c}}\Big) \Big(\sum_{d=1}^{n}\dfrac{\epsilon_{n+2}.p_{d}}{\rho-\sigma_{d}}\Big)
\Bigg\}  (k_{n+1}.k_{n+2})^{-1} \epsilon_{n+1}.k_{n+2} \sum_{c=1}^{n}
\dfrac{\epsilon_{n+2}.p_{c}}{\rho-\sigma_{c}} \sum_{b=1}^{n}\dfrac{k_{n+1}.p_{b}}
{\rho-\sigma_{b}}
\Bigg]\, . \non\\ 
\ee
These terms were missed in the analysis of \cite{1607.02700}. They cancel similar contributions \eqref{eatotalnondeg}
coming from non-degenerate solutions.

\medskip

\noindent{\bf Residue at $\infty$}: Finally we can examine the residue 
of the integral \eqref{eammaster}
at $\rho=\infty$. This is given by
\be \label{efindeg}
&&\f{k_{n+1}.k_{n+2}}{\tau}\, \BM_n\,
\Big(\sum_{b=1}^{n} (k_{n+1}+k_{n+2}).p_{b}\Big)^{-1} \Big(\sum_{a=1}^{n}
k_{n+1}.p_{a}\Big)^{-1}
\Big(\sum_{a=1}^{n}
k_{n+2}.p_{a}\Big)^{-1}
\non\\
&&
\Bigg[\Bigg\{\Big(\sum_{c=1}^{n}
\epsilon_{n+1}.p_{c}\Big) \Big(\sum_{d=1}^{n}\epsilon_{n+2}.p_{d}\Big)
+ \epsilon_{n+1}\cdot \epsilon_{n+2} \ 
{1\over \, k_{n+1}.k_{n+2}} 
\Big(\sum_{a=1}^{n}
k_{n+1}.p_{a}\Big)
\Big(\sum_{a=1}^{n}
k_{n+2}.p_{a}\Big)\Bigg\}^2
 \non \\
&& - 2  \ \Bigg\{\Big(\sum_{c=1}^{n}
\epsilon_{n+1}.p_{c}\Big) \Big(\sum_{d=1}^{n}\epsilon_{n+2}.p_{d}\Big)
\Bigg\}  (k_{n+1}.k_{n+2})^{-1}  \non\\
&&\Big(\epsilon_{n+2}.k_{n+1} \sum_{c=1}^{n}
\epsilon_{n+1}.p_{c}\sum_{b=1}^{n}k_{n+2}.p_{b}+ \epsilon_{n+1}.k_{n+2} \ \sum_{c=1}^{n}
\epsilon_{n+2}.p_{c} \
\sum_{b}^{n}k_{n+1}.p_{b}\Big)
\non\\
&& + \epsilon_{n+1}\cdot \epsilon_{n+2} \   (k_{n+1}.k_{n+2})^{-2} \
\Big(\sum_{a=1}^{n}
k_{n+1}.p_{a}\Big)
\Big(\sum_{a=1}^{n}
k_{n+2}.p_{a}\Big)\non\\
&&\Big(\epsilon_{n+2}.k_{n+1} \sum_{c=1}^{n}
\epsilon_{n+1}.p_{c}
- \epsilon_{n+1}.k_{n+2} \ \sum_{c=1}^{n}
\epsilon_{n+2}.p_{c}\Big) \,  \left(
\sum_{b=1}^{n}k_{n+1}.p_{b}-\sum_{b=1}^{n}k_{n+2}.p_{b} \right) \non\\
&& - (k_{n+1}.k_{n+2})^{-2} \Big(\epsilon_{n+2}.k_{n+1} \sum_{c=1}^{n}
\epsilon_{n+1}.p_{c}
- \epsilon_{n+1}.k_{n+2} \ \sum_{c=1}^{n}
\epsilon_{n+2}.p_{c}\Big)^2 \sum_{b=1}^{n}k_{n+1}.p_{b}\sum_{b=1}^{n}k_{n+2}.p_{b}
\Bigg] \, .\non\\
\ee
Using momentum conservation law $\sum_{a=1}^n p_a = -\tau\, (k_{n+1}+k_{n+2})$ we can
see that this is of order $\tau$. Therefore this contribution is subsubleading and
can be ignored.

\subsection{Total contribution}
The full amplitude corresponding to two soft gravitons is obtained by adding the contributions \refb{eatotalnondeg} 
from the non-degenerate
solutions and the contributions \refb{eadefdd1} and 
\refb{e4.66} from the degenerate solutions. This is given by
\be \label{e4.68}
\BM_{n+2}&=& \tau^{-2}\, S^{(0)}_{n+1} \, S^{(0)}_{n+2} \BM_n + \tau^{-1}\,
S^{(0)}_{n+1} \, S^{(1)}_{n+2} \BM_n
+ \tau^{-1}\, S^{(0)}_{n+2} \, S^{(1)}_{n+1} \BM_n \non\\ &&
+ \tau^{-1} \, \BM_n\, \sum_{a=1}^n \{
p_a.(k_{n+1}+k_{n+2})\}^{-1} \MM(p_a; \ve_{n+1}, k_{n+1}, \ve_{n+2}, k_{n+2})\, .
\ee
This agrees with the result of \cite{1707.06803} reviewed in \refb{efullgenintro} 
for two soft gravitons. In \cite{1707.06803} it was also shown that
 \refb{e4.68}  is invariant under the gauge transformations of the
soft graviton polarizations.

Note that \refb{e4.66} cancels part of the contribution from \refb{eatotalnondeg}. This suggests
that there may be a better way of organising the calculation instead of representing it as a sum of contributions from
degenerate and non-degenerate solutions.


\sectiono{Multiple soft graviton theorem} \label{smultiple}

In this section we shall generalize the analysis of the previous section to the case where
arbitrary number of gravitons become soft. 

\subsection{Degenerate and non-degenerate solutions} \label{sdegnon}

We assume that there are $n+m$ number of gravitons and 
$m$ of them become soft. We parametrize the $m$ soft momenta as
\be
p_a^\mu=\tau\ k_a^\mu\quad,\qquad a=n+1,\cdots, n+m\, ,
\ee 
and take the soft limit by taking $\tau\to 0$ at fixed $k_a$. 
The momentum conservation now takes the form
\be
p_1^\mu+\cdots+p_n^\mu+\tau\, (k_{n+1}^\mu+\cdots +k_{n+m}^\mu)=0\, .
\ee
In this case 
the full solution space of the scattering equations gets 
divided into different sectors. 
These different sectors correspond to the case when 
a group of $r_1$ punctures associated with soft gravitons 
come within a distance of order $\tau$ of each other, 
another group of $r_2$ punctures associated with soft gravitons come 
within a distance of order $\tau$ of each other 
and so on. 
A detailed analysis of the number of solutions of each type can be found in
appendix \ref{sa}.
In this
section our goal will be to prove that for the subleading multiple soft graviton 
amplitude, 
only two sectors contribute -- non-degenerate solutions where 
all the punctures are finite distance away from each other and degenerate solutions
where two of the punctures come within a distance $\tau$ of each other and all other
punctures are finite distance away from each other.

To prove this, we note that the CHY formula for the amplitude is given by
\be
\BM_{n+m}=\int D\sigma  \prod_{q=1}^md\sigma_{n+q}\Big[ 
\prod_{a\neq i,j,k}\delta \Big( \sum_{\substack{b=1\\b\neq a}}^{n+m} 
\dfrac{p_{a}\cdot p_{b}}{\sigma_{a}-\sigma_{b}} \Big) \Big] \, I_{n+m} \, .
 \label{chy_multi}
\ee
We now analyze the product of delta functions and focus on the last $m$ of them corresponding to the soft gravitons. These $m$ delta functions are given by 
\be
\delta \Big(\tau \sum_{b=1}^{n} \dfrac{k_{n+q}\cdot p_{b}}{\sigma_{n+q}-\sigma_{b}}+\tau^2 
\sum_{\substack{u=1\\u\neq q}}^{m} \dfrac{k_{n+q}\cdot k_{n+u}}{\sigma_{n+q}-\sigma_{n+u}} \Big)\, ,
\quad 1\le q\le m\, .
\ee
Each of these delta functions gives a factor of $\f{1}{\tau}$ to the amplitude irrespective of how
many of the $m$ $\sigma_a$'s come within a distance of order $\tau$ of each other. 
These are the only source of singular $\tau$ dependence coming 
from the delta functions. 
Therefore we get a net factor of $\tau^{-m}$ from the delta functions.

Next, we consider the $\tau$ factors coming from the measure when $r$ of the $m$ 
$\sigma_a$'s
associated with the soft gravitons come within a distance of order $\tau$ of each
other. Without loss of generality we can label these as 
$\sigma_{n+m-r+q}$ for $1\le q\le r$.
We now make following redefinitions of the coordinates 
\be \label{earepar}
\sigma_{a}&=& \sigma_{a}' \quad \hbox{for $1\le a \le n+m-r+1$}\, ,\non\\
\sigma_{n+m-r+q}&=& \sigma_{n+m-r+1}'+\tau \, \xi_q\, , \qquad (q=2,\cdots, r)\, .
\ee
For the above coordinate transformation, we have
\be
\prod_{c=1}^{n+m} d\sigma_{c} \ \ =\tau^{r-1} \left(\prod_{c=1}^{n+m-r+1} d\sigma_{c}'
\right) \,
\left(\prod_{q=2}^r d\, \xi_q\right)\, .
\ee
We shall prove shortly that $I_{n+m}$ does not give rise to any 
singular behavior in the $\tau\rightarrow 0$ limit for finite $\sigma'_c,\xi_q$. 
Assuming this to be the case, we see that this contribution is leading only 
for $r=1$ and can receive subleading contribution only for 
$r= 1,2$. In other words, for the subleading soft graviton amplitude, 
the contributions come only from those 
solutions for which none of the punctures go 
close to each other (non degenerate solutions) or two of the punctures 
go close to each other (degenerate solutions). 

Let us now prove that $I_{n+m}$ has a finite limit as $\tau\to 0$ with $\sigma'_c$,
$\xi_q$ fixed (i.e., when $r$ punctures go close to each other). We define
\be
p\equiv n+m-r\, ,
\ee
and express $I_{n+m}$ as
\be
I_{n+m} = 4\, (-1)^{n+m} \, (\sigma_s-\sigma_t)^{-2} \, \hat P^2\, ,
\ee
where $\hat P$ is the Pfaffian of the matrix
\be \label{eamatrix2}
\renewcommand\arraystretch{1.3}
&& \hskip -.6in \left[
\begin{array}{c|c|c|c|c||c|c|c}
 (A_{p})_{a b} & \f{\tau p_a.k_{p+1}}{\sigma_a -\sigma_{p+1}} &\cdots& \f{\tau p_a.k_{p+r}}{\sigma_a -\sigma_{p+r}}& -(C^{T}_{p})_{ad}&
 -\f{\epsilon_{p+1}.p_a}{\sigma_{p+1}-\sigma_a}&\cdots& -\f{
 \epsilon_{p+r}.p_a}{\sigma_{p+r}-\sigma_a}\\
  \hline
\f{\tau k_{p+1}.p_{b}}{\sigma_{p+1} -\sigma_{b}} & 0  &\cdots& \f{\tau^2 k_{p+1}.k_{p+r}}{\sigma_{p+1} -\sigma_{p+r}}&
 -\f{\tau\epsilon_{d}.k_{p+1}}{\sigma_{d}-\sigma_{p+1}}&-C_{p+1, p+1}&\cdots&-\f{\tau\epsilon_{p+r}.k_{p+1}}{\sigma_{p+r}-\sigma_{p+1}}\\
  \hline  
  \vdots & \vdots & \cdots\vdots\cdots & \vdots &  \vdots & \vdots & \cdots\vdots\cdots & \vdots\\
  \hline
  \f{\tau k_{p+r}.p_{b}}{\sigma_{p+r} -\sigma_{b}} & \f{\tau^2 k_{p+r}.k_{p+1}}{\sigma_{p+r} -\sigma_{p+1}}&\cdots& 0
  &-\f{\tau\epsilon_{d}.k_{p+r}}{\sigma_{d}-\sigma_{p+r}}&-\f{\tau\epsilon_{p+1}.k_{p+r}}{\sigma_{p+1}-\sigma_{p+r}}&\cdots\vdots\cdots &-C_{p+r, p+r}\\
  \hline
 (C_{p})_{cb}& \f{\tau \epsilon_{c}.k_{p+1}}{\sigma_{c}-\sigma_{p+1}}&\cdots&\f{\tau\epsilon_{c}.k_{p+r}}{\sigma_{c}-\sigma_{p+r}}& (B_{p})_{cd}&
 \f{\epsilon_{c}.\epsilon_{p+1}}{\sigma_{c}-\sigma_{p+1}}&\cdots&\f{\epsilon_{c}.\epsilon_{p+r}}{\sigma_{c}-\sigma_{p+r}}\\
  \hline\hline
\f{\epsilon_{p+1}.p_{b}}{\sigma_{p+1} -\sigma_{b}} & C_{p+1,p+1}  &\cdots& \f{\tau \epsilon_{p+1}.k_{p+r}}{\sigma_{p+1} -\sigma_{p+r}}&
 \f{\epsilon_{p+1}.\epsilon_{d}}{\sigma_{p+1}-\sigma_{d}}&0&\cdots&\f{\epsilon_{p+1}.\epsilon_{p+r}}{\sigma_{p+1}-\sigma_{p+r}}\\
  \hline  
  \vdots & \vdots & \cdots\vdots\cdots & \vdots &  \vdots & \vdots & \cdots\vdots\cdots & \vdots\\
  \hline
  \f{\epsilon_{p+r}.p_{b}}{\sigma_{p+r} -\sigma_{b}} & \f{\tau \epsilon_{p+r}.k_{p+1}}{\sigma_{p+r} -\sigma_{p+1}}&\cdots& C_{p+r,p+r}
  &\f{\epsilon_{p+r}.\epsilon_{d}}{\sigma_{p+r}-\sigma_{d}}&\f{\epsilon_{p+r}.\epsilon_{p+1}}{\sigma_{p+r}-\sigma_{p+1}}
  &\cdots\vdots\cdots &0\\
\end{array}
\right]
\ee
where the matrices $A$, $B$ and $C$ are defined as in \eqref{eadefabc} for $m+n=p+r$
particles and $A_p, B_p$ and $C_p$ denote the first $p\times p$ 
blocks of these matrices.
Using the parametrization \refb{earepar}, the matrix \refb{eamatrix2} may be expressed as
\be \label{eamatrix1}
\renewcommand\arraystretch{1.3}
&& \hskip -.6in \left[
\begin{array}{c|c|c|c|c||c|c|c}
 (A_{p})_{a b} & \f{\tau p_a.k_{p+1}}{\sigma_a -\sigma_{p+1}} &\cdots& \f{\tau p_a.k_{p+r}}{\sigma_a -\sigma_{p+r}}& -(C^{T}_{p})_{ad}&
 -\f{\epsilon_{p+1}.p_a}{\sigma_{p+1}-\sigma_a}&\cdots& -\f{
 \epsilon_{p+r}.p_a}{\sigma_{p+r}-\sigma_a}\\
  \hline
\f{\tau k_{p+1}.p_{b}}{\sigma_{p+1} -\sigma_{b}} & 0  &\cdots& -
\f{\tau k_{p+1}.k_{p+r}}{\xi_r}&
 -\f{\tau\epsilon_{d}.k_{p+1}}{\sigma_{d}-\sigma_{p+1}}&-C_{p+1, p+1}&\cdots&-\f{\epsilon_{p+r}.k_{p+1}}{\xi_r}\\
  \hline  
  \vdots & \vdots & \cdots\vdots\cdots & \vdots &  \vdots & \vdots 
  & \cdots\vdots\cdots & \vdots\\
  \hline
  \f{\tau k_{p+r}.p_{b}}{\sigma_{p+r} -\sigma_{b}} & \f{\tau
  k_{p+r}.k_{p+1}}{\xi_r}&\cdots& 0
  &-\f{\tau\epsilon_{d}.k_{p+r}}{\sigma_{d}-\sigma_{p+r}}&
  \f{\epsilon_{p+1}.k_{p+r}}{\xi_r}
  &\cdots\vdots\cdots &-C_{p+r, p+r}\\
  \hline
 (C_{p})_{cb}& \f{\tau \epsilon_{c}.k_{p+1}}{\sigma_{c}-\sigma_{p+1}}&\cdots&\f{\tau\epsilon_{c}.k_{p+r}}{\sigma_{c}-\sigma_{p+r}}& (B_{p})_{cd}&
 \f{\epsilon_{c}.\epsilon_{p+1}}{\sigma_{c}-\sigma_{p+1}}&\cdots&
 \f{\epsilon_{c}.\epsilon_{p+r}}{\sigma_{c}-\sigma_{p+r}}\\
  \hline\hline
\f{\epsilon_{p+1}.p_{b}}{\sigma_{p+1} -\sigma_{b}} & C_{p+1,p+1}  
&\cdots& -\f{\epsilon_{p+1}.k_{p+r}}{\xi_r}&
 \f{\epsilon_{p+1}.\epsilon_{d}}{\sigma_{p+1}-\sigma_{d}}&0&\cdots&
 -\f{\epsilon_{p+1}.\epsilon_{p+r}}{\tau\xi_r}\\
  \hline  
  \vdots & \vdots & \cdots\vdots\cdots & \vdots &  \vdots & \vdots 
  & \cdots\vdots\cdots & \vdots\\
  \hline
  \f{\epsilon_{p+r}.p_{b}}{\sigma_{p+r} -\sigma_{b}} & 
  \f{\epsilon_{p+r}.k_{p+1}}{\xi_r}&\cdots& C_{p+r,p+r}
  &\f{\epsilon_{p+r}.\epsilon_{d}}{\sigma_{p+r}-\sigma_{d}}&
  \f{\epsilon_{p+r}.\epsilon_{p+1}}{\tau\xi_r}
  &\cdots\vdots\cdots &0\\
\end{array}
\right]
\ee

\medskip

We now note the following features of the matrix shown in \eqref{eamatrix1}:
\begin{enumerate}
 \item $(A_{p})_{ab}$, $(B_{p})_{ab}$ and $(C_{p})_{a b}$ 
 have finite $\tau\to 0$ limit. 
 \item There are four blocks of the matrix formed by the vertical double line and the horizontal double line. We note that 
 \begin{enumerate}
 \item The  upper left block  has $r$ rows and $r$ columns that are proportional to $\tau$. 
 \item The upper right and lower left blocks have finite
 $\tau\rightarrow 0$ limit. 
 \item In the lower right block, the diagonal matrix elements vanish, whereas 
 the off-diagonal matrix elements are proportional to $1/\tau$.
 \end{enumerate}
\end{enumerate}

We now note that the inverse powers of $\tau$ appear only in the elements of the lower 
right block of size $r\times r$.  Let us suppose first that $r$ is even. Then we get a maximum
contribution of $\tau^{-r/2}$ from the Pfaffian of the lower right block, 
and the coefficient of this term is proportional
to the Pfaffian of the upper left block of
\eqref{eamatrix1}. This matrix has $r$ rows and columns with every element proportional to
$\tau$, and therefore its Pfaffian will be proportional to $\tau^{r/2}$, cancelling the 
$\tau^{-r/2}$ factor. Therefore this term is finite as $\tau\to 0$.

If $r$ is odd, then the singular terms in the lower right block 
can give a maximum contribution of $\tau^{-(r-1)/2}$.
This is given by a sum of terms, one of which is proportional to the Pfaffian of the matrix given
by the lower right block of size $(r-1)\times (r-1)$, multiplied by the Pfaffian of the matrix
obtained by eliminating the last $(r-1)$ rows and columns of $\hat\Psi$. The other terms are
 related to this one
by rearrangement of the last $r$ rows and columns and can be analyzed similarly.
It is easy to see that the matrix obtained by eliminating the last $(r-1)$ rows and columns
of \refb{eamatrix1}
has $r$ rows and columns proportional to each other in the $\tau\to 0$ limit, 
and therefore its Pfaffian gives a factor of 
$\tau^{(r-1)/2}$. This cancels the $\tau^{-(r-1)/2}$ factor, giving a finite $\tau\to 0$
limit.

Next we need to consider the possibility that we may not choose the maximally singular 
terms from the lower right block. One such term corresponds to choosing the Pfaffian of the
matrix given by the last $k\times k$ block for some even integer $k<r$, multiplied by
the Pfaffian of the matrix obtained by eliminating the last $k$ rows and columns, but from
the latter we do not pick any term that has $1/\tau$ factor. The Pfaffian of the $k\times
k$ matrix goes as $\tau^{-k/2}$, whereas the matrix obtained by
eliminating the last $k$ rows and columns has $r$ rows (and columns) given by linear combinations
of $(r-k)$ independent vectors in the $\tau\to 0$ limit. Therefore its Pfaffian goes as $\tau^{k/2}$. 
This again shows that the
product has a finite $\tau\to 0$ limit. The other terms of this kind are related to the
one discussed above by rearrangement of the last $r$ 
rows and columns and are therefore also finite
as $\tau\to 0$.

This finishes our proof that $I_{n+m}$ remains finite in the $\tau\to 0$ limit.

\subsection{Contribution from non-degenerate solutions}
In this section we shall 
compute the contribution to the amplitude from non-degenerate solutions to the
scattering equations. The amplitude is given by
\be \label{eno}
&&\BM_{n+m}(p_{1},...,p_{n},\tau \,  k_{n+1},\cdots,\tau \,  k_{n+m})\non\\
&=& \int D\sigma\int \Big[\prod\limits_{q=1}^m d\sigma_{n+q} \Big] \Big[ \prod_{\substack{a=1\\a\neq i,j,k}}^{n}\delta \Big( \sum_{\substack{b=1\\b\neq
a}}^{n}\frac{p_{a}.p_{b}}{\sigma_{ab}}+\tau \,  \sum_{\substack{v=1}}^{m}\dfrac{p_{a}.k_{n+v}}{\sigma_{a(n+v)}} \Big)\Big]\non\\
&&\prod\limits_{r=1}^m\delta\Big( \tau \,  \sum_{b=1}^{n}\frac{k_{n+r}.p_{b}}{\sigma_{(n+r)b}}+\tau^{2}\, \sum_{\substack{u=1\\u\neq
r}}^{m}\dfrac{k_{n+r}.k_{n+u}}{\sigma_{(n+r)(n+u)}} \Big)\, I_{n+m}\, .
\ee
We now expand the various factors in this expression up to order $\tau$ assuming that
$(\sigma_a-\sigma_b)\sim 1$ for all pairs $a,b$ with $a\ne b$, $1\le a,b\le n+m$.
Expansion of the first $(n-3)$ delta functions in \refb{eno} takes the form:
\be
&&\prod_{\substack{a=1\\a\neq i,j,k}}^{n}\delta \Big( \sum_{\substack{b=1\\b\neq
a}}^{n}\frac{p_{a}.p_{b}}{\sigma_{ab}}+\tau \sum_{\substack{v=1}}^{m}
\dfrac{p_{a}.k_{n+v}}{\sigma_{a(n+v)}} \Big)\non\\
&=& \prod_{\substack{a=1\\a\neq i,j,k}}^{n}\delta \Big( 
\sum_{\substack{b=1\\b\neq
a}}^{n}\frac{p_{a}.p_{b}}{\sigma_{ab}}\Big)+\tau \sum_{l=1}^{n}
\sum_{\substack{v=1}}^{m}\dfrac{p_{l}.k_{n+v}}{\sigma_{l(n+v)}}
\delta' \Big( \sum_{\substack{c=1\\c\neq
l}}^{n}\frac{p_{l}.p_{c}}{\sigma_{lc}}\Big)\prod_{\substack{a=1\\a\neq i,j,k,l}}^{n}
\delta \Big( \sum_{\substack{b=1\\b\neq
a}}^{n}\frac{p_{a}.p_{b}}{\sigma_{ab}}\Big) + O(\tau^2)\non\\
&\equiv& \delta^{(0)}+\tau \, \delta^{(1)} + O(\tau^2)\, .
\ee
Expansion of the last $m$ delta functions in \refb{eno} takes the form:
\be
&&\prod\limits_{r=1}^m\delta\Big( \tau \sum_{b=1}^{n}\frac{k_{n+r}.p_{b}}{\sigma_{(n+r)b}}+\tau^{2}\sum_{\substack{u=1\\u\neq
r}}^{m}\dfrac{k_{n+r}.k_{n+u}}{\sigma_{(n+r)(n+u)}} \Big)\non\\
&=&\tau^{-m}\, 
\prod\limits_{r=1}^m \Big[\delta(f_{n+r}^{n})+\tau \sum_{\substack{u=1\\u\neq
r}}^{m}\frac{k_{n+r}.k_{n+u}}{\sigma_{(n+r)(n+u)}}\delta'(f_{n+r}^{n})
+O(\tau^{2})\Big]\non\\
&=&\tau^{-m}\, \left[\prod\limits_{r=1}^m\delta(f_{n+r}^{n})+\tau 
\sum_{\substack{v=1}}^{m}\sum_{\substack{u=1\\u\neq
v}}^{m}\frac{k_{n+v}.k_{n+u}}{\sigma_{(n+v)(n+u)}}\delta'(f_{n+v}^{n})\prod
\limits_{\substack{r=1\\r\not =v}}^m\delta(f_{n+r}^{n})+O(\tau^{2})\right]\, ,
\ee
where $f^n_{n+r}$ has been defined in \refb{eadefvar}.
Finally, expansion of the integrand $I_{n+m}$ gives, 
\be
I_{n+m}&=&I_{n+m}\Bigl|_{\tau=0}+\tau \, 
\f{\p I_{n+m}}{\p \tau}\Bigl|_{\tau=0}+O(\tau^{2})\, \, 
\equiv\,\, I_{n+m}^{(0)}+\tau \, I_{n+m}^{(1)}+O(\tau^{2})\, .
\ee
Then, up to subleading order, the non degenerate contribution becomes
\be \label{eeva1}
&& \f{1}{\tau^{m}}\int D\sigma \int \Big[\prod\limits_{q=1}^m d\sigma_{n+q}\Big] \big[\delta^{(0)}+\tau \, \delta^{(1)}\big] \, \Bigl[I_{n+m}^{(0)}+\tau \, I_{n+m}^{(1)}\Bigl]\non\\
&&\times \left[\prod\limits_{r=1}^m\delta(f_{n+r}^{n})+\tau \, 
\sum_{\substack{v=1}}^{m}\sum_{\substack{u=1\\u\neq
v}}^{m}\frac{k_{n+v}.k_{n+u}}{\sigma_{(n+v)(n+u)}}
\delta'(f_{n+v}^{n})\prod\limits_{\substack{r=1\\r\not =v}}^m\delta(f_{n+r}^{n})\right]\non\\
&=& \f{1}{\tau^{m}}\int D\sigma \, \delta^{(0)}\int 
\Big[\prod\limits_{q=1}^m d\sigma_{n+q}\Big]
\prod\limits_{r=1}^m\delta(f_{n+r}^{n})\ I_{n+m}^{(0)}\non\\
&&+\f{1}{\tau^{m-1}}\int D\sigma \int \Big[\prod\limits_{q=1}^m 
d\sigma_{n+q}\Big]\ \delta^{(1)}
\prod\limits_{r=1}^m\delta(f_{n+r}^{n})\, I_{n+m}^{(0)}\non\\
&&+\f{1}{\tau^{m-1}}\int D\sigma \ \delta^{(0)}\int 
\Big[\prod\limits_{q=1}^md\sigma_{n+q}\Big]
\prod\limits_{r=1}^m\delta(f_{n+r}^{n})\, I_{n+m}^{(1)}\non\\
&&+\f{1}{\tau^{m-1}}\int D\sigma  \, \delta^{(0)} \int 
\Big[\prod\limits_{q=1}^md\sigma_{n+q}\Big]
\sum_{\substack{v=1}}^{m}\sum_{\substack{u=1\\u\neq
v}}^{m}\frac{k_{n+v}.k_{n+u}}{\sigma_{(n+v)(n+u)}}\delta'(f_{n+v}^{n})
\prod\limits_{\substack{r=1\\r\not =v}}^m\delta(f_{n+r}^{n})\, I_{n+m}^{(0)} \, . \non\\
\label{nondegeneratemulti }
\ee

For evaluating $I^{(0)}_{n+m}$ and $I^{(1)}_{n+m}$
it will be convenient to introduce $m$ soft parameters $\tau_{1},\tau_{2},...,\tau_{m}$
generalizing the approach followed in \eqref{edefhatpsinew},  
and label the 
momenta of $(n+m)$ particles as $p_{1},p_{2},...,p_{n},
\tau_{1}\, k_{n+1},\tau_{2}\, k_{n+2},...,\tau_{m}\, k_{n+m}$. 
Then to linear order in $\tau_i$, $I_{n+m}$ is given by 
\be
I_{n+m}=4(-1)^{n+m}(\sigma_{s}-\sigma_{t})^{-2}\hat{P}^{2}
\ee
where $\hat{P}$ is the Pfaffian of the matrix

\be \label{multimatrix}
\hat{\Psi} = \left[\begin{array}{cccc|cccc}
A_{ab} &  {\tau_1 p_a.k_{n+1}\over \sigma_a - \sigma_{n+1}} 
&...
&  {\tau_m p_a.k_{n+m}\over \sigma_a - \sigma_{n+m}}
& - C^T_{ad}
& {p_a. \eps_{n+1}\over \sigma_a-\sigma_{n+1}}&...
& {p_a. \eps_{n+m}\over \sigma_a-\sigma_{n+m}} \cr 
{\tau_1 k_{n+1}.p_b\over
\sigma_{n+1}-\sigma_b} & 0 &...& 0 & {\tau_1 k_{n+1}.\eps_d\over \sigma_{n+1}-\sigma_d}
& - C_{(n+1)(n+1)}&...  & - {\tau_1 \eps_{n+m}. k_{n+1}\over \sigma_{n+m}
-\sigma_{n+1}}\cr
.&.&...&.&.&.&...&.\cr
.&.&...&.&.&.&...&.\cr
{\tau_m k_{n+m}.p_b\over
\sigma_{n+m}-\sigma_b} & 0 &...& 0 & {\tau_m k_{n+m}.\eps_d\over \sigma_{n+m}-\sigma_d}
& - {\tau_m \eps_{n+1}. k_{n+m}\over \sigma_{n+1}
-\sigma_{n+m}}&...& - C_{(n+m)(n+m)}  \cr
\hline
C_{cb} & {\tau_1 \eps_c.k_{n+1}\over \sigma_c - \sigma_{n+1}}&... 
& {\tau_m \eps_c.k_{n+m}\over \sigma_c - \sigma_{n+m}} & B_{cd} &
{\eps_c .\eps_{n+1}\over \sigma_{c} - \sigma_{n+1}} &...&
{\eps_c .\eps_{n+m}\over \sigma_c - \sigma_{n+m}} \cr
{\eps_{n+1}.p_b\over \sigma_{n+1}-\sigma_b} & C_{(n+1)(n+1)} &...& {\tau_m 
\eps_{n+1}.k_{n+m}\over \sigma_{n+1} - \sigma_{n+m}} & {\eps_{n+1}.\eps_d\over
\sigma_{n+1}-\sigma_{d}} & 0 &...& {\eps_{n+1} .\eps_{n+m}\over \sigma_{n+1} - \sigma_{n+m}} \cr
.&.&...&.&.&.&...&.\cr
.&.&...&.&.&.&...&.\cr
{\eps_{n+m}.p_b\over \sigma_{n+m}-\sigma_b} & {\tau_1 
\eps_{n+m}.k_{n+1}\over \sigma_{n+m} - \sigma_{n+1}}&... & C_{(n+m)(n+m)} & {\eps_{n+m}.\eps_d\over
\sigma_{n+m}-\sigma_{d}} & {\eps_{n+m} .\eps_{n+1}\over \sigma_{n+m} - \sigma_{n+1}}&... & 0 
\end{array} 
\right] \, .\non\\
\ee
The values of $C_{(n+r)(n+r)}$ for $r=1,2,...,m$ are:
\be \label{ecnrnr}
C_{(n+r)(n+r)}=-\sum_{a=1}^{n}\dfrac{\epsilon_{n+r}.p_{a}}{\sigma_{n+r}-\sigma_{a}}-\sum_{\substack{q=1\\q\neq r}}^{m}\tau_{q}\dfrac{\epsilon_{n+r}.k_{n+q}}{\sigma_{n+r}-\sigma_{n+q}}
\ee
With this, \refb{multimatrix} gives 
\be 
\hat P|_{\tau_i=0} = (-1)^{mn + m(m+1)/2} \, \tilde P \, 
\prod\limits_{q=1}^m \Big(\sum_{b=1}^{n}
\dfrac{\epsilon_{n+q}.p_{b}}{\sigma_{n+q}-\sigma_{b}}\Big)\, ,
\ee
where $\tilde P$ is the Pfaffian of \refb{multimatrix} for $m=0$. Therefore we have
\be
I_{n+m}^{(0)} =(-1)^m\prod\limits_{q=1}^m \Big(\sum_{b=1}^{n}\dfrac{\epsilon_{n+q}.p_{b}}{\sigma_{n+q}-\sigma_{b}}\Big)^{2}\ I_{n}\, .
\ee

To evaluate $I_{n+m}^{(1)}$ we use the analogue of \eqref{eadefin1new}:
\be \label{eainpm1}
I_{n+m}^{(1)}&\equiv & \sum_{r=1}^{m}\dfrac{\p I_{n+m}}{\p \tau_{r}}\Big|_{\tau_{1}=\tau_{2}=...=\tau_{m}=0}= 8(-1)^{n+m}(\sigma_{s}-\sigma_{t})^{-2}\hat{P}\sum_{r=1}^{m}\dfrac{\p \hat{P}}{\p \tau_{r}}\Big|_{\tau_{1}=\tau_{2}=...=\tau_{m}=0}\, . \non\\
\ee
The strategy for evaluating 
$\hat{P}\dfrac{\p \hat{P}}{\p \tau_{r}}\Big|_{\lbrace\tau_{q}=0\rbrace}$ will be to
first set $\tau_{q}=0$ for 
$q\ne r$ in the matrix \eqref{multimatrix}  and 
then expand the Pfaffian successively 
about the rows $(n+1)$ to $(n+m)$ except the $(n+r)$-th row. 
This gives
\be \label{epms}
\hat P = \pm \, \Big(\prod_{\substack{q=1\\q\neq r}}^{m}C_{(n+q)(n+q)}\Big)\hat P^{(n)}_{r}\, ,
\ee
where $\hat P^{(n)}_r$ is the Pfaffian of the matrix $\hat\Psi$ for $(n+1)$ graviton 
scattering with the first $n$ gravitons carrying momenta $p_1,\cdots , p_n$ and polarizations
$\ve_1,\cdots \, \ve_n$ and 
the last one carrying soft momentum $\tau_{r}k_{n+r}$ 
and polarization $\varepsilon_{n+r}$. The overall sign in \refb{epms} will not be needed
for our analysis. Using \refb{ecnrnr} and \refb{epms} we get
\be
\hat{P}\dfrac{\p \hat{P}}{\p \tau_{r}}\Big|_{\tau_{1}=...=\tau_{m}=0}
&=& \Big(\prod_{\substack{q=1\\q\neq r}}^{m}C_{(n+q)(n+q)}\Big)\Big|_{\tau_{1}=...=\tau_{m}=0}\hat P^{(n)}_r\dfrac{\p}{\p \tau_{r}}\Big[\Big(\prod_{\substack{q=1\\q\neq r}}^{m}C_{(n+q)(n+q)}\Big)\hat P^{(n)}_r\Big]\Big|_{\tau_{1}=...=\tau_{m}=0}\non\\
&=& \prod_{\substack{q=1\\q\neq r}}^{m}\Big(C_{(n+q)(n+q)}\Big)^{2}\hat P^{(n)}_r\dfrac{\p \hat P^{(n)}_r}{\p \tau_{r}}\Big|_{\tau_{1}=...=\tau_{m}=0}\non\\
&&+\tilde{P}^{2}\sum_{\substack{u=1\\u\neq r}}^{m}
\left[\dfrac{\epsilon_{n+u}.k_{n+r}}{\sigma_{n+u}-\sigma_{n+r}}\right]
\left[\sum_{a=1}^{n}\dfrac{\epsilon_{n+u}.p_{a}}{\sigma_{n+u}-\sigma_{a}}\right]
\prod_{\substack{q=1\\q\neq u}}^{m}\Big(C_{(n+q)(n+q)}\Big)^{2}\Big|_{\tau_{1}
=...=\tau_{m}=0}\, . \non\\
\ee
Hence the expression for $I_{n+m}^{(1)}$ becomes,
\be \label{earight}
I_{n+m}^{(1)}&=& 8\, (-1)^{n+m}(\sigma_{s}-\sigma_{t})^{-2}\sum_{r=1}^{m}
\prod_{\substack{q=1\\q\neq r}}^{m}\Big(C_{(n+q)(n+q)}\Big)^{2}\hat P^{(n)}_r
\dfrac{\p \hat P^{(n)}_r}{\p \tau_{r}}\Big|_{\tau_{1}=...=\tau_{m}=0}\non\\
&&+2\, (-1)^{m}I_{n}\sum_{r=1}^{m}\sum_{\substack{u=1\\u\neq r}}^{m}
\left[\dfrac{\epsilon_{n+u}.k_{n+r}}{\sigma_{n+u}-\sigma_{n+r}}\right]
\left[\sum_{a=1}^{n}\dfrac{\epsilon_{n+u}.p_{a}}{\sigma_{n+u}-\sigma_{a}}\right]
\prod_{\substack{q=1\\q\neq u}}^{m}\Big(C_{(n+q)(n+q)}\Big)^{2}\Big|_{\tau_{1}
=...=\tau_{m}=0}\, .
\non\\
\ee
We shall now proceed to evaluate the right hand side of \refb{eeva1}.

\medskip

\noindent{\bf First term:}

\medskip

The first term on the right hand side of \eqref{eeva1} may be expressed as
\be
\mathcal{F}_{0}&\equiv &  \f{1}{\tau^{m}}\int D\sigma \delta^{(0)}\int \Big[\prod\limits_{q=1}^m d\sigma_{n+q}\Big]
\prod\limits_{r=1}^m\delta(f_{n+r}^{n})\ I_{n+m}^{(0)}\non\\
&=& \dfrac{(-1)^{m}}{\tau^{m}}\int D\sigma \delta^{(0)}\ I_{n}\prod_{r=1}^{m}
\left\{\oint_{\lbrace A_{r,i} \rbrace} d\sigma_{n+r} \Big( \sum_{b=1}^{n}
\dfrac{k_{n+r}.p_{b}}{\sigma_{n+r}-\sigma_{b}} \Big)^{-1}
\Big( \sum_{d=1}^{n}\dfrac{\epsilon_{n+r}.p_{d}}{\sigma_{n+r}-\sigma_{d}} 
\Big)^{2}\right\}\, , \non \\
\ee
where for fixed $r$, $\lbrace A_{r,i} \rbrace$ are defined as the set of points satisfying
\be \label{edefari}
\sum_{b=1}^{n}\dfrac{k_{n+r}.p_{b}}{\sigma_{n+r}-\sigma_{b}} = 0  \hspace{15mm} at \hspace{5mm} \sigma_{n+r}=A_{r,i}\, .
\ee
We can now carry out integration over all $\lbrace \sigma_{n+r} \rbrace$ independently by deforming the contours to $\infty$. The contribution comes from residues at the poles at $\sigma_{a}$ -- which can be evaluated following the procedure described in section \ref{esfirst} -- and the contribution from $\infty$ can be evaluated after using
momentum conservation. The result up to subleading order is
\be
\mathcal{F}_{0}&=& \f{1}{\tau^{m}}\int D\sigma \delta^{(0)}\ I_{n}\ \prod_{r=1}^{m}\Big[\sum_{a=1}^{n}\dfrac{\big(\epsilon_{n+r}.p_{a}\big)^{2}}{k_{n+r}.p_{a}}+\tau\Big(\epsilon_{n+r}.\sum_{\substack{u=1 \\ u\neq r}}^{m}k_{n+u}\Big)^{2}\Big(k_{n+r}.\sum_{\substack{u=1 \\ u\neq r}}^{m}k_{n+u}\Big)^{-1}\Big]\non\\
&=& \f{1}{\tau^{m}}\Big[\prod_{r=1}^{m} S^{(0)}_{n+r}\Big]\BM_{n}+\f{1}{\tau^{m-1}}\sum_{v=1}^{m}\Big(\epsilon_{n+v}.\sum_{\substack{u=1 \\ u\neq v}}^{m}k_{n+u}\Big)^{2}\Big(k_{n+v}.\sum_{\substack{u=1 \\ u\neq v}}^{m}k_{n+u}\Big)^{-1}\Big[\prod_{\substack{r=1\\ r\neq v}}^{m} S^{(0)}_{n+r}\Big]\BM_{n}\, ,\non\\
\ee
where  $S^{(0)}_{n+r}$ has been defined in \refb{edefs0s1}.

\medskip

\noindent{\bf Second term}

\medskip

The second term on the right hand side of \eqref{eeva1} is given by
\be
\mathcal{F}_{1}&\equiv & \f{(-1)^{m}}{\tau^{m-1}}\int D\sigma 
\ I_{n}\ \sum_{l=1}^{n}\delta' \Big( \sum_{\substack{c=1\\c\neq
l}}^{n}\frac{p_{l}.p_{c}}{\sigma_{lc}}\Big)\prod_{\substack{a=1\\a\neq i,j,k,l}}^{n}\delta \Big( \sum_{\substack{b=1\\b\neq
a}}^{n}\frac{p_{a}.p_{b}}{\sigma_{ab}}\Big)\ \sum_{\substack{v=1}}^{m}\int 
\Big[\prod\limits_{q=1}^m d\sigma_{n+q}\Big]\dfrac{p_{l}.k_{n+v}}{\sigma_{l(n+v)}}\non\\
&& \Big[\prod\limits_{r=1}^m\, \delta
\Big(\sum_{e=1}^{n}\dfrac{k_{n+r}.p_{e}}{\sigma_{n+r}-\sigma_{e}}\Big)
\, \Big(\sum_{d=1}^{n}\dfrac{\epsilon_{n+r}.p_{d}}{\sigma_{n+r}-
\sigma_{d}}\Big)^{2} \Big]
\, .
\ee
The integrals over $\lbrace\sigma_{n+q}\rbrace$ for $q\neq v$ are of the form \eqref{ealead1} and can be analyzed as in section \ref{esfirst} to produce factors of $-S^{(0)}_{n+q}$.
The contribution from the pole at $\sigma_{n+q}=\infty$ is subsubleading and can
be ignored.
The remaining  
integration over $\sigma_{n+v}$ has the form given in \eqref{ea2first} and can 
be analyzed as in section \ref{essecond}. The final result, given in \refb{ea2second}, 
can in turn be related, using 
the equality of \refb{ea2second}, \refb{e3.42} and the first term on the right hand
side of \refb{e3.41}, to
\be
S^{(1)}_{n+v} \, \BM_n - \int D\sigma\, \delta^{(0)}\, S^{(1)}_{n+v} \, I^{(0)}\, ,
\ee
where $S^{(1)}_{n+v}$ has been defined in \refb{edefs0s1}.
Then the result up to subleading order is
\be
\mathcal{F}_{1}&=& \f{1}{\tau^{m-1}}\sum_{v=1}^{m}\Big[ \ \big[\prod_{\substack{r=1\\r\neq v}}^{m}S^{(0)}_{n+r}\big]\Big(S^{(1)}_{n+v}\, \BM_{n}-
\int D\sigma \, \delta^{(0)}\, S_{n+v}^{(1)}\, I_{n}\Big) \Big]\non\\
&=& \f{1}{\tau^{m-1}}\sum_{v=1}^{m} \ \big[\prod_{\substack{r=1\\r\neq v}}^{m}
S^{(0)}_{n+r}\big]\Big(S^{(1)}_{n+v}\, \BM_{n}\Big)-\f{1}{\tau^{m-1}}
\sum_{v=1}^{m} \ \big[\prod_{\substack{r=1\\r\neq v}}^{m}S^{(0)}_{n+r}\big]
\int D\sigma \, \delta^{(0)}\, S_{n+v}^{(1)}\, 
I_{n}
\, . \non\\
\ee

\medskip

\noindent{\bf Third term:} 

\medskip

We shall now consider the third term in the right hand side of \eqref{nondegeneratemulti }:
\be
\f{1}{\tau^{m-1}}\int D\sigma \ \delta^{(0)}\int \Big[\prod\limits_{q=1}^md\sigma_{n+q}\Big]
\prod\limits_{r=1}^m\delta(f_{n+r}^{n})\, I_{n+m}^{(1)}\, . \label{multi3}
\ee
Substituting the first term of $I_{n+m}^{(1)}$ given in \refb{earight}
into \eqref{multi3} and taking the sum 
over $r$ out of the integration we get,
\be \label{eneweq}
&& {1\over \tau^{m-1}}\, \sum_{r=1}^m \int D\sigma\, \delta^{(0)})
\left\{\prod_{q=1\atop q\ne r}^m \int dq_{n+q} \, \delta(f^n_{n+q})\, (C_{(n+q)(n+q)})^{2}
\right\} \non \\
&& \hskip 1in 
\int d\sigma_{n+r} \, \delta(f^n_{n+r}) \, 8\, (-1)^{m+n} \, (\sigma_s-\sigma_t)^{-2}\, 
\hat P^{(n)}_r {\p \hat P^{(n)}_r\over \p\tau_r}\bigg|_{\tau_1=\cdots =\tau_m=0}\, .
\ee
We can easily see that the integration over each
$\sigma_{n+q}$ for $q\ne r$ generates a factor of 
$-S_{n+q}^{(0)}$. On the other hand integration over 
$\sigma_{n+r}$ has exactly the structure of third term \refb{eadefa3} of 
single soft graviton case
with $I^{(1)}_{n+1}$ given in \refb{eadefin1}.
Using the equality of \refb{eadefa3}, \refb{efinA3}, \refb{eA3equiv} and
the second term on the right hand side of
\eqref{ang},  the expression
\refb{eneweq} can be written as:
\be
\mathcal{F}_{2}\equiv \f{1}{\tau^{m-1}}\sum_{r=1}^{m}\Big(\prod_{\substack{q=1\\q\neq r}}^{m}S_{n+q}^{(0)}\Big)\int D\sigma\, \delta^{(0)}\, S_{n+r}^{(1)}\, I_{n}\, .
\ee

On the other hand, substituting the second term on the right hand side of \eqref{earight}  into
\eqref{multi3} we get:
\be \label{eleft2}
&&2\f{(-1)^{m}}{\tau^{m-1}}\int D\sigma \delta^{(0)}I_{n}\sum_{r=1}^{m}\sum_{\substack{u=1\\u\neq r}}^{m}\prod_{\substack{q=1\\q\neq u}}^{m}\oint_{\{A_{q,i}\}} d\sigma_{n+q}\Big(\sum_{c=1}^{n}
\dfrac{k_{n+q}.p_{c}}{\sigma_{n+q}-\sigma_{c}}\Big)^{-1}
C_{(n+q)(n+q)}^{2}\non\\
&&\oint_{\lbrace A_{u,i}\rbrace} d\sigma_{n+u}\Big(\sum_{b=1}^{n}\dfrac{k_{n+u}.p_{b}}{\sigma_{n+u}-\sigma_{b}}\Big)^{-1}\Big(\dfrac{\epsilon_{n+u}.k_{n+r}}{\sigma_{n+u}-\sigma_{n+r}}\Big)\Big(\sum_{a=1}^{n}\dfrac{\epsilon_{n+u}.p_{a}}{\sigma_{n+u}-\sigma_{a}}\Big)\, .
\ee
We shall evaluate the integration over $\sigma_{n+u}$ by 
deforming it to $\infty$. 
During this deformation we shall encounter poles at $\sigma_{n+u}=\sigma_{n+r}$  for
$1\le r\le m$, $r\ne u$, 
and at $\sigma_{n+u}=\infty$ but there are no poles 
at $\lbrace\sigma_{a}\rbrace$ for $1\le a\le n$. 
The contribution to \eqref{eleft2} from the residue at the pole at
$\{\sigma_{n+r}: 1\le r\le m\}$ is given by:
\be
\mathcal{F}_{3}&\equiv & 2\f{(-1)^{m-1}}{\tau^{m-1}}\int 
D\sigma \delta^{(0)}I_{n}\sum_{r=1}^{m}\sum_{\substack{u=1\\u\neq r}}^{m}
(\epsilon_{n+u}.k_{n+r})\prod_{\substack{q=1\\q\neq u}}^{m}
\oint_{\{A_{q,i}\}} d\sigma_{n+q}\Big(\sum_{c=1}^{n}
\dfrac{k_{n+q}.p_{c}}{\sigma_{n+q}-\sigma_{c}}\Big)^{-1}\non\\
&&\Big(\sum_{d=1}^{n}\dfrac{\epsilon_{n+q}.p_{d}}{\sigma_{n+q}-
\sigma_{d}}\Big)^{2}
\Big(\sum_{b=1}^{n}\dfrac{k_{n+u}.p_{b}}{\sigma_{n+r}-\sigma_{b}}\Big)^{-1}
\Big(\sum_{a=1}^{n}\dfrac{\epsilon_{n+u}.p_{a}}{\sigma_{n+r}-\sigma_{a}}\Big)\non\\
&=& -\f{2}{\tau^{m-1}}\sum_{r=1}^{m}\sum_{\substack{u=1\\u\neq r}}^{m}
(\epsilon_{n+u}.k_{n+r})\Big[\prod_{\substack{q=1\\q\neq r,u}}^{m}S_{n+q}^{(0)}
\Big]\int D\sigma \delta^{(0)}I_{n} \ \oint_{\{A_{r,i}\}} d\sigma_{n+r}\Big(\sum_{c=1}^{n}
\dfrac{k_{n+r}.p_{c}}{\sigma_{n+r}-\sigma_{c}}\Big)^{-1}\non\\
&&\Big(\sum_{d=1}^{n}\dfrac{\epsilon_{n+r}.p_{d}}{\sigma_{n+r}-\sigma_{d}}\Big)^{2}
\Big(\sum_{b=1}^{n}\dfrac{k_{n+u}.p_{b}}{\sigma_{n+r}-\sigma_{b}}\Big)^{-1}
\Big(\sum_{a=1}^{n}\dfrac{\epsilon_{n+u}.p_{a}}{\sigma_{n+r}-\sigma_{a}}\Big)\, .
\ee
On the other hand the contribution to \eqref{eleft2} from the residue at the 
pole at $\sigma_{n+u}=\infty$ is given by:
\be
\mathcal{F}_{4}&\equiv & -\dfrac{2}{\tau^{m-1}}\BM_{n}
\sum_{\substack{u=1}}^{m}\Big(\prod_{\substack{q=1\\q\neq u}}^{m}
S_{n+q}^{(0)}\Big)\Big(\epsilon_{n+u}.
\sum_{\substack{r=1\\r\neq u}}^{m}k_{n+r}\Big)^{2}\Big(k_{n+u}.
\sum_{\substack{v=1\\v\neq u}}^{m}k_{n+v}\Big)^{-1}\, .
\ee

\medskip

\noindent {\bf Fourth term:}

\medskip

The fourth term on the right hand side of \eqref{nondegeneratemulti } is given by:
\be
&& \f{1}{\tau^{m-1}}\int D\sigma \, \delta^{(0)} \int \Big[\prod\limits_{q=1}^md\sigma_{n+q}\Big]
\sum_{\substack{v=1}}^{m}\sum_{\substack{u=1\\u\neq
v}}^{m}\frac{k_{n+v}.k_{n+u}}{\sigma_{(n+v)(n+u)}}\delta'(f_{n+v}^{n})\prod
\limits_{\substack{r=1\\r\not =v}}^m\delta(f_{n+r}^{n})I_{n+m}^{(0)}\non\\
&=& \f{(-1)^{m-1}}{\tau^{m-1}}\int D\sigma \delta^{(0)}\ I_{n}\  \sum_{v=1}^{m}
\sum_{\substack{u=1\\ u\neq v}}^{m}\prod_{\substack{r=1\\r\neq v}}^{m}\oint_{\{A_{r,i}\}}\,
 d\sigma_{n+r} \Big( \sum_{b=1}^{n}\dfrac{k_{n+r}.p_{b}}{\sigma_{n+r}-\sigma_{b}}
 \Big)^{-1} \, \Big(\sum_{d=1}^{n}\dfrac{\epsilon_{n+r}.p_{d}}{\sigma_{n+r}-\sigma_{d}}
 \Big)^{2}\non\\
&&\Big[\oint_{\lbrace A_{v,i}\rbrace} d\sigma_{n+v}\dfrac{k_{n+v}.k_{n+u}}{\sigma_{n+v}-\sigma_{n+u}}\Big(\sum_{a=1}^{n}\dfrac{k_{n+v}.p_{a}}{\sigma_{n+v}-\sigma_{a}}\Big)^{-2}\Big(\sum_{c=1}^{n}\dfrac{\epsilon_{n+v}
.p_{c}}{\sigma_{n+v}-\sigma_{c}}\Big)^{2} \Big]\, ,
\ee
where in the last line we have used the definition of derivative of delta function inside the contour integration as $\delta'(f)=-\f{1}{f^{2}}$. For evaluating the integral over $\sigma_{n+v}$, we shall deform the contour away from $\lbrace A_{v,i}\rbrace$ towards the infinity and pick the residues at $\sigma_{n+v}=\sigma_{n+u}$ as well as the contribution from infinity. There is no pole at $\sigma_{n+v}=\sigma_{a}$. The contribution from the pole at $\sigma_{n+u}$ is given by
\be \label{edefF5}
\mathcal{F}_{5}&\equiv & \f{(-1)^{m}}{\tau^{m-1}}\int D\sigma \delta^{(0)}\ 
I_{n}\  \sum_{v=1}^{m}\sum_{\substack{u=1\\ u\neq v}}^{m}
\prod_{\substack{r=1\\r\neq v}}^{m}\oint_{\lbrace A_{r,i}\rbrace} 
d\sigma_{n+r} \, \Big( \sum_{b=1}^{n}\dfrac{k_{n+r}.p_{b}}
{\sigma_{n+r}-\sigma_{b}}\Big)^{-1} \Big(\sum_{d=1}^{n}
\dfrac{\epsilon_{n+r}.p_{d}}{\sigma_{n+r}-\sigma_{d}}\Big)^{2}\non\\
&&\Big(\sum_{a=1}^{n}\dfrac{k_{n+v}.p_{a}}{\sigma_{n+u}-\sigma_{a}}\Big)^{-2}
\Big(\sum_{c=1}^{n}\dfrac{\epsilon_{n+v}
.p_{c}}{\sigma_{n+u}-\sigma_{c}}\Big)^{2} k_{n+v}.k_{n+u}\non\\
&=& \f{1}{\tau^{m-1}} \sum_{v=1}^{m}\sum_{\substack{u=1\\ u\neq v}}^{m}\Big[\prod_{\substack{r=1\\r\neq u,v}}^{m}S_{n+r}^{(0)}\Big]\int D\sigma
 \delta^{(0)} I_{n} \ \oint_{\lbrace A_{u,i}\rbrace} d\sigma_{n+u}
 \Big( \sum_{b=1}^{n}\dfrac{k_{n+u}.p_{b}}{\sigma_{n+u}-\sigma_{b}}
 \Big)^{-1}\Big(\sum_{d=1}^{n}\dfrac{\epsilon_{n+u}.p_{d}}
 {\sigma_{n+u}-\sigma_{d}}\Big)^{2}\non\\
&&\Big(\sum_{a=1}^{n}\dfrac{k_{n+v}.p_{a}}{\sigma_{n+u}-\sigma_{a}}\Big)^{-2}
\Big(\sum_{c=1}^{n}\dfrac{\epsilon_{n+v}
.p_{c}}{\sigma_{n+u}-\sigma_{c}}\Big)^{2} k_{n+v}.k_{n+u}\, .
\ee
On the other hand the contribution from infinity is given by, after using momentum conservation,
\be
\mathcal{F}_{6}&\equiv & \f{(-1)^{m-1}}{\tau^{m-1}}\int D\sigma \, \delta^{(0)}\ 
I_{n}\ \sum_{v=1}^{m}\prod_{\substack{r=1\\r\neq v}}^{m}\oint_{\lbrace A_{r,i}\rbrace} 
d\sigma_{n+r}\, \Big( \sum_{b=1}^{n}\dfrac{k_{n+r}.p_{b}}{\sigma_{n+r}-\sigma_{b}}\Big)^{-1}\Big(\sum_{d=1}^{n}\dfrac{\epsilon_{n+r}.p_{d}}{\sigma_{n+r}-\sigma_{d}}
\Big)^{2} \non\\
&&\big( k_{n+v}.\sum_{\substack{u=1\\u\neq v}}^{m}k_{n+u}\big)^{-1}
\big(\epsilon_{n+v}.\sum_{\substack{u=1\\u\neq v}}^{m} k_{n+u}\big)^{2}\non\\
&=& \f{1}{\tau^{m-1}}\sum_{v=1}^{m}\big( k_{n+v}.\sum_{\substack{u=1\\u
\neq v}}^{m}k_{n+u}\big)^{-1}\big(\epsilon_{n+v}.\sum_{\substack{u=1\\u\neq v}}^{m} 
k_{n+u}\big)^{2}\prod_{\substack{r=1\\r\neq v}}^{m}S^{(0)}_{n+r}\, \BM_{n}\, .
\ee
Adding the contributions from $\mathcal{F}_{0}$ to $\mathcal{F}_{6}$ we get the total contribution from non-degenerate solutions
\be \label{eficont}
\mathcal{F}&=&  \tau^{-m}\Big[\prod_{r=1}^{m} S^{(0)}_{n+r}\Big]\BM_{n}
+\tau^{-(m-1)}\sum_{v=1}^{m} \ \big[\prod_{\substack{r=1\\r\neq v}}^{m}
S^{(0)}_{n+r}\big]\Big(S^{(1)}_{n+v}\BM_{n}\Big)\non\\
&&-\f{2}{\tau^{m-1}}\sum_{r=1}^{m}\sum_{\substack{u=1\\u\neq r}}^{m}
(\epsilon_{n+u}.k_{n+r})\Big[\prod_{\substack{q=1\\q\neq r,u}}^{m}S_{n+q}^{(0)}\Big]
\int D\sigma \delta^{(0)} I_{n}\ 
\oint_{\{A_{r,i}\}} d\sigma_{n+r}\Big(\sum_{c=1}^{n}\dfrac{k_{n+r}.p_{c}}
{\sigma_{n+r}-\sigma_{c}}\Big)^{-1}\non\\
&&\Big(\sum_{d=1}^{n}\dfrac{\epsilon_{n+r}.p_{d}}{\sigma_{n+r}-\sigma_{d}}\Big)^{2}
\Big(\sum_{b=1}^{n}\dfrac{k_{n+u}.p_{b}}{\sigma_{n+r}-\sigma_{b}}\Big)^{-1}
\Big(\sum_{a=1}^{n}\dfrac{\epsilon_{n+u}.p_{a}}{\sigma_{n+r}-\sigma_{a}}\Big)
\non\\
&&+\f{1}{\tau^{m-1}} \sum_{r=1}^{m}\sum_{\substack{u=1\\ u\neq r}}^{m}
(k_{n+r}.k_{n+u})\Big[\prod_{\substack{q=1\\q\neq u,r}}^{m}S_{n+q}^{(0)}\Big]
\int D\sigma \delta^{(0)} I_{n}\non\\ &&
 \oint_{\lbrace A_{u,i}\rbrace} d\sigma_{n+u}
\Big( \sum_{b=1}^{n}\dfrac{k_{n+u}.p_{b}}{\sigma_{n+u}-\sigma_{b}}\Big)^{-1}
\Big(\sum_{d=1}^{n}\dfrac{\epsilon_{n+u}.p_{d}}{\sigma_{n+u}-\sigma_{d}}\Big)^{2}
\Big(\sum_{a=1}^{n}\dfrac{k_{n+r}.p_{a}}{\sigma_{n+u}-\sigma_{a}}\Big)^{-2}\Big
(\sum_{c=1}^{n}\dfrac{\epsilon_{n+r}
.p_{c}}{\sigma_{n+u}-\sigma_{c}}\Big)^{2} \, , \non\\
\ee
where in the last line we have relabelled the dummy indices from what they were in
\refb{edefF5}.

\subsection{Contribution from degenerate solutions}
We now evaluate the contribution to the amplitude 
from the degenerate solutions 
of the scattering equation.  As shown in section \ref{sdegnon}, we only have to consider the case
where two of the punctures come close to each other, and that this contribution begins at the
subleading order.
We can choose the two punctures out of $m$ punctures in ${m\choose 2}$ ways. We shall consider the contribution of two generic punctures $\sigma_{n+p}$ and $\sigma_{n+u}$ coming close to each other and then sum over all possible ${m\choose 2}$ terms.

When $|\sigma_{n+p}-\sigma_{n+u}|\sim \tau$, it is convenient to go to the $(\rho,\xi)$ coordinate system as in the double soft graviton case:
\be
\sigma_{n+p}=\rho -{\xi\over 2}, \qquad \sigma_{n+u}= \rho +{\xi\over 2}\, .
\ee
In terms of these variables, we have 
\be
\int d\sigma_{n+p}\, d\sigma_{n+u} \, \delta(f^n_{n+p}) \, \delta(f^n_{n+u})= -2\, \int d\rho \,
d\xi \, \delta(f^n_{n+p}+f^n_{n+u})\, \delta(f^n_{n+p}-f^n_{n+u})\, ,
\ee
with the understanding that $\rho$ integration needs to be done using the first delta function 
and $\xi$ integration using the second delta function. Therefore, 
the contribution of this solution to the amplitude becomes, to subleading order in $\tau$:
\be
\BM_{n+m}^{(p,u)}&=& -\f{2}{\tau^{m}}\int D\sigma \, \delta^{(0)}\prod
\limits_{\substack{r=1\\r\not=p,u}}^m d\sigma_{n+r} \,
\delta(f^n_{n+r})\int \, d\rho \, d\xi \,
\delta\bigg(\sum_{b=1}^n\dfrac{k_{n+p}.p_{b}}
{\rho-\f{\xi}{2}-\sigma_{b}}+\sum_{b=1}^n
\dfrac{k_{n+u}.p_{b}}{\rho+\f{\xi}{2}-\sigma_{b}}\bigg)\non\\
&& \delta\bigg(\sum_{b=1}^n\dfrac{k_{n+p}.p_{b}}
{\rho-\f{\xi}{2}-\sigma_{b}} - \sum_{b=1}^n\dfrac{k_{n+u}.p_{b}}
{\rho+\f{\xi}{2}-\sigma_{b}}-\dfrac{2\tau k_{n+p}.k_{n+u}}{\xi}\bigg)
\ I_{n+m} \non\\
&\simeq &-\f{2}{\tau^{m-1}}\int D\sigma \, \delta^{(0)} 
\prod\limits_{\substack{r=1\\r\not=p,u}}^m
d\sigma_{n+r} \, \delta(f^n_{n+r}) \, \int d\rho\ 
\delta\bigg(\sum_{b=1}^n\dfrac{k_{n+p}.p_{b}}{\rho-\f{\xi}{2}-\sigma_{b}}
+\sum_{b=1}^n\dfrac{k_{n+u}.p_{b}}{\rho+\f{\xi}{2}-\sigma_{b}}\bigg) \non \\ &&
\hskip 1in 
{\xi_1^2 \over  2 \, k_{n+p}.k_{n+u}}\, I_{n+m}\, ,
\ee
where in the second step we have explicitly performed the 
$\xi$ integration using the
second delta function, and
\be
\xi_1 \equiv 2 k_{n+p}.k_{n+u} \,  \left(
\sum_{b=1}^n\dfrac{k_{n+p}.p_{b}}{\rho-\sigma_{b}}-
\sum_{b=1}^n\dfrac{k_{n+u}.p_{b}}
{\rho-\sigma_{b}}
\right)^{-1}  \, .
\ee
Now
in the degeneration limit we can evaluate $I_{n+m}$ by regarding this as the 
square of the 
Pfaffian of the matrix $\hat\Psi$ given in \eqref{eamatrix2}. In computing the 
Pfaffian, it is convenient to first shift the $(n+p)^{th}$ and $(n+u)^{th}$ rows and 
columns to the $(n+m-1)^{th}$ and $(n+m)^{th}$ positions and also the
$(2n+m+p)^{th}$ and $(2n+m+u)^{th}$ rows and 
columns to the $(2n+2m-1)^{th}$ and $(2n+2m)^{th}$ positions,
and then evaluate the Pfaffian. Now, if we exchange two rows (say $\ell$ and $k$) 
and the corresponding columns ($\ell$ and $k$) , then the value of Pfaffian changes 
by a sign. However one can easily see that the combined effect of all the movements is to
not generate any sign.
Now the problem effectively reduces to that of computing 
$I_{n+m}$ for two soft gravitons, and using \refb{e4.58} we get,
\begin{eqnarray}
I_{n+m} &=& \Bigg[(\xi_1)^{-2} \epsilon_{n+p}\cdot \epsilon_{n+u} \ k_{n+p}\cdot k_{n+u} 
- (\xi_1)^{-2} \epsilon_{n+p}\cdot k_{n+u} \ \epsilon_{n+u}\cdot k_{n+p} 
\nonumber \\
&& \hskip 1in - C_{n+p,n+p} \, C_{n+u,n+u}
\bigg]^2  \, I_{n+m-2} + {\cal O}(\tau) \nonumber \\ 
&=& \Biggl[(\xi_1)^{-2} \epsilon_{n+p}\cdot \epsilon_{n+u} \ k_{n+p}\cdot k_{n+u} 
- (\xi_1)^{-2} \epsilon_{n+p}\cdot k_{n+u} \ \epsilon_{n+u}\cdot k_{n+p} 
\nonumber \\ 
&& \hskip -.2in - \bigg\{\sum_{c=1}^n {\epsilon_{n+p}\cdot p_c\over 
\rho -\sigma_c} - (\xi_1)^{-1} \, \epsilon_{n+p} \cdot k_{n+u} \bigg\}
\ \bigg\{\sum_{c=1}^n {\epsilon_{n+u}\cdot p_c\over 
\rho -\sigma_c} 
+ (\xi_1)^{-1}  \epsilon_{n+u}\cdot k_{n+p}\bigg\}
 \Biggl]^2  \, I_{n+m-2} \nonumber \\
&& + {\cal O}(\tau) \, ,
\end{eqnarray}
where $I_{n+m-2}$ is the integrand for the scattering amplitude of $(n+m-2)$ gravitons in which
we have removed the $s$-th and the $u$-th gravitons from the original set.
Therefore, to the desired order of expansion, 
the contribution to the $(n+m)$-point amplitude from the degenerate solution 
becomes,
\be \label{e5.46}
&& \BM_{n+m}^{(p,u)} \non\\
&&= -\f{4k_{n+p}.k_{n+u}}{\tau^{m-1}}
\int D\sigma\,  \delta^{(0)}\prod\limits_{\substack{r=1\\r\not=p,u}}^md\sigma_{n+r}\delta(f^n_{n+r})I_{n+m-2}\int 
d\rho \
\delta\Big(\sum_{b=1}^n\dfrac{k_{n+p}.p_{b}}{\rho-\sigma_{b}}+\sum_{b=1}^n\dfrac{k_{n+u}.p_{b}}{\rho-\sigma_{b}}\Big) \non\\
&&\Big(\sum_{a=1}^n
\dfrac{(k_{n+p}-k_{n+u}).p_{a}}{\rho-\sigma_{a}}\Big)^{-2}
\Bigg[\Big(\sum_{c=1}^n
\dfrac{\epsilon_{n+p}.p_{c}}{\rho-\sigma_{c}}\Big) \Big(\sum_{d=1}^n\dfrac{\epsilon_{n+u}.p_{d}}{\rho-\sigma_{d}}\Big)
-\ 
{ \epsilon_{n+p}\cdot \epsilon_{n+u}\over 4 \, k_{n+p}.k_{n+u}} 
\Big(\sum_{a=1}^n
\dfrac{(k_{n+p}-k_{n+u}).p_{a}}{\rho-\sigma_{a}}\Big)^{2}
 \non \\
&& + \Big(\epsilon_{n+u}.k_{n+p} \sum_{c=1}^n
\dfrac{\epsilon_{n+p}.p_{c}}{\rho-\sigma_{c}}
- \epsilon_{n+p}.k_{n+u} \ \sum_{c=1}^n
\dfrac{\epsilon_{n+u}.p_{c}}{\rho-\sigma_{c}}\Big) {1\over 2 k_{n+p}.k_{n+u}} \,  \left(
\sum_{b=1}^n\dfrac{k_{n+p}.p_{b}}{\rho-\sigma_{b}}-\sum_{b=1}^n\dfrac{k_{n+u}.p_{b}}
{\rho-\sigma_{b}}
\right)
\Bigg]^2\, . \non\\
\ee
In the above expression, we are only interested in the $\tau$ independent contribution of $I_{n+m-2}$. Hence, we can evaluate it at $\tau=0$. In terms of $I_n$, it is given by 
\be
I_{n+m-2}=(-1)^{m-2} \, \prod\limits_{\substack{r=1\\r\not=p,u}}^m \Big(\sum_{\substack{b=1}}^{n}\dfrac{\epsilon_{n+r}.p_{b}}{\sigma_{n+r}-\sigma_{b}}\Big)^2\ I_{n} \ +O(\tau)\, .
\ee
The integration over $\sigma_{n+r}$ for $r\ne p,u$ in \refb{e5.46} can now be performed by
the standard contour deformation, producing a factor of 
$(-1)^{m-2}\prod_{r=1,r\ne p,u}^m
S^{(0)}_{n+r}$. The remaining integral over $\rho$ has exactly the same form as \refb{e4.59}
and can be analyzed as in section \ref{s4.3}. Using \refb{eadefdd1}, 
\refb{e4.66} we see that
the final result for \refb{e5.46} is given by a sum of two terms:
\be \label{esecont}
\BM^{(p,u)(1)}_{n+m} &=& \tau^{-m+1} \prod\limits_{\substack{r=1\\r\not=p,u}}^m
S^{(0)}_{n+r} \, \sum_{\substack{a=1}}^{n} \{
p_a.(k_{n+p}+k_{n+u})\}^{-1} \, \MM(p_a; \ve_{n+p}, k_{n+p}, \ve_{n+u}, k_{n+u})\, 
\BM_n\, , \non\\
\ee
and 
\be \label{ethcont}
&& \BM^{(p,u)(2)}_{n+m} \non\\
&&
=- \f{k_{n+p}.k_{n+u}}{\tau^{m-1}}\int D\sigma \delta^{(0)}
\bigg(\prod\limits_{\substack{r=1\\r\not=p,u}}^m S^{(0)}_{n+r}\bigg)I_{n}\ointop_{\{A_{s,i}\}}
d\rho \
\Big(\sum_{b=1}^n\dfrac{k_{n+u}.p_{b}}{\rho-\sigma_{b}}\Big)^{-2} \Big(\sum_{a=1}^n
\dfrac{k_{n+p}.p_{a}}{\rho-\sigma_{a}}\Big)^{-1}  \non\\ && \hskip 2in
\Bigg\{\Big(\sum_{c=1}^n
\dfrac{\epsilon_{n+p}.p_{c}}{\rho-\sigma_{c}}\Big) \Big(\sum_{d=1}^n\dfrac{\epsilon_{n+u}.p_{d}}{\rho-\sigma_{d}}\Big)
\Bigg\}^2
 \non \\ && 
 - \f{k_{n+p}.k_{n+u}}{\tau^{m-1}}\int D\sigma \delta^{(0)}
 \bigg(\prod\limits_{\substack{r=1\\r\not=p,u}}^m S^{(0)}_{n+r}\bigg)I_{n}
 \ointop_{\{A_{u,i}\}}
d\rho \
\Big(\sum_{b=1}^n\dfrac{k_{n+u}.p_{b}}{\rho-\sigma_{b}}\Big)^{-1} 
\Big(\sum_{a=1}^n
\dfrac{k_{n+p}.p_{a}}{\rho-\sigma_{a}}\Big)^{-2} \non\\ && \hskip2in
\Bigg\{\Big(\sum_{c=1}^n
\dfrac{\epsilon_{n+p}.p_{c}}{\rho-\sigma_{c}}\Big) 
\Big(\sum_{d=1}^n\dfrac{\epsilon_{n+u}.p_{d}}{\rho-\sigma_{d}}\Big)
\Bigg\}^2
 \non \\
 && 
 + {2\over \tau^{m-1}}\int D\sigma \delta^{(0)}
\bigg(\prod\limits_{\substack{r=1\\r\not=p,u}}^m S^{(0)}_{n+r}\bigg)I_{n}
\ointop_{\{A_{s,i}\}}
d\rho \ \Big(\sum_{b=1}^n\dfrac{k_{n+u}.p_{b}}{\rho-\sigma_{b}}\Big)^{-1} \Big(\sum_{a=1}^n
\dfrac{k_{n+p}.p_{a}}{\rho-\sigma_{a}}\Big)^{-1} \non\\ && \hskip 2in
\Bigg\{\Big(\sum_{c=1}^n
\dfrac{\epsilon_{n+p}.p_{c}}{\rho-\sigma_{c}}\Big)^2 \Big(\sum_{d=1}^n\dfrac{\epsilon_{n+u}.p_{d}}{\rho-\sigma_{d}}\Big)
\Bigg\}  \epsilon_{n+u}.k_{n+p}  \non\\ &&
+ {2\over \tau^{m-1}}\int D\sigma \delta^{(0)}
\bigg(\prod\limits_{\substack{r=1\\r\not=p,u}}^m S^{(0)}_{n+r}\bigg)I_{n} \ointop_{\{A_{u,i}\}}
d\rho \ \Big(\sum_{b=1}^n\dfrac{k_{n+u}.p_{b}}{\rho-\sigma_{b}}\Big)^{-1} \Big(\sum_{a=1}^n
\dfrac{k_{n+p}.p_{a}}{\rho-\sigma_{a}}\Big)^{-1}  \non\\ && \hskip 2in
\Bigg\{\Big(\sum_{c=1}^n
\dfrac{\epsilon_{n+p}.p_{c}}{\rho-\sigma_{c}}\Big) \Big(\sum_{d=1}^n\dfrac{\epsilon_{n+u}.p_{d}}{\rho-\sigma_{d}}\Big)^2
\Bigg\}  \epsilon_{n+p}.k_{n+u} \, . \ee
$\MM$ is the same function as defined in \refb{edefMM}.

\subsection{Total contribution}

We can now add the expressions given in \eqref{esecont} and \eqref{ethcont}, sum over all
possible choices $p,u$ in the range $1\le p<u\le m$, and add this to
\eqref{eficont}
to get the final result.  The result is
\be \label{eafinform}
\BM_{n+m}&=& \tau^{-m}\Big[\prod_{r=1}^{m} S^{(0)}_{n+r}\Big]\BM_{n}
+\tau^{-(m-1)}\sum_{v=1}^{m} \ \big[\prod_{\substack{r=1\\r\neq v}}^{m}
S^{(0)}_{n+r}\big]\Big(S^{(1)}_{n+v}\BM_{n}\Big) \non\\ && \hskip -.5in
+\tau^{-(m-1)} \sum_{p,u=1\atop p<u}^m
\bigg(\prod\limits_{\substack{r=1\\r\not=p,u}}^m S^{(0)}_{n+r}\bigg)
\sum_{\substack{a=1\\a\not=p,u}}^{n+m} \{
p_a.(k_{n+p}+k_{n+u})\}^{-1} \, \MM(p_a; \ve_{n+p}, k_{n+p}, \ve_{n+u}, k_{n+u})\, 
\BM_n\, . \non \\
\ee
This agrees with the result of \cite{1707.06803} reviewed in \refb{efullgenintro}.

\bigskip

\noindent{\bf Acknowledgments:}  
We would like to thank Anirban Basu, Freddy Cachazo, Rajesh Gopakumar,
Dileep Jatkar, Alok Laddha, 
Satchitananda Naik
and Arnab Priya Saha
for discussions. 
The work of SC, SPK and MV was supported
in part by Infosys Scholarship for Senior Students.
The work of AS was supported
in part by the J.C.~Bose fellowship of the Department of Science and Technology,
India.
The work of MV was also supported in part by the SPM fellowship of CSIR. 
We also thank the people and Government of India for their continuous support for theoretical physics.

\appendix
\sectiono{Counting different solutions for $m$ soft gravitons} \label{sa}

In this appendix, we shall analyze the solutions of the scattering equations for the case when 
there are $m$ number of soft gravitons. For multiple soft gravitons, the solutions to the 
scattering equation fall into different classes. A given class corresponds to the case 
when a group of $r_1$ punctures 
carrying soft momenta come within a distance of order 
$\tau$ of each other, 
another group of $r_2$ punctures carrying soft momenta come 
within a distance of order $\tau$ of each other 
and so on. We shall now derive the number of solutions of the scattering equations 
for this situation. The scattering equations for the first $n$ gravitons (which are 
finite energy gravitons) 
is given by
\be
\sum_{\substack{b=1{}\\b\not=a}}^{n}\f{p_a\cdot p_b}{\sigma_a-\sigma_b}
+\tau \sum\limits_{v=1}^m\f{p_a\cdot k_{n+v}}{\sigma_a-\sigma_{n+v}}=0
\quad\qquad (a=1,\cdots,n)\, .\label{1stset}
\ee
The scattering equations for the gravitons in the $i$-th group $\SSS_i$
containing $r_i$ punctures that come with a distance of order 
$\tau$ can be written as
\be
\sum_{\substack{b=1}}^{n}\f{k_{n+u}\cdot p_b}{\sigma_{n+u}-\sigma_b}+\tau
\sum_{\substack{v\in \SSS_i\\v\not=u}}\f{k_{n+u}\cdot k_{n+v}}{\sigma_{n+u}-\sigma_{n+v}}
+\tau
\sum_{\substack{v=1\\v\not\in \SSS_i}}^m
\f{k_{n+u}\cdot k_{n+v}}{\sigma_{n+u}-\sigma_{n+v}}=0 \, ,
\qquad \forall u\in \SSS_i\, ,\label{3rdset}
\ee
where we have removed an overall factor of $\tau$ from the equation.

In the $\tau\rightarrow 0$ limit, we can ignore the second term in the 
left hand side of  \eqref{1stset}. The equation \eqref{1stset} then 
reduces to the scattering equations of $n$ finite energy particles and 
hence the number of solutions for the set $(\sigma_1,\cdots,\sigma_n)$ is $(n-3)!$. 
To obtain the number of solutions for the rest of the punctures, we note that
the first two terms in \eqref{3rdset} are of order unity  since 
the denominators of the second term is of the order $\tau$, while the
third term is of order $\tau$. Hence, we can ignore the third term. 
With this, the scattering equation for the punctures in each group $\SSS_i$
decouple from each other and we can focus on solutions of the 
punctures of each group separately given the solution for the set 
$(\sigma_1,\cdots,\sigma_n)$. 

\vspace*{.06in}For ease of notation, we consider the case of first set $\SSS_1$ 
and assume that its punctures are labelled as $(\sigma_{n+1},\cdots,\sigma_{n+r_1})$. 
We redefine the variables as
\be
\sigma_{n+a}&=& \sigma_{n+1}+\tau \, \xi_a \, ,\qquad a=2,\cdots, r_1\, .\label{redefvar}
\ee
In the $\tau\rightarrow0$ limit, all the $r_1$ punctures $(\sigma_{n+1},\cdots,
\sigma_{n+r_1})$ come within a distance of order $\tau$ of each other, and 
therefore $\xi_a\sim 1$. 
Our goal now is to find out how many solutions exist for the variables 
$(\sigma_{n+1},\xi_2,\cdots,\xi_{r_1})$. 
We shall now prove that the number of solutions for this set is $(n-2)(r_1-1)!$. 
To do this, we write the scattering equations 
for the group $\SSS_1$ in terms of the variables in \eqref{redefvar} to leading order in $\tau$:
\be
\sum_{b=1}^{n}\f{ k_{n+1}\cdot p_b}{\sigma_{n+1}-\sigma_b}- \f{k_{n+1}\cdot k_{n+2}}{\xi_2}-\cdots- \f{k_{n+1}\cdot k_{n+r_1}}{\xi_{r_1}}&=&0 \label{soft1eq}\\
\sum_{b=1}^{n}\f{ k_{n+2}\cdot p_b}{\sigma_{n+1}-\sigma_b}+ \f{k_{n+2}\cdot k_{n+1}}{\xi_2}+\f{k_{n+2}\cdot k_{n+3}}{\xi_2-\xi_3}+\cdots+ \f{k_{n+2}\cdot k_{n+r_1}}{\xi_{2}-\xi_{r_1}}&=&0\label{soft2eq}\\
\hspace*{-4in}&\vdots&\non\\[.8cm]
\hspace*{-.3in}\sum_{b=1}^{n}\f{ k_{n+r_1-1}\cdot p_b}{\sigma_{n+1}-\sigma_b}+ \f{k_{n+r_1-1}\cdot k_{n+1}}{\xi_{r_1-1}}+\cdots+ \f{k_{n+r_1-1}\cdot k_{n+r_1-2}}{\xi_{r_1-1}-\xi_{r_1-2}}+ \f{k_{n+r_1-1}\cdot k_{n+r_1}}{\xi_{r_1-1}-\xi_{r_1}}&=&0\label{soft4eq}\\
\sum_{b=1}^{n}\f{ k_{n+r_1}\cdot p_b}{\sigma_{n+1}-\sigma_b}+ \f{k_{n+r_1}\cdot k_{n+1}}{\xi_{r_1}}+\cdots+ \f{k_{n+r_1}\cdot k_{n+r_1-1}}{\xi_{r_1}-\xi_{r_1-1}}&=&0\label{soft3eq}
\ee
Adding all the equations and using momentum conservation, we find that 
all the $\xi_a$'s drop 
out of the
equation and we get  a polynomial equation of degree $n-2$ for $\sigma_{n+1}$. 
Hence it has $n-2$ solutions. We can now recursively prove that the number of 
solutions for the set $(\xi_2,\cdots,\xi_{r_1})$ is $(r_1-1)!$. We consider a situation 
in which the last momentum $k_{n+r_1}$ is softer than the other $r_1-1$ soft momenta. 
We then replace $k_{n+r_1}$ by $\omega\, \ell_{n+r_1}$ for a new soft parameter 
$\omega$ and fixed $\ell_{n+r_1}$, and  take the limit $\omega\rightarrow0$.\footnote{This 
trick was used in \cite{1306.6575,1609.00008} for counting the number of solutions to the 
scattering equations.}
In this limit,  
the terms involving $k_{n+r_1}$ in   \eqref{soft1eq}-\eqref{soft4eq}
can be ignored and therefore they reduce to 
 the analog of eqs.\refb{soft1eq}-\refb{soft3eq}  for a cluster of $(r_1-1)$
punctures. These by assumption have $(r_1-2)!$ solutions for the variables 
$(\xi_2,\cdots,\xi_{r_1-1})$. On the other hand, after we factor out
an overall factor of $\omega$ from
the last equation \refb{soft3eq}, it reduces to a polynomial equation 
for $\xi_{r_1}$ of degree $r_1-1$ for a given solution of 
$(\sigma_1,\cdots,\sigma_{n+1},\xi_2,\cdots,\xi_{r-1})$.  
Therefore $\xi_{r_1}$ has $(r_1-1)$ solutions. 
Thus, there are a total of $(r_1-1)!$ solutions for the set $(\xi_2,\cdots,\xi_{r})$. For finite $\omega$
the solutions will change, but we expect their number to remain the same. 
Since for fixed $\sigma_{1},
\cdots \sigma_n$, $\sigma_{n+1}$ has $(n-2)$ solutions, 
this proves our claim that the number of solutions for the set 
$(\sigma_{n+1},\cdots,\sigma_{n+r_1})$ is $(n-2)(r_1-1)!$.

\vspace*{.06in}We can repeat the analysis for 
the remaining group of punctures with the result that the number of solutions for the punctures in the  group $\SSS_i$
 is $(n-2)(r_i-1)!$. The total number of solutions is, thus, given by
\be
(n-3)!\prod\limits_{i}[(r_i-1)!(n-2)]
\ee
where the product runs over all clusters containing punctures carrying soft momentum.

To see that it gives the correct number of total solutions $(n+m-3)!$ for the $n+m$ number of gravitons, we need to sum over all the possibilities. This is given by,
\be
\mathcal{N} &\equiv& (n-3)!\sum_{\substack{\lbrace m_{r}\rbrace}\atop\sum_{r=1}^{m}r\, m_{r}=m
}\Big\lbrace \prod_{r=1}^{m} \big[ (r-1)!(n-2) \big]^{m_{r}} \Big\rbrace \dfrac{m!}{\prod_{r=1}^{m}(r!)^{m_{r}}m_{r}!} \label{sumN}
\ee
where $m_r$ is the multiplicity for the $r$ punctures coming together. 
The factor $\dfrac{m!}{\prod_{r=1}^{m}(r!)^{m_{r}}m_{r}!}$ is the number of ways of 
dividing $m$ objects into groups of $r$ objects with multiplicity $m_r$. 
To evaluate the constrained sum in \eqref{sumN}, it is 
convenient to define the following unconstrained sum (generating function)
\be
\mathcal{N}(y)&\equiv & (n-3)!\prod_{r=1}^{\infty}\sum_{m_{r}=0}^{\infty}\Big\lbrace \big[ (r-1)!(n-2) \big]^{m_{r}}\dfrac{m!}{(r!)^{m_{r}}m_{r}!} y^{rm_{r}} \Big\rbrace\non\\
&=& m!(n-3)!\prod_{r=1}^{\infty}\exp\Big[\dfrac{(n-2)}{r} y^{r}\Big]\non\\
&=& m!(n-3)!(1-y)^{-(n-2)}\non\\
&=&m!(n-3)!\sum\limits_{k=0}^\infty \f{(n+k-3)!}{k!(n-3)!}y^k\, .
\label{generatingfunction}
\ee
The coefficient of $y^m$ in  \eqref{generatingfunction} is the quantity in \eqref{sumN}. 
This coefficient is precisely $(n+m-3)!$ -- the expected number of solutions of
the scattering equations for $n+m$ particles. This shows that we have not left out any 
solutions.

\end{document}